\newif\ifarxiv
\arxivtrue

\newif\ifcomments
\ifarxiv
\commentsfalse
\else
\commentsfalse
\fi

\newif\ifappendixincluded
\ifarxiv
\appendixincludedtrue
\else
\appendixincludedfalse
\fi

\ifarxiv
\documentclass[acmsmall,nonacm,screen]{acmart}
\else
\documentclass[acmsmall]{acmart}
\fi

\setcitestyle{nosort}

\usepackage{booktabs}   %
\usepackage{subcaption} %

\newcommand{\Hvar}{x}

\definecolor{jade}{HTML}{00a86b}
\definecolor{emerald}{HTML}{045307}
\newcommand{\keyword}[1]{{\color{purple}{\textlog{#1}}}}

\newcommand{\Hifelse}[3]{\keyword{if}~#1~\keyword{then}~#2~\keyword{else}~#3}

\newcommand{\Halloc}[2]{\keyword{alloc}~#1~#2}
\newcommand{\Hload}[1]{\mathord{\textbf{!}}#1}
\newcommand{\Hstore}[2]{#1 \leftarrow #2}

\newcommand{\Hfree}[2]{\keyword{free}~#1~#2}

\newcommand{\Hfndef}[2]{\keyword{fn}~\mathsf{#1}(#2)\eqdef}
\newcommand{\Hcall}[2]{\mathsf{#1}(#2)}
\newcommand{\Hlocal}[2]{\keyword{local}~#1[#2]}

\newcommand{\Hlet}[2]{\keyword{let}~#1 := #2~\keyword{in}}

\newcommand{\Hid}[1]{\mathsf{#1}}

\ifcomments
\usepackage[draft]{fixme} %
\else
\usepackage[final]{fixme} %
\fi

\usepackage[utf8]{inputenc}
\usepackage[english]{babel}
\usepackage{xparse}
\usepackage{xspace}
\usepackage{thmtools}
\usepackage{xcolor}
\usepackage{mathpartir}
\usepackage[]{microtype}
\usepackage{multirow}
\usepackage{makecell}
\usepackage{marvosym}
\usepackage{wasysym}
\usepackage{pifont}
\usepackage{mathtools}
\usepackage{stmaryrd}
\usepackage{scalerel}
\usepackage{tensor}
\usepackage{xifthen}
\usepackage{iris}
\usepackage{pftools}
\usepackage{soul}
\usepackage{tikz}
\usepackage{minted}
\usepackage{threeparttable}
\usepackage{tabularx}
\usepackage[most]{tcolorbox}
\usepackage{fontawesome5}
\usepackage{enumitem}
\usetikzlibrary{calc, shapes, arrows, automata, patterns, backgrounds, tikzmark, positioning}

\makeatletter
\addto\extrasenglish{%
  \renewcommand*\chapterautorefname{\S\@gobble}
  \renewcommand*\sectionautorefname{\S\@gobble}
  \renewcommand*\subsectionautorefname{\S\@gobble}
  \renewcommand*\appendixautorefname{\S\@gobble}
  }
\makeatother
\newcommand*{\myeqref}[2][]{%
  \hyperref[{#2}]{#1(\ref*{#2})}%
}

\renewcommand*{\lineref}[1]{\hyperref[#1]{line~\ref*{#1}}}
\newcommand*{\linerangeref}[2]{\hyperref[#1]{lines~\ref*{#1}}\hyperref[#2]{-\ref*{#2}}}
\newcommand{\mypageref}[1]{\hyperref[#1]{page~\pageref*{#1}}}
\newcommand*{\stepref}[1]{\hyperref[#1]{line~\ref*{#1}}}

\newcommand{\appendixref}{\ifappendixincluded\autoref{sec:appendix:rec}, \autoref{sec:appendix:asm}, and \autoref{sec:appendix:pseudo}\else the appendix~\cite{Appendix}\fi}
\newcommand{\appendixrefshort}{\ifappendixincluded\autoref{sec:appendix:rec}--\autoref{sec:appendix:pseudo}\else\cite{Appendix}\fi}

\declaretheorem[name=Definition,style=definition]{definition}
\declaretheorem[name=Theorem,sibling=definition]{theorem}
\setminted{escapeinside=||,mathescape=true, linenos=true, numbersep=5pt, framesep=2mm, fontsize=\footnotesize}

\DeclareUnicodeCharacter{2264}{$\le$}
\DeclareUnicodeCharacter{2208}{$\in$}
\DeclareUnicodeCharacter{2203}{$\exists$}
\DeclareUnicodeCharacter{25C1}{$\triangleleft$}
\DeclareUnicodeCharacter{2097}{${}_l$}
\DeclareUnicodeCharacter{2260}{$\neq$}
\DeclareUnicodeCharacter{2205}{$\emptyset$}
\DeclareUnicodeCharacter{222A}{$\cup$}
\DeclareUnicodeCharacter{228E}{$\uplus$}
\DeclareUnicodeCharacter{2200}{$\forall$}
\DeclareUnicodeCharacter{25A1}{$\always$}
\DeclareUnicodeCharacter{03B3}{$\gamma$}
\DeclareUnicodeCharacter{03BB}{$\lambda$}
\DeclareUnicodeCharacter{2286}{$\subseteq$}
\DeclareUnicodeCharacter{03A0}{$\Pi$}
\DeclareUnicodeCharacter{2217}{$\ast$}

\definecolor{airforceblue}{rgb}{0.36, 0.54, 0.66}
\definecolor{brickred}{rgb}{0.8, 0.25, 0.33}
\definecolor{ao}{rgb}{0.0, 0.0, 1.0}
\definecolor{cobalt}{rgb}{0.0, 0.28, 0.67}
\definecolor{darkergreen}{rgb}{0,0.7,0.3}
\definecolor{magenta}{rgb}{1.0,0.0,1.0}

\FXRegisterAuthor{al}{aal}{AL}%
\FXRegisterAuthor{ms}{ams}{MS}%
\FXRegisterAuthor{nm}{anm}{NM}%

\makeatletter
\providecommand*{\Dashv}{%
  \mathrel{%
    \mathpalette\@Dashv\vDash
  }%
}
\newcommand*{\@Dashv}[2]{%
  \reflectbox{$\m@th#1#2$}%
}
\makeatother
\makeatletter %
\def\arcr{\@arraycr}
\makeatother

\newcommand\ie{\emph{i.e.}, }
\newcommand\eg{\emph{e.g.}, }

\newcommand{\Repeat}[1]{\overline{#1}}

\definecolor{stringcolor}{HTML}{BA2121}

\makeatletter
\def\@parfont{\bfseries\itshape}
\makeatother

\RenewDocumentCommand \hoare {m m m m O{}}{
	\curlybracket{#1}\spac#2\spac@\spac#3\spac\curlybracket{#4}_{#5}%
}
\NewDocumentCommand \tgthoare {m m m m O{}}{
	\curlybracket{#1}\spac#2\spac@^{\textlog{t}}\spac#3\spac\curlybracket{#4}_{#5}%
}
\NewDocumentCommand \srchoare {m m m m O{}}{
	\curlybracket{#1}\spac#2\spac@^{\textlog{s}}\spac#3\spac\curlybracket{#4}_{#5}%
}
\NewDocumentCommand\srcwp{m m m}%
  {\textlog{SRC}\spac#1\spac@\spac#2\spac{\left\{#3\right\}}}
\NewDocumentCommand\tgtwp{m m m}%
  {\textlog{SRC}\spac#1\spac@\spac#2\spac{\left\{#3\right\}}}

\definecolor{royalblue}{rgb}{0.06, 0.40, 0.63}
\definecolor{redorange}{rgb}{0.82, 0.39, 0.27}
\definecolor{emerald}{rgb}{0.00, 0.61, 0.56}
\definecolor{yelloworange}{rgb}{0.88, 0.62, 0.28}
\definecolor{carnationpink}{rgb}{0.84, 0.50, 0.65}

\newcommand{\thelogic}{Hotpot\xspace}

\newcommand*{\inferHS}[4]{\inferrule*[lab=\textlabel{#1}{#1}, right={#2}]{#3}{#4}}

\newcommand{\powerset}[1]{\mathcal{P}(#1)}

\newcommand{\bnfor}{\mathrel{\,|\,}}
\newcommand{\ite}[3]{\mathsf{if}\,#1\,\mathsf{then}\,#2\,\mathsf{else}\,#3}
\newcommand{\eqdefcoind}{\eqdef_{\mathsf{coind}}}

\newcommand{\Bool}{\mathbb{B}}

\newcommand{\lengthOf}[1]{\vert#1\vert}

\newcommand{\gap}{\cdot}

\newcommand{\sendevent}[1]{#1!}
\newcommand{\recvevent}[1]{#1?}

\newcommand{\intval}{z}
\newcommand{\boolval}{b}
\newcommand{\nolang}[1]{{\color{black}\mathnormal{#1}}}

\newcommand{\thelangRec}{\ensuremath{\imp{\Imp}}}

\newcommand{\thelangImp}{\thelangRec}
\newcommand{\thelangImpSub}{\imp{\texttt{r}}}
\newcommand{\thelangImpSubLong}{\imp{\texttt{rec}}}
\newcommand{\prov}{\mathimp{p}}
\newcommand{\Prov}{\mathimp{Id}}

\newcommand{\loc}{\mathimp{\ell}}
\newcommand{\locS}{\mathimp{\ell'}}

\newcommand{\compOffset}{\mathimp{offset}}
\newcommand{\compBlockId}{\mathimp{blockid}}
\renewcommand{\Loc}{\mathimp{Loc}}
\newcommand{\RunExpr}{\mathimp{RT\,Expr}}
\newcommand{\rexpr}{\mathimp{E}}
\newcommand{\RetExt}[2]{\mathimp{Ret(\nolang{#1}, \nolang{#2})}}
\newcommand{\WaitingImp}[1]{\mathimp{Wait(\nolang{#1})}}
\newcommand{\LibraryRec}{\mathimp{Library}}
\newcommand{\trueImp}{\mathimp{true}}

\newcommand{\gapImp}{\imp{\gap}}

\newcommand{\heap}{\mathimp{h}}
\newcommand{\heapA}{\mathimp{\heap_1}}
\newcommand{\heapB}{\mathimp{\heap_2}}

\newcommand{\heapS}{\mathimp{\heap'}}
\renewcommand{\val}{\mathimp{v}}
\newcommand{\valA}{\mathimp{\val_1}}
\renewcommand{\valB}{\mathimp{\val_2}}
\newcommand{\valC}{\mathimp{\val_3}}
\newcommand{\valS}{\mathimp{\val'}}
\newcommand{\ExprImp}{\mathimp{Expr}}
\newcommand{\ValImp}{\mathimp{Val}}
\newcommand{\exprimp}{\mathimp{\expr}}
\newcommand{\exprimpA}{\mathimp{\expr_1}}
\newcommand{\exprimpB}{\mathimp{\expr_2}}
\newcommand{\exprimpC}{\mathimp{\expr_3}}

\newcommand{\Binop}{\mathimp{BinOp}}
\newcommand{\binop}{\mathop{\mathimp{\oplus}}}
\newcommand{\impctx}{\mathimp{K}}

\newcommand{\fnname}{\mathsf{f}}
\newcommand{\FnName}{\mathimp{FnName}}

\newcommand{\vals}{\mathimp{\Repeat{\val}}}

\newcommand{\ECallH}{\mathimp{Call}}
\newcommand{\ECall}[3]{\mathimp{\ECallH(\nolang{#1}, \nolang{#2}, \nolang{#3})}}
\newcommand{\ECallRecv}[3]{\mathimp{\recvevent{\ECallH}(\nolang{#1}, \nolang{#2}, \nolang{#3})}}
\newcommand{\ECallSend}[3]{\mathimp{\sendevent{\ECallH}(\nolang{#1}, \nolang{#2}, \nolang{#3})}}
\newcommand{\EReturnH}{\mathimp{Return}}
\newcommand{\EReturn}[2]{\mathimp{\EReturnH(\nolang{#1}, \nolang{#2})}}
\newcommand{\EReturnRecv}[2]{\mathimp{\recvevent{\EReturnH}(\nolang{#1}, \nolang{#2})}}
\newcommand{\EReturnSend}[2]{\mathimp{\sendevent{\EReturnH}(\nolang{#1}, \nolang{#2})}}

\newcommand{\thelangAsm}{\ensuremath{\asm{\Asm}}}
\newcommand{\thelangAsmSub}{\asm{\texttt{a}}}
\newcommand{\thelangAsmSubLong}{\asm{\texttt{asm}}}
\newcommand{\aval}{\mathasm{v}}
\newcommand{\avalA}{\asm{\aval_1}}
\newcommand{\avalB}{\asm{\aval_2}}

\newcommand{\avalS}{\asm{\aval'}}
\newcommand{\regs}{\mathasm{r}}

\newcommand{\mem}{\mathasm{m}}

\newcommand{\memS}{\mathasm{\mem'}}
\newcommand{\pc}{\mathasm{pc}}
\newcommand{\reg}[1]{\mathasm{x#1}}
\newcommand{\SP}{\mathasm{sp}}
\newcommand{\AsmRegs}{\mathasm{Regs}}
\newcommand{\AsmRegName}{\mathasm{RName}}
\newcommand{\Instr}{\mathasm{Instr}}
\newcommand{\instr}{\mathasm{c}}
\newcommand{\instrs}{\mathasm{\Repeat{c}}}
\newcommand{\instrmap}{\mathasm{I}}
\newcommand{\ExecState}{\mathasm{Exec\,State}}
\newcommand{\rinstr}{\mathasm{I}}
\newcommand{\unmapped}{\asm{\#}}
\newcommand{\LibraryAsm}{\mathasm{Library}}

\newcommand{\WaitingAsm}{\mathasm{Wait}}

\newcommand{\RunningInstrAsm}[3]{\mathasm{Run(\nolang{#1}, \nolang{#2}, \nolang{#3})}}
\newcommand{\WaitingSyscallInstrAsm}[2]{\mathasm{WaitSyscall(\nolang{#1}, \nolang{#2})}}
\newcommand{\HaltedAsm}{\mathasm{Halted}}

\newcommand{\regMeta}{\mathasm{x}}
\newcommand{\regMetaA}{\asm{\regMeta_1}}
\newcommand{\regMetaB}{\asm{\regMeta_2}}

\newcommand{\storeRegAsm}[3]{\mathsf{upd}(#1, #2.\spac #3);}
\newcommand{\readMemAsm}[4]{\mathsf{ldr}(#1, #2, #3.\spac #4);}
\newcommand{\storeMemAsm}[4]{\mathsf{str}(#1, #2, #3. \spac #4);}
\newcommand{\syscallAsm}{\mathsf{syscall};}
\newcommand{\jumpAsm}{\mathsf{jump}}
\newcommand{\nextAsm}{\mathsf{next}}

\newcommand{\insaddr}{\mathasm{i}}

\newcommand{\addrraw}{\mathasm{a}}

\newcommand{\Amov}[2]{\keyword{mov}~#1,~#2}

\newcommand{\Asyscall}{\keyword{syscall}}
\newcommand{\Aret}{\keyword{ret}}
\newcommand{\avals}{\asm{\Repeat{\aval}}}
\newcommand{\rlookup}[2]{\asm{\nolang{#1}(\nolang{#2})}}
\newcommand{\rset}[3]{\asm{\nolang{#1}[\nolang{#2} \mapsto \nolang{#3}]}}

\newcommand{\EJumpH}{\mathasm{Jump}}
\newcommand{\EJump}[2]{\mathasm{\EJumpH(\nolang{#1}, \nolang{#2})}}
\newcommand{\EJumpRecv}[2]{\mathasm{\recvevent{\EJumpH}(\nolang{#1}, \nolang{#2})}}
\newcommand{\EJumpSend}[2]{\mathasm{\sendevent{\EJumpH}(\nolang{#1}, \nolang{#2})}}

\newcommand{\ESyscallSend}[2]{\mathasm{\sendevent{\mathasm{Syscall}}(\nolang{#1}, \nolang{#2})}}

\newcommand{\ESyscallRetRecv}[2]{\mathasm{\recvevent{\mathasm{SyscallRet}}(\nolang{#1}, \nolang{#2})}}

\newcommand{\modules}[1]{\mathsf{Module}(#1)}
\newcommand{\module}{M}
\newcommand{\events}{E}
\newcommand{\event}{e}
\newcommand{\eventS}{\event'}
\newcommand{\silent}{\tau}
\newcommand{\labels}{\taggedevents \uplus \{\silent\}}
\newcommand{\lbl}{\kappa}

\newcommand{\mstates}{S}
\newcommand{\mstate}{\sigma}
\newcommand{\mstateinit}{\mstate^0}
\newcommand{\mstateP}{\Sigma}
\newcommand{\mstepraw}{\rightarrow}
\newcommand{\mstep}[1]{\xrightarrow{#1}}
\newcommand{\mstepSmall}[1]{\xrightarrow{\smash{\raisebox{-.2ex}{\ensuremath{\scriptstyle #1}}}}}
\newcommand{\msteps}[1]{\xrightarrow{\smash{\raisebox{-.2ex}{\ensuremath{\scriptstyle #1}}}}^{*}}

\newcommand{\refinesraw}{\preceq}

\newcommand{\refines}[2]{#1 \refinesraw #2}

\newcommand{\refinesS}[2]{#1 \refinesraw #2}

\newcommand{\specprog}[1]{{#1}_\thelangSpecSubLong{}}
\newcommand{\thelangSpecSub}{\texttt{p}}
\newcommand{\thelangSpecSubLong}{\texttt{pseudo}}
\newcommand{\specsyn}[1]{\modsyn{\thelangSpecSub}{#1}}

\newcommand{\thelangSpecOrig}{\ensuremath{\mathsf{Spec}}}
\newcommand{\thelangSpec}{\ensuremath{\mathsf{Pseudo}}}

\newcommand{\modlinkraw}[1]{\oplus_{#1}}
\newcommand{\modlink}[3]{#2 \oplus_{#1} #3}

\newcommand{\modprepost}[2]{\lceil #2 \rceil_{#1}}

\newcommand{\modseqleft}{\mathsf{L}}
\newcommand{\modseqright}{\mathsf{R}}
\newcommand{\modseqnone}{\mathsf{E}}

\newcommand{\modseqdir}{d}
\newcommand{\modseqdirs}{\mathsf{D}}

\newcommand{\linkstates}{S}
\newcommand{\linkstate}{s}
\newcommand{\linkstateinit}{\linkstate^0}
\newcommand{\linkfn}{\rightsquigarrow}
\newcommand{\inlinkfn}[3]{#2\spac #1 \spac#3}

\newcommand{\taggedevents}{\events_{?!}}

\newcommand{\satisfiable}[1]{\mathsf{sat}(#1)}

\newcommand{\prepostframe}{F}

\newcommand{\prerel}{\leftharpoondown}
\newcommand{\prerelitoa}{\prerel_{\mathimp{\thelangImpSub} \rightleftharpoons \mathasm{\thelangAsmSub}}}
\newcommand{\postrel}{\rightharpoonup}
\newcommand{\postrelitoa}{\postrel_{\mathimp{\thelangImpSub} \rightleftharpoons \mathasm{\thelangAsmSub}}}

\newcommand{\seplogic}{\mathcal{L}}
\newcommand{\seplogicitoa}{\seplogic_{\mathimp{\thelangImpSub} \rightleftharpoons \mathasm{\thelangAsmSub}}}
\newcommand{\sepprop}{\mathit{Prop}_{\!\seplogic}}

\newcommand{\modsyn}[2]{\Sem{#2}_{#1}}
\newcommand{\synlink}[3]{#2 \cup_{#1} #3}

\newcommand{\Asm}{\asm{Asm}}
\newcommand{\asm}[1]{\mathbf{{\color{royalblue}#1}}}
\newcommand{\mathasm}[1]{\asm{#1}}

\newcommand{\asmprog}[1]{\asm{#1}}
\newcommand{\asmsynlink}[2]{\asm{\synlink{\thelangAsmSub}{\nolang{#1}}{\nolang{#2}}}}
\newcommand{\asmsynlinkraw}{\asm{\cup_{\thelangAsmSub}}}
\newcommand{\asmlink}[2]{\asm{\modlink{\thelangAsmSub}{\nolang{#1}}{\nolang{#2}}}}

\newcommand{\asmlinkraw}{\asm{\oplus_{\thelangAsmSub}}}

\newcommand{\asmdom}[1]{\asm{\lengthOf{\nolang{#1}}}}
\newcommand{\asmsyn}[1]{\asm{\modsyn{\thelangAsmSub}{\nolang{#1}}}}
\newcommand{\AsmEvents}{\mathasm{Events}}
\newcommand{\asmevent}{\mathasm{\event}}
\newcommand{\AsmVal}{\mathasm{Val}}
\newcommand{\AsmHeap}{\mathasm{Memory}}

\newcommand{\compile}[1]{{\downarrow}\ \!{#1}}

\newcommand{\moditoa}[1]{\modprepost{\mathimp{\thelangImpSub} \rightleftharpoons \mathasm{\thelangAsmSub}}{#1}}

\newcommand{\Imp}{\mathsf{Rec}}
\newcommand{\imp}[1]{\mathsf{{\color{redorange}#1}}}
\newcommand{\mathimp}[1]{\imp{#1}}
\newcommand{\moduleImp}{\imp{\module}}

\newcommand{\moduleImpA}{\imp{\moduleImp_1}}

\newcommand{\moduleImpB}{\imp{\moduleImp_2}}

\newcommand{\impprog}[1]{\imp{#1}}
\newcommand{\impsynlink}[2]{\imp{\synlink{\thelangImpSub}{\nolang{#1}}{\nolang{#2}}}}
\newcommand{\implink}[2]{\imp{\modlink{\thelangImpSub}{\nolang{#1}}{\nolang{#2}}}}

\newcommand{\impsynlinkraw}{\imp{\cup_{\thelangImpSub}}}

\newcommand{\implinkraw}{\imp{\modlinkraw{\thelangImpSub}}}
\newcommand{\impsyn}[1]{\imp{\modsyn{\thelangImpSub}{\nolang{#1}}}}
\newcommand{\ImpEvents}{\mathimp{Events}}
\newcommand{\impevent}{\mathimp{\event}}
\newcommand{\ImpVal}{\mathimp{Val}}
\newcommand{\ImpHeap}{\mathimp{Memory}}

\newcommand{\mainId}{\Hid{main}}

\newcommand{\echoId}{\Hid{echo}}
\newcommand{\getcId}{\Hid{getc}}
\newcommand{\putcId}{\Hid{putc}}
\newcommand{\readId}{\Hid{read}}
\newcommand{\writeId}{\Hid{write}}

\newcommand{\libImp}{\imp{R}}

\newcommand{\libAsm}{\asm{A}}

\newcommand{\link}{\oplus}

\newcommand{\heapinv}[2]{\mathimp{inv(\nolang{#1}, \nolang{#2})}}

\newcommand{\proofstepkern}{-8.5mu}%
\newcommand{\proofcenter}[1]{\rlap{\ensuremath{\mkern\proofstepkern\mathclap{#1}}}}
\newcommand{\proofstep}[1]{%
  \noalign{\vspace{2pt}}%
  \multicolumn{2}{r@{}}{\proofcenter{\Downarrow}} & \mathrlap{\quad\vcenter{\hbox{\small #1}}}%
  \\[1pt]}
\newcommand{\proofstepN}[3]{%
  \noalign{\vspace{#2}}%
  \multicolumn{2}{r@{}}{\proofcenter{\Downarrow}} & \mathrlap{\quad\vcenter{\hbox{\small #1}}}%
  \\\noalign{\vspace{#3}}}
\newcommand{\proofvdots}{\multicolumn{2}{r@{}}{\proofcenter{\vdots}} &}
\makeatletter
\newcounter{proofstepctr}
\newcommand{\stepnum}[1]{\stepcounter{proofstepctr}%
  \mbox{\small
    \ifthenelse{\equal{#1}{}}{}{%
      \protected@write\@auxout{}{\string\newlabel{#1}{{\theproofstepctr}{\thepage}{}{#1}{}}}%
      \hypertarget{#1}{}}%
    (\theproofstepctr)}}
\makeatother

\renewcommand{\expr}{\epsilon}
\newcommand{\exprs}{\mathcal{E}}
\newcommand{\lang}{\textlog{L}}
\newcommand{\ctx}{K}
\newcommand{\ctxs}{\textdom{Ctx}}

\newcommand{\lcons}[1]{#1 \mathop{::}}
\newcommand{\piname}{lane\xspace}
\newcommand{\pinames}{lanes\xspace}
\newcommand{\PiName}{Lane\xspace}
\newcommand{\PiNames}{Lanes\xspace}

\newcommand{\Abscalls}{Abstract calls\xspace}
\newcommand{\abscalls}{abstract calls\xspace}
\newcommand{\abscall}{abstract call\xspace}

\newcommand{\absjump}{abstract jump\xspace}

\newcommand{\simgenname}{\textlog{mwp}\xspace}

\newcommand{\ext}{external\xspace}

\newcommand{\ExtCalls}{External Calls\xspace}

\newcommand{\tokennames}{tokens\xspace}

\newcommand{\ts}{\rho}

\newcommand{\tgtsub}{\textlog{i}}
\newcommand{\srcsub}{\textlog{s}}

\newcommand{\eqdefind}{\eqdef_{\mathsf{ind}}}

\NewDocumentCommand\simgenaux{O{} m m m}%
  {\textlog{mwp}_{#2}^{#1}\spac#3\spac{\left\{#4\right\}}}

\NewDocumentCommand\simgen{m m m}{\simgenaux[\ts]{#1}{#2}{#3}}
\NewDocumentCommand\simgenlane{m m m}{\simgenaux[\ts_{\lane}]{#1}{#2}{#3}}
\NewDocumentCommand\simtgt{m m m}{\simgenaux[\tgtsub]{#1}{#2}{#3}}
\NewDocumentCommand\simsrc{m m m}{\simgenaux[\srcsub]{#1}{#2}{#3}}

\newcommand{\susp}[2]{#1 \Mapsfrom #2\xspace}

\newcommand{\switchSome}[3]{\smash[b]{{}\text{\textinterrobang}#3 \prescript{#1}{}{\Rightarrow}^{#2}}\,}
\newcommand{\switchSomeS}[2]{{}\text{\textinterrobang}^{#1}#2;}
\newcommand{\switchSomeX}[2]{{}\text{\textinterrobang}^{#1}#2}
\newcommand{\switchout}[3]{\smash[b]{{}!#3 \prescript{#1}{}{\Rightarrow}^{#2}}\,}
\newcommand{\switchoutsilent}[2]{\smash[b]{{}\silent \prescript{#1}{}{\Rightarrow}^{#2}}\,}
\newcommand{\switchoutSraw}[2]{{}!^{#1}#2}
\newcommand{\switchoutS}[2]{\switchoutSraw{#1}{#2};}
\newcommand{\switchoutX}[2]{{}!^{#1}#2}
\newcommand{\switchinraw}[3]{\smash[b]{#3 \leftarrow ?^{#1}#2}}
\newcommand{\switchin}[3]{\switchinraw{#1}{#2}{#3};}
\newcommand{\switchinSraw}[2]{{}?^{#1}#2}
\newcommand{\switchinS}[2]{\switchinSraw{#1}{#2};}

\newcommand{\switchsilent}[1]{\tau^{#1};}

\newcommand{\switchinto}[3]{\smash[b]{{}?#3 \prescript{#1}{}{\Rightarrow}^{#2}}\,}

\newcommand{\maybelater}[1][\ts]{{\later}_{\mkern-4mu#1}\spac}
\newcommand{\dslater}{\mathop{{\triangleright}}}

\newcommand{\exprrelname}[1]{\mathrel{\leq\!\!\!{\raisebox{.08em}{\ensuremath{\cdot}}}}_{#1}}
\newcommand{\exprrel}[3]{#2 \exprrelname{#1} #3}
\NewDocumentCommand\simgenexpraux{O{} m m m}%
  {\textlog{wp}_{#3}\spac#2\spac{\left\{#4\right\}}}

\NewDocumentCommand\simgenexpr{m m m}{\simgenexpraux[\ts]{#1}{#2}{#3}}
\NewDocumentCommand\simtgtexpr{O{} m m m}{\simgenexpraux[\tgtsub]{#1}{#2}{#3}}
\NewDocumentCommand\simsrcexpr{O{} m m m}{\simgenexpraux[\srcsub]{#1}{#2}{#3}}
\newcommand{\post}{\Phi}

\NewDocumentCommand\simgenexprsimpl{O{} m m}%
  {\textlog{wp}_{#1}\spac#2\spac{\left\{#3\right\}}}

\newcommand{\simgenexprname}{\textlog{wp}}

\newcommand{\SI}[1]{\mathit{SI}_{#1}}

\newcommand{\dimsumsim}[4]{(#1, #2) \preceq_{\text{co}} (#3, #4)}

\newcommand{\simbinraw}{\preceq_{\text{S}}}
\newcommand{\simbin}[4]{(#1, #2) \preceq_{\textsc{S}} (#3, #4)}

\newcommand{\lane}{\Pi}
\newcommand{\laneS}{\Pi'}
\newcommand{\laneA}{\Pi_1}
\newcommand{\laneB}{\Pi_2}
\newcommand{\laneC}{\Pi_3}

\newcommand{\lanepost}[1]{\Phi_{#1}}

\newcommand{\intokenraw}[2]{\text{\faArrowDown}^{#1}_{#2}}
\newcommand{\outtokenraw}[2]{\text{\faArrowUp}^{#1}_{#2}}

\newcommand{\tgtside}[1]{\bullet \preceq \_}
\newcommand{\srcside}[1]{\_ \preceq \bullet}
\newcommand{\tgtPi}[1]{[\bullet \preceq \_]}
\newcommand{\srcPi}[1]{[\_ \preceq \bullet]}
\newcommand{\srcintoken}[1]{\intokenraw{}{\!\!\preceq\bullet}}
\newcommand{\tgtintoken}[1]{\intokenraw{}{\!\!\bullet \preceq}}
\newcommand{\srcouttoken}[1]{\outtokenraw{}{\!\!\preceq\bullet}}
\newcommand{\tgtouttoken}[1]{\outtokenraw{}{\!\!\bullet \preceq}}
\newcommand{\tgtgvar}{\gvar_{\tgtside{}}}
\newcommand{\srcgvar}{\gvar_{\srcside{}}}

\newcommand{\rightgvar}{\gvar_{\linkrightsub{}}}

\newcommand{\linkrightsub}[1]{\modlink{#1}{\_}{\bullet}}
\newcommand{\linkLPi}[2]{#1[\modlink{#2}{\bullet}{\_}]}
\newcommand{\linkRPi}[2]{#1[\modlink{#2}{\_}{\bullet}]}

\newcommand{\linkrouting}[2]{\text{\faArrows*}^{#1}_{\oplus_{#2}}}

\newcommand{\linkouttoken}[2]{\outtokenraw{#1}{\oplus_{#2}}}
\newcommand{\linkintoken}[2]{\intokenraw{#1}{\oplus_{#2}}}

\newcommand{\gvar}{\gamma}

\newcommand{\lanes}[1]{\mathsf{\PiName}(#1)}

\newcommand{\mstatefrac}[2]{#1 \leadsto_{\frac{1}{2}} #2}
\newcommand{\mstatefull}[1]{#1 \leadsto_1\!-}

\newcommand{\linkRecLPi}[1]{#1[\implink{\bullet}{\_}]}
\newcommand{\linkRecRPi}[1]{#1[\implink{\_}{\bullet}]}

\newcommand{\linkRecouttoken}[1]{\text{\faArrowUp}^{#1}_{\mathimp{\oplus_{\thelangImpSub}}}}
\newcommand{\linkRecintoken}[1]{\text{\faArrowDown}^{#1}_{\mathimp{\oplus_{\thelangImpSub}}}}

\newcommand{\fnext}[1]{#1 \not{\hookrightarrow}_{\thelangImpSub}\,}
\newcommand{\fninternal}[2]{#1 \hookrightarrow_{\thelangImpSub} #2}
\NewDocumentCommand\recwaitraw{o}{\imp{\text{\faClock[regular]}}\IfValueT{#1}{^{#1}}}
\NewDocumentCommand\recwait{o}{\IfValueTF{#1}{\recwaitraw[#1]}{\recwaitraw};}

\newcommand{\hptsto}{\mathop{\mapsto_{\heap}}}
\newcommand{\hptstoBlock}{\mathop{\mapsto_{\heap}^{\ast}}}

\newcommand{\bicall}[3]{#3\!\shortleftarrow\! #1(#2)_{\thelangImpSub};}
\newcommand{\bicallS}[2]{#1(#2)_{\thelangImpSub};}
\newcommand{\bicallSraw}[2]{#1(#2)_{\thelangImpSub}}

\newcommand{\mptsto}{\mathop{\mapsto_{\mem}}}

\newcommand{\rptsto}{\mathop{\mapsto_{\regs}}}

\newcommand{\insext}[1]{#1 \not{\hookrightarrow}_{\insaddr}\,}
\newcommand{\insinternal}[2]{#1 \hookrightarrow_{\insaddr} #2}
\newcommand{\inssinternal}[2]{#1 \hookrightarrow_{\insaddr}^{\!\!*} #2}

\newcommand{\READ}{\mathasm{READ}}
\newcommand{\ESyscallH}{\mathasm{\mathasm{Syscall}}}
\newcommand{\ESyscallRetH}{\mathasm{\mathasm{SyscallRet}}}

  \NewDocumentCommand\simgenexprasm{m m}{\simgenexpraux[\ts]{#1}{#2}{{-}}}

\newcommand{\linkAsmLPi}[1]{#1[\asmlink{\bullet}{\_}]}
\newcommand{\linkAsmLPiSubst}[2]{#1[\asmlink{#2}{\_}]}
\newcommand{\linkAsmRPi}[1]{#1[\asmlink{\_}{\bullet}]}

\NewDocumentCommand\asmwait{o}{\asm{\text{\faClock[regular]}}\IfValueT{#1}{^{#1}}}

\newcommand{\looparrowup}{\mathord{\rotatebox[origin=c]{90}{$\looparrowright$}}}
\newcommand{\bijump}{\looparrowup_\thelangAsmSub\,\textlog{jmp}}

\newcommand{\fnextSpec}[1]{#1 \not{\hookrightarrow}_{\thelangSpecSub}\,}
\newcommand{\fninternalSpec}[2]{#1 \hookrightarrow_{\thelangSpecSub} #2}
\NewDocumentCommand\specwaitraw{o}{{\text{\faClock}}\IfValueT{#1}{^{#1}}}
\NewDocumentCommand\specwait{o}{\IfValueTF{#1}{\specwaitraw[#1]}{\specwaitraw};}
\newcommand{\speccall}[3]{#3\!\shortleftarrow\! #1(#2)_{\thelangSpecSub};}
\newcommand{\speccallS}[2]{#1(#2)_{\thelangSpecSub};}
\newcommand{\pseudokeyword}{\keyword}
\newcommand{\Plet}[2]{\pseudokeyword{let}~#1 := #2;}
\newcommand{\Pupd}[2]{#1 := #2}
\newcommand{\Pifthen}[2]{\pseudokeyword{if}\!~#1\!~#2}

\newcommand{\Pstore}[2]{#1 \leftarrow #2}

\newcommand{\PcallN}[2]{#1(#2)}
\newcommand{\Pret}[1]{\pseudokeyword{return}~#1}
\newcommand{\Pvar}[1]{\mathit{#1}}
\newcommand{\Pcmd}[1]{\pseudokeyword{#1}}
\newcommand{\Pfname}[1]{\mathit{#1}}
\newcommand{\Ploop}[1]{\pseudokeyword{loop}\{~#1~\}}
\newcommand{\Pexists}[1]{\pseudokeyword{$\exists$}~\!#1}
\newcommand{\Pload}[1]{\mathord{\textbf{!}}#1}
\newcommand{\Passert}[1]{\pseudokeyword{assert}(#1)}
\newcommand{\Pvalid}[1]{\pseudokeyword{valid}(#1)}

\newcommand{\valSpec}{\mathsf{v}}
\newcommand{\valsSpec}{\mathsf{\Repeat{\valSpec}}}

\newcommand{\gotolbl}{\mathit{X}}
\newcommand{\bigotoS}[1]{{#1}}
\newcommand{\rettok}{R}

\newcommand{\echoRec}{\imp{\echoId_{\thelangImpSubLong}}}
\newcommand{\getcRec}{\imp{\getcId_{\thelangImpSubLong}}}
\newcommand{\putcRec}{\imp{\putcId_{\thelangImpSubLong}}}
\newcommand{\mainRec}{\imp{\mainId_{\thelangImpSubLong}}}
\newcommand{\mainPseudo}{\specprog{\mainId}}
\newcommand{\putcPseudo}{\specprog{\putcId}}
\newcommand{\readAsm}{\asmprog{read}_{\thelangAsmSubLong}}

\newcommand{\PiSpec}{\mathbb{P}}
\newcommand{\PiRec}{\imp{\mathbb{R}}}
\newcommand{\PiAsm}{\asm{\mathbb{A}}}

\newcommand{\reclib}[1]{\imp{\textlog{P}_{#1}}}

\newcommand{\speclib}{\textlog{P}}
\newcommand{\LibrarySpec}{\mathsf{Library}}

\newcommand{\echolib}{\ensuremath{\mathimp{Echo}}}
\newcommand{\echobuflib}{\ensuremath{\mathimp{Echo}_{\textlog{buf}}}}
\newcommand{\rtoaclosed}{\text{\faLock}_{\thelangImpSub \rightleftharpoons \thelangAsmSub}}
\newcommand{\rtoaexchange}{\text{\faSync}}
\newcommand{\exchanged}{\mathimp{\Repeat{\loc}}}
\newcommand{\itoabij}{\mathop{{\rightarrowtail}}}

\newcommand{\rtoaPi}[1]{#1{}[\moditoa{\bullet}]}
\newcommand{\rtoaPiStandalone}{\moditoa{\bullet}}
\newcommand{\rtoawait}[1]{\text{\faClock[regular]}^{#1}_{\thelangImpSub \rightleftharpoons \thelangAsmSub}}

\newcommand{\rtoasub}{{\thelangImpSub \rightleftharpoons \thelangAsmSub}}

\newcommand{\wrapsub}{\nolang{X}}
\newcommand{\wembed}[1]{\langle #1 \rangle}
\newcommand{\wembedtok}{\langle \blacksquare \rangle}
\newcommand{\wembedclosed}[1]{\langle \rangle_{#1}}
\newcommand{\wrapPi}[1]{#1[\modprepost{\wrapsub}{\bullet}]}

\newcommand{\wrapwait}[1]{\text{\faClock[regular]}^{#1}_{\modprepost{\wrapsub}{\cdot}}}

\newcommand{\memtrader}{\text{\faCompressArrows*}_{\textlog{m}}}

\newcommand{\meminv}{\textlog{inv}_{\textlog{m}}}

\renewcommand{\heapinv}{\textlog{inv}_{\textlog{h}}}
\newcommand{\rtoainj}{\textlog{inv}_\rtoasub}
\newcommand{\rtoabij}{\itoabij}

\newcommand{\fninv}{\textlog{inv}_{\textlog{fn}}}
\newcommand{\reginv}{\textlog{inv}_{\textlog{r}}}
\newcommand{\instrinv}{\textlog{inv}_{\textlog{i}}}

\renewcommand{\vs}{\vdash \upd}

\newcommand{\mainPseudoTwo}{\specorigprog{\Hid{main\_buf}}}
\newcommand{\mainPseudoThree}{\specorigprog{\Hid{main\_syscall}}}
\newcommand{\mainPseudoThreeS}{\specprog{\Hid{main\_syscall}}}
\newcommand{\thelangSpecOrigSubLong}{\texttt{spec}}
\newcommand{\specorigprog}[1]{{#1}_\thelangSpecOrigSubLong{}}
\newcommand{\recbullet}{\imp{\bullet}}
\newcommand{\asmbullet}{\asm{\blacktriangle}}
\newcommand{\specbullet}{\blacklozenge}
\newcommand{\pseudobullet}{\specbullet}

\newcommand{\recoplus}{\mathbin{\imp{\oplus}_\thelangImpSub}}
\newcommand{\asmoplus}{\mathbin{\asm{\oplus}_\thelangAsmSub}}

\newcommand{\putcPseudoTwo}{\specprog{\Hid{putc\_unbuf}}}
\newcommand{\getcAsm}{\asmprog{\getcId}_{\thelangAsmSubLong}}
\newcommand{\putcAsm}{\asmprog{\putcId}_{\thelangAsmSubLong}}

\newcommand{\reclinecolor}{redorange}
\newcommand{\asmlinecolor}{royalblue}
\newcommand{\pseudolinecolor}{black}

\newcommand{\bilabelSRev}[1]{\nu{#1}.~}

\newcommand{\switchinChoice}[4]{\switchinraw{#1}{#2}{#3}:{(#4)};}

\begin{document}

\title{Modular Reasoning about External Code in Separation Logic}
\title{Multi-language Reasoning in Separation Logic}
\title{Multi-Iris: Multi-language Program Logics}

\ifarxiv
\title{Multi-Language Program Logics (Extended Version)}
\else
\title{Multi-Language Program Logics}
\fi

\author{Alexander Loitzl}
\orcid{0009-0002-7417-2537}
\affiliation{%
  \institution{Institute of Science and Technology Austria (ISTA)}
  \city{Klosterneuburg}
  \country{Austria}
}
\email{alexander.loitzl@ista.ac.at}

\author{Niklas Mück}
\orcid{0009-0006-9622-0762}
\affiliation{%
  \institution{MPI-SWS}
  \department{Saarland Informatics Campus}
  \city{Saarbrücken}
  \country{Germany}
}
\email{mueck@mpi-sws.org}

\author{Michael Sammler}
\correspondingauthor
\orcid{0000-0003-4591-743X}
\affiliation{%
  \institution{Institute of Science and Technology Austria (ISTA)}
  \city{Klosterneuburg}
  \country{Austria}
}
\email{michael.sammler@ista.ac.at}

\begin{abstract}
Real-world programs are rarely written in a single language: For example, C programs call assembly routines, and high-level languages like OCaml link with low-level C libraries. Yet program logics---one of the most successful techniques for modular program verification---almost exclusively target single-language programs.

We present \emph{\thelogic{}}, the first framework for building \emph{multi-language program logics}. \thelogic{} enables compilation-independent, cross-language reasoning about languages with heterogeneous views of shared state.
\thelogic{} rests on four key ideas:
\emph{abstract calls} to specify calls to unknown functions,
\emph{lanes and the switching modality} to move between languages inside the program logic,
uniform integration of \emph{refinement reasoning} via lanes,
and \emph{exchanges} to translate between separation logic assertions of different languages.
We demonstrate that \thelogic{} allows reusing specifications across implementations in different languages, supports reasoning about higher-order cross-language function calls, and integrates with verified compilation.
\thelogic{} is built on top of Iris and DimSum and mechanized in the Rocq Prover.
 \end{abstract}




\maketitle

\section{Introduction}\label{sec:introduction}
Many real-world programs are not written in a single programming language but are composed of components written in multiple different languages~\cite{MultiLangPrevalence}.
For example, C programs commonly call assembly routines---\eg to communicate with the operating system or to use hardware-specific optimizations.
Programs in high-level languages like OCaml or Python often link with libraries implemented in lower-level languages for efficiency (\eg the popular NumPy Python library implemented in C), or to integrate with the platform (\eg by linking to the C standard library).

While multi-language programs are common in practice, large parts of the PL literature focus on single-language programs. In particular, consider program logics.
Program logics~\cite{HoareLogic}, especially those based on separation logic~\cite{SeparationLogic1, SeparationLogic2}, are a popular and widely used technique for verifying programs. %
A key ingredient of the success of program logics is the modular reasoning they enable: Hoare triples enable modular reasoning about functions, separation logic with its frame rule allows modular reasoning about memory and concurrency, and there is a steady stream of new program logics with novel modular reasoning principles~\cite{Perennial, PulseCore, Aneris, OutcomeLogicProb}.

While the work on program logics has developed many modular reasoning principles, they focus on single-language programs: In most work, a client can modularly link with an arbitrary function implementing a specification, \emph{as long as the function is implemented in the same language as the client}.
However, this rules out \emph{multi-language} programs.
As a concrete example, consider some C code that calls a function from a library.
The library function might either be implemented in C or---\eg for performance---in assembly.
For most program logics, this language choice makes a big difference: Either they can verify the interaction of the C code and the library function (if they are in the same language) or they cannot (if they are in different languages).
Our goal is to address this limitation:
In this paper, we show how to develop \textbf{\emph{multi-language program logics}} that enable modular reasoning \emph{across multiple languages}.

More concretely, in this paper we aim to develop multi-language program logics that satisfy the following four desiderata:
A multi-language program logic should be a \emph{program logic} that supports \emph{cross-language reasoning}, is \emph{compilation-independent}, and applies to \emph{heterogeneous languages with shared state}. Let us explain these four desiderata in more detail.

\emph{First}, a multi-language program logic should provide all the usual benefits \emph{modern program logics} provide, like modular function specifications, abstract predicates, small-footprint reasoning, higher-order reasoning, custom ghost state, and more.
This distinguishes multi-language program logics from other approaches for multi-language reasoning (as are common, \eg in compositional compiler verification~\cite{MultilanguageCompiler, Pilsner, CompCertO, DimSum}) that use a different reasoning style than program logics.

\emph{Second}, a multi-language program logic should combine program logics for the individual languages into one overarching program logic that allows \emph{cross-language reasoning} and switching between the different language-local program logics.
Supporting cross-language reasoning is in contrast to program logics that can reason about the interaction with an external environment, \eg via a foreign function interface (FFI), but do not actually allow reasoning \emph{across} the interface~\cite{VeriFastIO, VSTFFI, AdamsLightbulb}.

\emph{Third}, we focus on multi-language program logics that are \emph{independent of the compilation strategy}.
This independence enables reusing proofs across different target languages and avoids complicating the verification with details of the compilation.
This is in contrast to approaches that reason about multi-language programs after compilation to a common target language~\cite{SemanticSoundness, SMACKCrossLanguage}.

\textit{Fourth}, our focus is on multi-language programs composed of \emph{heterogeneous languages that share state}. This is a common scenario: Think, for example, of a C program with a block-based memory model that interacts with assembly with a flat memory model. Or an OCaml program with its typed representation of the memory that interacts with a C program without typed memory.
What makes these cases interesting is that the different languages share state (often memory), but they have a different view on the same state.
There are multi-language settings where this is not the case: Either because the languages are homogeneous and use the same notion of memory (for example, in CompCert-based approaches~\cite{CompCompCert, CompCertO} or WebAssembly~\cite{IrisWasm}).
Or because the languages are heterogeneous, but do not share state (\eg the ML-language and IO-language considered by \citet{GITrees}).
However, in this paper we address the more general setting of heterogeneous languages with shared state since many interesting multi-language programs fall into this category.

Despite the prevalence of multi-language programs in practice, we are only aware of a single program logic that fulfills all these desiderata: Melocoton~\cite{Melocoton}, which considers the interaction of OCaml and C.
Melocoton provides program logics for OCaml and C and shows that one can combine them to verify mixed OCaml-C programs.
While this is an impressive result dealing with many of the details of the OCaml-C-FFI (like garbage collection), it only considers linking two specific languages, OCaml and C.

In this paper, we study how to design multi-language program logics more generally. Concretely, we present \textbf{\thelogic},
the first framework for building multi-language program logics that support compilation-independent, cross-language reasoning about heterogeneous multi-language programs. %

\thelogic is based on two pillars:
On the one hand, it is based on the Iris framework~\cite{Iris1, Iris3, IrisGroundUp} that has proven to be an expressive foundation for building program logics.
However, so far most Iris-based verification has focused on single-language verification, not multi-language verification.

On the other hand, \thelogic is based on the DimSum framework for multi-language semantics~\cite{DimSum}.
DimSum's signature feature is its decentralized approach to multi-language semantics, enabling multi-language programs to be composed flexibly from components in different languages.
As example instantiations, DimSum provides a C-like language \thelangRec{} with a block-based memory model and an assembly language \thelangAsm{} with a flat memory model, together with a verified compiler from \thelangRec{} to \thelangAsm{}.
However, DimSum does not provide a program logic. Instead, refinements between programs are proven monolithically.
\thelogic provides a program logic that is proven sound against DimSum's multi-language semantics and its results can be linked with DimSum's verified compiler.

In this paper, we show how \thelogic{} enables the development of expressive multi-language program logics on the foundations of Iris and DimSum.

\subsection{Overview of \thelogic}
\label{sec:overview}
Let us now give an overview of the challenges of building multi-language program logics and the ideas that \thelogic uses to address these challenges. (\autoref{sec:keyideas} describes the challenges and key ideas in more depth.) As a running example, consider the following $\echoRec$ program (written in DimSum's \thelangRec{} language) that first calls $\getcId$ to get a character and then $\putcId$ to print the character:%
\footnote{$\getcId$ and $\putcId$ are inspired by the \href{https://en.cppreference.com/c/header/stdio}{corresponding C standard library functions}.}
\begin{align*}
  \echoRec() \eqdef {} &
  \Hlet{\Hvar}{\Hcall{\getcId}{}}~
  \Hcall{\putcId}{\Hvar}
\end{align*}
We identify four key challenges for the verification of programs like $\echoRec$:

\paragraph{Challenge \#1: Modular verification}
There are many ways in which $\getcId$ and $\putcId$ could be implemented. For example, $\getcId$ and $\putcId$ could be implemented in the same language as wrappers around more primitive I/O functions. Or they could directly be implemented in another language like assembly. Or they could be fully external for the verification without any implementation.
The challenge is to support \emph{modular} verification of $\echoRec$, \ie we want to verify $\echoRec$ once and reuse the verification with all these possible implementations of $\getcId$ and $\putcId$.

\emph{Key idea \#1: \Abscalls.}
To enable modular verification, we introduce a specification construct for \emph{\abscalls}.
\Abscalls allow us to specify and verify that $\echoRec$ first calls $\getcId$ and then $\putcId$ (with the result of $\getcId$), independently of the concrete implementation of $\getcId$ and $\putcId$.
\Abscalls are similar to the staged specifications of \citet{HSSL}, though applied to multi-language programs instead of higher-order functions.

\paragraph{Challenge \#2: Switching between languages}
After we have verified $\echoRec$, we want to use it with different implementations of $\getcId$ and $\putcId$.
This is where the multi-language aspects come in: If the implementation of $\getcId$ is in a different language than $\echoRec$, the program logic needs to \emph{switch} between these languages when $\echoRec$ calls $\getcId$.

\emph{Key idea \#2: Lanes and switching modality.}
To reason about switching between languages, \thelogic identifies each component of a multi-language program with a \emph{\piname} and provides a modality for switching between \pinames.
Thanks to these lanes, \thelogic can locally express switches between languages without any global assumptions about the composition of the multi-language program.

\paragraph{Challenge \#3: External functions}
There is another case to consider:
$\getcId$ and $\putcId$ might be \emph{external functions}, \ie we want to treat them as primitive I/O operations without any implementation.
How can we integrate such external functions into the verification?

\emph{Key idea \#3: Refinement reasoning via \pinames.}
The basic idea of \thelogic follows DimSum: External functions are handled via refinement.
Concretely, \thelogic proves that each external call of the verified implementation is matched by a call in a specification program.
This allows us to prove arbitrary safety\footnote{Following DimSum, \thelogic only supports safety properties.}
properties about the implementation (\eg $\echoRec$ only calls $\putcId$ with characters returned by $\getcId$).
What is novel is how \thelogic integrates this refinement reasoning into the program logic. Concretely, we observe that we can unify the reasoning about multiple languages and refinement by \emph{treating the specification program as another \piname} that one can switch to.
Thanks to this unification, \thelogic can use the previously introduced concepts like \abscalls and the switching modality uniformly for multi-language reasoning and refinement reasoning.

\paragraph{Challenge \#4: View reconciliation for separation logic}
The final challenge comes from the desire to verify the interaction of heterogeneous languages with shared state.
The program logics for these languages are based on different assertions.
For example, a C-like language with a block-based memory will have a points-to predicate based on abstract locations while an assembly language with a flat memory model will have a points-to predicate based on integer addresses.
However, these different assertions relate the same underlying state (often the memory).
Translating between these different assertions is what \citet{Melocoton} call the \emph{view reconciliation problem}, \ie that different languages have different views of the same shared resource like memory.

\emph{Key idea \#4: Exchanges via embedding and traders.}
To address the view reconciliation problem, \thelogic introduces \emph{exchanges} for exchanging points-to predicates of different languages.
These exchanges are built on the more primitive concepts of embedding and traders that create connections between the assertions of different separation logics.
Crucially, all these concepts are proven sound against DimSum's mechanism for translating between languages (called \emph{wrappers}).
This means that the results proven by \thelogic{} can be soundly combined with DimSum's verified compiler.

\subsection{Contributions}
Our overarching contribution is \thelogic%
\footnote{\thelogic: Heterogeneous Ownership Transfer for Programs Of many Tongues.}, the first framework for building multi-language program logics, based on DimSum and Iris.
This includes the following contributions:
\begin{itemize}
\item Definition of \thelogic{} based on DimSum and Iris (\autoref{sec:model}).
\item \Abscalls for modular specification of (potentially) external calls (\autoref{sec:key:abscalls}).
\item \PiNames and the switching modality for cross-language reasoning inside \thelogic (\autoref{sec:key:switching}, \autoref{sec:model:lanes}).
\item Refinement reasoning based on \pinames (\autoref{sec:key:extcalls}, \autoref{sec:model:refinement-lanes}).
\item Embedding, traders, and exchanges for addressing view reconciliation (\autoref{sec:key:wrapper}, \autoref{sec:model:wrapper}, \autoref{sec:rec-asm-wrapper}).
\item Examples demonstrating that \thelogic{} allows reusing specifications across implementations in different languages, supports higher-order cross-language reasoning, and integrates with DimSum's verified compiler (\autoref{sec:case-study}).
\end{itemize}

All contributions of this paper are formalized in the Rocq Prover~\cite{Rocq} based on the Iris framework~\cite{IrisGroundUp} and DimSum~\cite{DimSum}.
The Rocq development is provided as supplementary material~\cite{Artifact}.

\paragraph{Limitations}
The main limitation of \thelogic is that it supports only first-order ghost state, but not higher-order ghost state.
This is because Iris' natural number-based step-indexing is incompatible with DimSum's use of angelic non-determinism.
We plan to address this problem in the future by porting \thelogic to transfinite Iris~\cite{TransfiniteIris}.
A benefit of this limitation is that all constructions introduced in this paper are expressed using first-order ghost state, which might make it simpler to apply these ideas to other program logics that do not support higher-order ghost state.

Additionally, \thelogic inherits DimSum's restriction to safety properties.
We use \thelangRec{} and \thelangAsm{} as examples in this paper since they exercise many aspects of multi-language reasoning (\eg view reconciliation) and leave other languages and concurrency to future work.

\section{Key Ideas} \label{sec:keyideas}
We illustrate the key ideas of \thelogic{} using the $\echoRec$ function from~\autoref{sec:introduction} as a running example.
\begin{align*}
  \echoRec() \eqdef {} & \Hlet{\Hvar}{\Hcall{\getcId}{}}~\Hcall{\putcId}{\Hvar}
\end{align*}
$\echoRec$ is implemented in the \thelangRec{} language provided by DimSum. For this discussion, it is sufficient to see \thelangRec{} as a language with recursion, booleans, integers, function pointers, and an abstract heap where local variables are allocated.
$\echoRec$ ``echoes'' a character by first reading it using $\getcId$ and then outputting it using $\putcId$.
$\echoRec$ is interesting because there are many possible ways to implement $\getcId$ and $\putcId$:
For example, one can implement them in a \thelangRec{} library using more primitive functions (\autoref{sec:key:abscalls}), or in another language similar to \thelangRec{} (\autoref{sec:key:switching}), or they can be treated as external functions (\autoref{sec:key:extcalls}), or they could be implemented directly as system calls in assembly (\autoref{sec:key:wrapper}).

\begin{figure}[t]
  \small
  \newcommand{\rowsep}{\noalign{\vspace{2pt}{\color{black!30}\hrule}\vspace{2pt}}}%
  \begin{tcolorbox}[enhanced, colback=white, colframe=black, boxrule=0.4pt,
      arc=1pt, left=2pt, right=2pt, top=2pt, bottom=2pt]
    \textbf{\thelogic{}'s base logic}\par
    \vspace{1pt}{\color{black!30}\hrule}\vspace{2pt}
    {\renewcommand{\arraystretch}{1.3}%
    \renewcommand{\tabularxcolumn}[1]{m{#1}}%
    \noindent
    \begin{tabularx}{\linewidth}{@{}>{$}l<{$} >{\itshape\raggedright\arraybackslash}X r@{}}
\prop \mathbin{\circ} \propB \mid \All x.\, \prop(x) \mid \Exists x.\, \prop(x)
  \mid \bilabelSRev{\gotolbl}\prop(\gotolbl) \mid \ldots
  \quad (\circ \in \{\ast, \wand, \wedge, \vee\})
        & Base connectives & see \hyperref[par:key:table]{below} \\
      \rowsep
      \switchout{\lane}{\laneS}{\event}\prop \mid
      \switchin{\lane}{\laneS}{\event}\prop \mid
      \switchoutS{\lane}{\event}\prop \mid
      \switchinS{\lane}{\event}\prop \mid
      \switchsilent{\lane}\prop \mid
      \susp{\lane}{\prop} \mid
      \ldots
        & Switching assertions & see \autoref{sec:key:switching}, \autoref{sec:model:lanes} \\
      \rowsep
      \outtokenraw{\lane}{X}(s) \mid \intokenraw{\lane}{X}(s,\event) \mid
      \text{\faClock[regular]}^{\lane}_{X};\prop
      \mid \ldots
        & Assertions for lane state & see \autoref{sec:key:switching}, \autoref{sec:model} \\
    \end{tabularx}\par}
  \end{tcolorbox}

  \noindent\textit{Language-specific sublogics}\par\nointerlineskip
  \vspace{4pt}
  \newlength{\instgap}\setlength{\instgap}{3pt}%
  \newlength{\instD}\setlength{\instD}{0.35\linewidth}%
  \newlength{\instW}\setlength{\instW}{\dimexpr(\linewidth-\instD-3\instgap-1pt)/3\relax}%
  \newlength{\instWA}\setlength{\instWA}{0.85\instW}%
  \newlength{\instWO}\setlength{\instWO}{\dimexpr\instW+0.5\instW-0.5\instWA\relax}%
  \newcommand{\instrow}{\rule[-0.9ex]{0pt}{3.1ex}}%
  \newcommand{\instcolon}{\llap{\makebox[\instgap]{\bfseries\large :}}\,}%
  \newcommand{\instsep}{\noalign{\hbox{\tikz\draw[black!30, line width=0.4pt, dash pattern=on 1.5pt off 1.5pt] (0,0) -- (\linewidth,0);}}}%
  \newif\ifinstrule\instruletrue
  \newcommand{\instbox}[5][]{%
    \begin{tcolorbox}[enhanced, nobeforeafter, box align=top,
        width=#2, colback=white, colframe=black, boxrule=0.4pt, arc=1pt,
        left=2pt, right=2pt, top=2pt, bottom=2pt, #1]
      \strut\textbf{#3}\hfill#4\par
      \vspace{1pt}{\color{black!30}\ifx\relax#3\relax\instrulefalse\fi\ifinstrule\hrule\else\vspace{0.4pt}\hrule height 0pt\fi}\vspace{2pt}
      $\begin{array}[t]{@{}l@{}} #5 \end{array}$
    \end{tcolorbox}}%
  \noindent\makebox[\linewidth][l]{%
  \instbox{\instWO}{\thelangRec{}}{\autoref{sec:key:abscalls}, \autoref{sec:rec-asm-pseudo}}{
    \instrow \loc \hptsto \val \\ \instsep
    \instrow \fninternal{\fnname}{\mathimp{fn}} \mid \fnext{\fnname} \\ \instsep
    \instrow \simgenexprsimpl[\PiRec]{\exprimp}{\Phi} \\ \instsep
    \instrow \bicall{\fnname}{\vals}{\val}\prop(\val) \mid \recwait \prop}\hspace{\instgap}%
  \instbox{\instWA}{\thelangAsm{}}{\autoref{sec:key:wrapper}, \autoref{sec:rec-asm-pseudo}}{
    \instrow \addrraw \mptsto \aval \mid \reg{} \rptsto \aval \\ \instsep
    \instrow \inssinternal{\insaddr}{\instrs} \mid \insext{\insaddr} \\ \instsep
    \instrow \simgenexprsimpl[\PiAsm]{\instr}{{-}} \\ \instsep
    \instrow \bijump \mid \asmwait}\hspace{\instgap}%
  \instbox{\instWO}{Pseudo}{\autoref{sec:key:switching}, \autoref{sec:rec-asm-pseudo}}{
    \instrow \\ \instsep
    \instrow \fninternalSpec{\fnname}{\mathsf{fn}} \mid \fnextSpec{\fnname} \\ \instsep
    \instrow \simgenexprsimpl[\PiSpec]{\expr}{\Phi} \\ \instsep
    \instrow \speccall{\fnname}{\valsSpec}{\valSpec}\prop(\valSpec) \mid \specwait \prop}\hspace{2pt}%
  \instrulefalse
  \instbox[frame hidden, left=0pt, right=0pt]{\instD}{{\color{black!50}\instcolon\textmd{see \appendixrefshort{} for full \ifappendixincluded\else language \fi definitions}}}{}{
    \instrow \instcolon \textit{Points-to assertions} \\ \instsep
    \instrow \instcolon \textit{Internal/external code assertions} \\ \instsep
    \instrow \instcolon \textit{Weakest precondition} \\ \instsep
    \instrow \instcolon \textit{\Abscalls and waiting modality}}}\par
  \vspace{\instgap}
  \noindent\makebox[\linewidth][l]{%
  \instbox{\linewidth}{\thelangRec{}-to-\thelangAsm{}}{\autoref{sec:key:wrapper}, \autoref{sec:rec-asm-wrapper}}{
    \rtoaexchange(\exchanged)\,\, (\textit{Exchange}) \,\,\mid\,\, \rtoaclosed \,\, (\textit{Resource lock}) \,\,\mid\,\, \loc \itoabij \addrraw \,\, (\textit{Injection}) \,\, \mid\,\, \rtoawait{\lane} \,\, (\textit{Waiting modality}) \,\,\mid \ldots}}

  \caption{The \thelogic{} logic: language-independent assertions (top) and the language-specific sublogics (bottom).}
  \label{fig:structure}
\end{figure}

\paragraph{Structure of \thelogic{}}\phantomsection\label{par:key:table}
Before we describe how \thelogic{} addresses this challenge, we first give a brief overview of the components of the \thelogic{} program logic depicted in \autoref{fig:structure}.

First, \thelogic{} provides a language-independent base logic, which contains the connectives of higher-order separation logic\footnote{\thelogic{} does \emph{not} support higher-order ghost state since DimSum is incompatible with (finite) step-indexing, see \autoref{sec:introduction}.} inherited from Iris~\cite{IrisGroundUp}, assertions for switching between languages and tokens for tracking the state of DimSum combinators. The latter two are introduced in \autoref{sec:key:switching}.

Second, \thelogic{} allows the user to define sublogics containing the program logic assertions for concrete languages. In this paper, we consider three languages: \thelangRec{} introduced above, the assembly language \thelangAsm{} (\autoref{sec:key:wrapper}), and \thelangSpec{}, a pseudocode language for writing specifications (\autoref{sec:key:switching}).
Considered individually, each sublogic behaves like a standard program logic for that language, with points-to assertions denoting ownership of memory locations ($\loc \hptsto \val$, $\addrraw \mptsto \aval$), assertions stating that a function has a specific implementation in this language ($\fninternal{\fnname}{\mathimp{fn}}$, $\inssinternal{\insaddr}{\instrs}$, $\fninternalSpec{\fnname}{\mathsf{fn}}$) or is not implemented in this language ($\fnext{\fnname}$, $\insext{\insaddr}$, $\fnextSpec{\fnname}$), and Iris-style weakest preconditions ($\simgenexprsimpl[\PiRec]{\exprimp}{\Phi}$, $\simgenexprsimpl[\PiAsm]{\instr}{{-}}$, $\simgenexprsimpl[\PiSpec]{\expr}{\Phi}$). These assertions enable standard Iris-based verification inside that language. Additionally, each sublogic has a few \thelogic{} constructs for reasoning about calls to other languages. We will introduce these constructs in the following sections.

A proof in \thelogic{} is not confined to a single sublogic, but can pick an arbitrary set of sublogics. Crucially, a proof with one set of sublogics is automatically valid in any larger set of sublogics. This means that we can verify a \thelangRec{} function like $\echoRec$ purely in the \thelangRec{} logic, and later reuse this proof when linking $\echoRec$ with \thelangSpec{} or with \thelangAsm{}.
Additionally, we can add new sublogics for relating the assertions of different languages like the  \thelangRec{}-to-\thelangAsm{} sublogic, which will be presented in \autoref{sec:key:wrapper}.
All of this is achieved thanks to Iris' flexible ghost state mechanism.
The title of each of the following subsections indicates which sublogics it uses.

\subsection[Challenge \#1: Modular Specifications (Rec)]{Challenge \#1: Modular Specifications (\thelangRec{})}
\label{sec:key:abscalls}
Our first challenge is to provide a specification for $\echoRec$ that abstracts over, and can be reused across, all the different implementations of $\getcId$ and $\putcId$.

\paragraph{Specifying $\echoId$: \Abscalls}
The standard recipe for separation logic verification would be to first come up with a specification (\eg Hoare triples) for $\getcId$ and $\putcId$ and then use these specifications in the verification of $\echoRec$.
However, this approach poses a problem: What specifications can we assume for $\getcId$ and $\putcId$ if their implementations are unknown?
In fact, calling $\getcId$ and $\putcId$ might require arbitrary separation logic resources that we do not know about.

To address this problem, we introduce the \emph{\abscall} construct $\bicall{\fnname}{\vals}{\valS} \rettok$.
This construct allows us to flip the problem on its head:
Instead of resolving the calls to $\getcId$ and $\putcId$ during the verification of $\echoId$, \abscalls enable the specification of $\echoId$ to shift the responsibility of resolving the calls to $\getcId$ and $\putcId$ to the caller of $\echoId$.%
\footnote{This basic idea is inspired by \citet{VeriFastIO} and \citet{HSSL}, though the details differ. See \autoref{sec:related-work}.}
Concretely, we provide the following specification for $\echoRec$:
\begin{mathpar}
  \inferH{spec-echo}{
    \fninternal{\echoId}{\echoRec}\\
    \bicall{\getcId}{}{\val}
    \bicall{\putcId}{\val}{\valS}
    \rettok(\valS)
}{\bicall{\echoId}{}{\val}\rettok(\val)}
\and
  \inferH{wp-call}{
    \bicall{\fnname}{\vals}{\val}{\rettok(\val)}
}{\simgenexprsimpl[\PiRec]{\fnname(\vals)}{\val. \rettok(\val)}}
\end{mathpar}
The specification \ruleref{spec-echo} looks different from standard separation logic specifications, so let us walk through it step by step.
First, we notice that this specification is not phrased as a Hoare triple, but instead as an inference rule (inside the logic of \thelogic{}) where the conclusion is an \abscall to $\echoId$.
Such \abscalls are created by \ruleref{wp-call}, which turns the verification of a function call to $\fnname$ into an \abscall to $\fnname$.
To resolve an \abscall to $\echoId$ using \ruleref{spec-echo}, we have to prove the premises of the rule.
The first premise $\fninternal{\echoId}{\echoRec}$ states that this rule only applies if the function $\echoId$ is actually implemented by $\echoRec$.%
\footnote{It is important to distinguish the \emph{name} of the function $\echoId$ from its \emph{implementation} $\echoRec$.}
The actual specification is the second premise:
It states that an \abscall to $\echoId$ corresponds to an \abscall to $\getcId$ followed by an \abscall to $\putcId$ with the result of $\getcId$ as the argument.
The \abscalls enable this specification of $\echoRec$ to capture the precise behavior of the implementation without assuming any details about how $\getcId$ and $\putcId$ are implemented.
We will see in the rest of \autoref{sec:keyideas} how these \abscalls can be resolved and how \ruleref{spec-echo} can be reused across many different implementations of $\getcId$ and $\putcId$.

\paragraph{Specifying $\getcId$ and $\putcId$: Separation logic}
To see how \abscalls integrate with separation logic assertions, consider the following \thelangRec{} implementations of $\getcId$ and $\putcId$:
\begin{align*}
  \getcRec() \eqdef {} & {\Hlocal{\Hvar}{1};\Hcall{\readId}{\Hvar, 1}};\Hload{\Hvar} &
  \putcRec(\val) \eqdef {} & {\Hlocal{\Hvar}{1};\Hstore{\Hvar}{\val}; \Hcall{\writeId}{\Hvar, 1}}
\end{align*}
$\getcRec$ calls $\readId$ to read a single value into the (stack-allocated) local variable $\Hvar$ of size 1.
In \thelangRec, local variables are allocated in memory and thus the variable $\Hvar$ represents a memory location.
$\putcRec$ outputs its argument via the $\writeId$ function.
For both $\readId$ and $\writeId$, the second argument denotes the number of values read into or written from the location passed as the first argument.
Using \abscalls, we can specify $\getcRec$ and $\putcRec$ as follows:
\begin{mathpar}
  \inferH{spec-getc}{
    \fninternal{\getcId}{\getcRec}\\
    \All \loc, \val. \loc \hptsto \val \wand
    \bicallS{\readId}{\loc, 1}
    \Exists \valS. \loc \hptsto \valS \ast
    \rettok(\valS)
}{\bicall{\getcId}{}{\val}\rettok(\val)}
\and
  \inferH{spec-putc}{
    \fninternal{\putcId}{\putcRec}\\
    \All \loc. \loc \hptsto \val \wand
    \bicall{\writeId}{\loc, 1}{\valS}
    \loc \hptsto \val \ast
    \rettok(\valS)
}{\bicall{\putcId}{\val}{\valS}\rettok(\valS)}
\end{mathpar}
Let us start with \ruleref{spec-getc}:
It treats $\readId$ as abstract (similar to how $\getcId$ and $\putcId$ are abstract in \ruleref{spec-echo}).
This means the obligation to resolve the call to $\readId$ is shifted to the caller of $\getcRec$.
For the caller to be able to do this resolution, \ruleref{spec-getc} needs to state that $\readId$ is called with a valid memory location $\loc$.
This information is encoded in the specification by providing a points-to predicate $\loc \hptsto \val$ to the caller using the magic wand ($\wand$).
The caller can use this points-to predicate to verify the call to $\readId$ (in fact, we will see this in \autoref{sec:key:extcalls}).
Afterwards, the caller has to give back the points-to predicate with the read value $\valS$, which is then returned from $\getcRec$.
The specification for $\putcRec$ is similar, except that it calls $\writeId$ instead of $\readId$ and expects that $\writeId$ does not update the value stored at $\loc$.
These examples show how \abscalls can be combined with other separation logic constructs in specifications.

\paragraph{Specifying $\mainId$: Loops}
Consider the following $\mainId$ function that runs $\echoId$ in a loop:
\begin{align*}
  \mainRec() \eqdef {} & {\Hcall{\echoId}{}};\Hcall{\mainId}{}
\end{align*}
Usually, a diverging function like $\mainId$ would be characterized by a triple with a trivial postcondition.
However, here we do not just care about the fact that $\mainId$ diverges, but also that it calls $\echoId$ in the loop.
\Abscalls allow us to specify this behavior of $\mainId$ using the (greatest) fixpoint operator $\bilabelSRev{\gotolbl}P$.
Concretely, we specify $\mainId$ as follows, capturing that it calls $\echoId$ in a loop:
\begin{mathpar}
  \inferH{spec-main}{
    \fninternal{\mainId}{\mainRec}
    \\
    \bilabelSRev{\gotolbl}
    \bicall{\echoId}{}{\val}
    \bigotoS{\gotolbl}
}{\bicall{\mainId}{}{\val}\rettok}
\end{mathpar}
\paragraph{What have we seen?}
This section presented \abscalls and how they can be used to specify calls to unknown functions by shifting the responsibility of resolving the call to the caller. The next sections show how such \abscalls can be resolved across languages.
\subsection{Background: Multi-Language Semantics in DimSum}\label{sec:key:dimsum}
Before we dive into multi-language reasoning with \thelogic{}, let us first provide some background for its underlying multi-language semantics provided by DimSum.

\paragraph{A \thelangSpec{} program}
To talk about multi-language programs in DimSum, we introduce a second language: \thelangSpec{}. \thelangSpec{} is a pseudocode language for mathematical descriptions of algorithms based on DimSum's \thelangSpecOrig{} language.
\thelangSpec{} is useful for two use cases:
First, \thelangSpec{} can be used as a specification language on the specification side of the refinement to show that the external I/O behavior of an implementation follows some pseudocode algorithm.
Second, \thelangSpec{} specifications can be used on the implementation side for expressing multi-language programs, where one does not want to link with a concrete implementation of an algorithm, but with an abstract implementation.
In this and the next section we focus on the second point, while \autoref{sec:key:extcalls} discusses the first.

As a concrete example, consider the following \thelangSpec{} implementation of $\putcId$:
{ \small
\begin{align*}
  \putcPseudo(\val) \eqdef {}
  &\Pupd{\Pvar{buf}}{\Pvar{buf} \dplus [\val]};
  \Pifthen{*}\{\Plet{\loc}{\Pcmd{alloc}(|\Pvar{buf}|)}~ \Pstore{\loc}{\Pvar{buf}};~ \PcallN{\Pfname{\writeId}}{\loc, |\Pvar{buf}|};~ \Pupd{\Pvar{buf}}{[]}\}~ \Pret{0};
\end{align*}
}
$\putcPseudo$ expresses an abstract algorithm for buffered writes:
It adds its argument to the buffer $\Pvar{buf}$, which is then non-deterministically (denoted by $\mathit{if~*}$) flushed using $\writeId$.
This \thelangSpec{} implementation abstracts over concrete implementations of $\putcId$ (\eg in \thelangRec{}) in two ways:
First, it uses non-determinism to overapproximate different concrete flushing strategies.
Second, it uses a mathematical list type to model the buffer, abstracting over its concrete implementation.
To link $\putcPseudo$ with $\echoRec$, we rely on DimSum's multi-language semantics, which we review next.

\paragraph{DimSum's multi-language semantics}
In DimSum, program semantics are defined as (non-deterministic) state transition systems called \emph{modules} that synchronize on \emph{events} (similar to process algebra). A module is invoked by synchronizing with an incoming event, \eg a call to one of its functions. While executing, it may interact with the environment by emitting outgoing events, after which it enters a \emph{waiting} state, waiting to synchronize on an incoming event.

To build multi-language programs, DimSum provides \emph{semantic linking} operators ($\link_X$) that combine two modules over the same set of events into a new module. The combined module forwards events between its components and the environment.
Since linking only relies on events, the two modules can be written in different languages, as long as they agree on the kind of events they exchange. (We will discuss heterogeneous languages with different event types in \autoref{sec:key:wrapper}.)

\paragraph{Linking \thelangRec{} and \thelangSpec{} using DimSum}
Now, let us return to our running example and combine the \thelangRec{} functions $\echoRec$ and $\getcRec$ with the \thelangSpec{} function $\putcPseudo$ via semantic linking. The $\impsyn{\reclib{}}$ operator turns the syntactic \thelangRec{} program $\reclib{}$ into a \thelangRec{} module, similarly $\specsyn{\cdot}$ for \thelangSpec{}. For \thelangRec{}, DimSum provides the linking operator $\implink{\moduleImpA}{\moduleImpB}$ that synchronizes the outgoing events of the module $\moduleImpA$ with incoming events of the module $\moduleImpB$ and vice versa.
Its events can be calls $\ECall{\fnname}{\vals}{\heap}$ of the function $\fnname$ with arguments $\vals$ and current heap $\heap$\footnote{Note that in the rest of this and the next section, we omit the heap for simplicity. See \autoref{sec:rec-asm-pseudo} for how heaps are integrated.}, or returns $\EReturn{\val}{\heap}$ with return value $\val$ and current heap $\heap$.
While $\putcPseudo$ and $\echoRec$ are written in different languages, they share the same events and thus can be directly (semantically) linked.
Putting everything together, we consider the following multi-language program that links $\echoRec$ and $\getcRec$ in \thelangRec{} with $\putcPseudo$ in \thelangSpec{}:
\[
  \echobuflib \eqdef \implink{\impsyn{\impsynlink{\echoRec}{\getcRec}}}{\specsyn{\putcPseudo}}
\]
Note that $\impsynlinkraw$ denotes ordinary syntactic linking of \thelangRec{} programs in contrast to $\implinkraw$ which allows us to link \thelangRec{} and \thelangSpec{} modules.

\subsection[Challenge \#2: Switching between Languages (Rec + Pseudo)]{Challenge \#2: Switching between Languages (\thelangRec{} + \thelangSpec{})}\label{sec:key:switching}
Let us now see how \thelogic{} enables verification of $\echobuflib$. Crucially, we want to reuse \ruleref{spec-echo} and resolve its \abscall to $\putcId$ with a \thelangSpec{} specification of $\putcPseudo$.

\paragraph{A \thelangSpec{} specification for $\putcId$}
Using the \thelangSpec{} sublogic of \thelogic{} and reusing the \thelangRec{} points-to assertions,  we can prove the following \thelogic{} specification for $\putcPseudo$:
\begin{mathpar}
  \inferH{spec-putc-pseudo}{
    \fninternalSpec{\putcId}{\putcPseudo}\\
    \textlog{buffer}(b) \\
    \raise1.5ex\hbox{$\bigwedge$}
    \begin{array}[b]{l}
    \textlog{buffer}(b \dplus [\val]) \wand
      \rettok(0)\cr
    \All \loc. \loc \hptsto (b \dplus [\val])\wand
      \speccallS{\writeId}{\loc, \lengthOf{b} + 1}
      \loc \hptsto (b \dplus [\val]) \ast
      (\textlog{buffer}([]) \wand
      \rettok(0))
    \end{array}}{\speccall{\putcId}{\val}{\valS}\rettok(\valS)}
\end{mathpar}
The state of $\putcId$'s buffer is tracked by the abstract predicate $\textlog{buffer}(b)$.
To prove an \abscall to $\putcId$ implemented by $\putcPseudo$, one has to give up ownership of $\textlog{buffer}(b)$, stating that the buffer currently contains $b$.
Then, one needs to prove both branches of the non-deterministic choice (expressed using the non-separating conjunction $\wedge$). In the first branch, the argument $\val$ is added to the buffer and $\putcId$ returns.
In the second branch, the buffer is flushed: $\putcPseudo$ produces a location $\loc$ containing the updated buffer, performs an \abscall to $\writeId$, requires the ownership of $\loc$ to be given back, and finally continues in $\rettok$ with $\textlog{buffer}([])$, stating that the buffer is now empty.

\paragraph{Reasoning about linking in \thelogic{}}
Let us now see how \thelogic{} enables modular reasoning about $\echobuflib$.
Concretely, we want to resolve the \abscall to $\putcId$ in \ruleref{spec-echo} using \ruleref{spec-putc-pseudo}.
For this, we need to \emph{switch} from \thelangRec{} to \thelangSpec{}. How can we handle this formally?

First, \thelogic{} introduces the concept of a \emph{\piname}:
Each component of a multi-language program (formally, each module) is assigned a \piname.
We use $\PiRec$ to refer to the \piname of $\thelangRec{}$ and $\PiSpec$ for the \piname of $\thelangSpec{}$.%
\footnote{Here we assume each language has at most one \piname. \thelogic{} supports having the same language in multiple lanes.}
Intuitively, a \piname describes where (the module of) the language is situated in the multi-language program.
The multi-language program above has two \pinames:
The left side of the linking operator $\implinkraw$ and its right side.
The left side contains the \thelangRec{} module, thus $\PiRec \eqdef \linkRecLPi{\Pi}$, while the right side contains the \thelangSpec{} module, thus $\PiSpec \eqdef \linkRecRPi{\Pi}$.
This linking operator appears inside a larger context (\eg inside another linking operator) represented by the \piname $\Pi$.
For this section, this \piname $\Pi$ is not relevant.
When verifying a specific function, the \piname is kept abstract, allowing the specification (\eg \ruleref{spec-echo}) to be reused in different multi-language programs. %

Second, to switch between \pinames, we introduce the \emph{switching modality} $\switchout{\Pi}{\Pi'}{\event}P$.
This modality states that we emit the event $\event$ in the \piname $\Pi$ and then continue with proving $P$ in \piname $\Pi'$.
There are a few variants of this modality:
$\switchinto{\Pi}{\Pi'}{\event} P$ handles incoming instead of outgoing events.
$\switchoutS{\Pi}{\event}P \eqdef \switchout{\Pi}{\Pi}{\event}P$ denotes the special case where the \piname stays the same (and similar for $\switchinS{\Pi}{\event} P$).
Incoming events use a left arrow $\leftarrow$ to bind variables that are determined by the incoming event.

\begin{figure}
  \centering
\begin{mathpar}
  \inferhref{switch-rec-call}{key-switch-rec-call}{\fnext{\fnname} \\ \switchoutS{\PiRec}{\ECallH(\fnname, \vals)}
      \recwait
      \switchin{\PiRec}{\EReturnH}{\valS}\rettok(\valS)}
    {\bicall{\fnname}{\vals}{\val}\rettok}
  \and
  \inferhref{switch-redirect}{key-switch-redirect}{\susp{\Pi'}{P'} \\ \susp{\Pi}{P} \wand \switchout{\Pi}{\Pi'}{\event}{P'}}
    {\switchoutS{\Pi}{\event}P}
    \and
      \inferhref{switch-rec-link-l-to-r}{key-switch-rec-link-l-to-r}{\fnname \in \textlog{dom}(\linkRecRPi{\Pi}) \\ \linkRecouttoken{} \\ \linkRecintoken{}(\ECallH(\fnname, \vals)) \wand P}
    {\switchout{\linkRecLPi{\Pi}}{\linkRecRPi{\Pi}}{\ECallH(\fnname, \vals)}P}
    \and
    \inferhref{switch-rec-link-r-in}{key-switch-rec-link-r-in}{\linkRecintoken{}(\ECallH(\fnname, \vals)) \\ \linkRecouttoken{} \wand P~\fnname~\vals }
    {\switchin{\linkRecRPi{\Pi}}{\ECallH}{\fnname, \vals} P}
  \and
  \inferhref{pseudo-wait-call}{key-pseudo-wait-call}{
    \switchin{\PiSpec}{\ECallH}{\fnname, \vals}
    \speccall{\fnname}{\vals}{\val}
    \switchoutS{\PiSpec}{\EReturnH(\val)}
    \specwait
    P}
    {\specwait P}
\end{mathpar}

  \caption{Switching rules (simplified)}
  \label{fig:key:switch-rules}
\end{figure}

\begin{figure}
  \centering
  \setcounter{proofstepctr}{0}%
  \[
  \begin{array}{ l@{\quad} >{\displaystyle}r @{\ \vdash\ } >{\displaystyle}l }
  \proofvdots \\[-.2em]
  \stepnum{putcproof:reccall} & \ldots \ast \fnext{\putcId} \ast \linkRecouttoken{} \ast \susp{\PiSpec}{\specwait \bot} & \bicall{\putcId}{\val}{\valS} \rettok(\valS) \\
\proofstep{\ruleref{key-switch-rec-call}}
  \stepnum{} & \ldots \ast \fnext{\putcId} \ast \linkRecouttoken{} \ast \susp{\PiSpec}{\specwait \bot} & \switchoutS{\PiRec}{\ECallH(\putcId, [\val])} \recwait \switchin{\PiRec}{\EReturnH}{\valS} \rettok(\valS) \\
\proofstepN{\ruleref{key-switch-redirect}}{2pt}{-1pt}
  \stepnum{putcproof:redirect} & \ldots \ast \linkRecouttoken{} \ast \susp{\PiRec}{\recwait \switchin{\PiRec}{\EReturnH}{\valS} \rettok(\valS)} & \switchout{\PiRec}{\PiSpec}{\ECallH(\putcId, [\val])} \specwait \bot \\
\proofstep{\ruleref{key-switch-rec-link-l-to-r}}
  \stepnum{putcproof:wait} & \ldots \ast \linkRecintoken{}(\ECallH(\putcId, [\val])) \ast \susp{\PiRec}{\ldots} & \specwait \bot \\
\proofstep{\ruleref{key-pseudo-wait-call}}
  \stepnum{putcproof:pseudoin} & \ldots \ast \linkRecintoken{}(\ECallH(\putcId, [\val])) \ast \susp{\PiRec}{\ldots} &
    \switchin{\PiSpec}{\ECallH}{\fnname, \vals}
    \speccall{\fnname}{\vals}{\val}\ldots \\
\proofstep{\ruleref{key-switch-rec-link-r-in}}
  \stepnum{putcproof:speccall} & \ldots \ast \linkRecouttoken{} \ast \susp{\PiRec}{\ldots} &
    \speccall{\putcId}{\val}{\valS} \switchoutS{\PiSpec}{\EReturnH(\valS)}
    \specwait
    \bot \\
\proofstepN{\ruleref{spec-putc-pseudo}}{2pt}{0pt}
  \end{array}
  \]
  \caption{Calling $\putcPseudo$ from $\echoRec$}
  \label{fig:key:putcproof}
\end{figure}

Operationally, a switch corresponds to a module in one \piname{} emitting an event that is received by the module in another \piname.
After emitting an event, a module typically enters its waiting state, reflected by waiting modalities (\eg $\recwaitraw$, $\specwaitraw$). While control is elsewhere, the \piname is suspended, reflected by the \emph{suspended token} $\susp{\Pi}{P}$: once control returns to $\Pi$, its proof continues with $P$. Tokens track the state of a \piname: $\outtokenraw{\Pi}{X}$ states that it can emit events, while $\intokenraw{\Pi}{X}(\event)$ states that it has to synchronize with the incoming event $\event$.

\paragraph{Calling $\putcPseudo{}$ from $\echoRec{}$}
Let us see these assertions in action by using the rules in \autoref{fig:key:switch-rules} to resolve the \abscall from $\echoRec$ to $\putcPseudo$.
At a high level, the resolution of the \abscall happens in three steps:
First, we start in \thelangRec{}.
Second, we switch through the linking operator.
Finally, we end up in \thelangSpec{}.
The rules in \autoref{fig:key:switch-rules} can be classified into these three categories:
First, \ruleref{key-switch-rec-call} turns the \thelangRec{} \abscall into a switching modality.
Second, \ruleref{key-switch-redirect}, \ruleref{key-switch-rec-link-l-to-r},
and \ruleref{key-switch-rec-link-r-in} handle the routing of the event through the linking. We call such rules \emph{routing rules}.
Finally, \ruleref{key-pseudo-wait-call} ensures that \thelangSpec{} can receive the incoming call.

Let us now walk through the proof more concretely.
The proof steps are shown in \autoref{fig:key:putcproof}.
We start at \stepref{putcproof:reccall} where we need to prove the \abscall to $\putcId$. This state is reached after resolving the \abscall to $\getcId$ in \ruleref{spec-echo}.
As the first step, we apply the rule \ruleref{key-switch-rec-call}.
This rule states that an \abscall to a non-\thelangRec{} function $\fnname$ (denoted by $\fnext{\fnname}$) first emits an outgoing $\ECallH$ event ($\switchoutSraw{\PiRec}{\ECallH(\fnname, \vals)}$), then waits for possible incoming mutually recursive calls ($\recwaitraw$)
and finally receives an incoming $\EReturnH$ ($\switchinraw{\PiRec}{\EReturnH}{\valS}$).
After applying this rule, we need to resolve the outgoing $\ECallH$.

This is where we leave \thelangRec{} and enter the reasoning about the linking operator $\implinkraw$.
First, we use one of the most important \thelogic{} rules: \ruleref{key-switch-redirect} (here in a specialized form, see \autoref{sec:model:lanes}).
This rule allows us to redirect a switch from its current target \piname $\Pi$ to an arbitrary other \piname $\Pi'$.
In our case, this requires giving up $\susp{\PiSpec}{\specwait\bot}$, stating that the \thelangSpec{} \piname $\PiSpec$ is currently suspended in the waiting state $\specwaitraw$.
After applying \ruleref{key-switch-redirect} (\stepref{putcproof:redirect}), the \thelangRec{} \piname $\PiRec$ is suspended and the switch targets $\PiSpec$.
Next, we use the routing rule \ruleref{key-switch-rec-link-l-to-r} to perform the switch from the left side of the semantic linking operator to its right side. (Recall that in this example we have $\PiRec \eqdef \linkRecLPi{\Pi}$ and $\PiSpec \eqdef \linkRecRPi{\Pi}$.)
The routing rules are guarded by tokens---here $\linkRecouttoken{}$ and $\linkRecintoken{}(\event)$---that ensure that the linking operator can route the event.
In this case, \ruleref{key-switch-rec-link-l-to-r} requires the $\linkRecouttoken{}$ token, which states that the linking operator can currently handle outgoing events.
It gets transformed into the $\linkRecintoken{}(\ECallH(\fnname, \vals))$ token, stating that the module inside the linking operator must synchronize on the incoming call.
The rule \ruleref{key-switch-rec-link-r-in} transforms $\linkRecintoken{}$ back into $\linkRecouttoken{}$.

Finally (\stepref{putcproof:wait}), we arrive in \thelangSpec{}. We apply \ruleref{key-pseudo-wait-call}, which states that the waiting state $\specwaitraw$ can receive a call, invoke the corresponding function and return the result, before entering the waiting state again.
To resolve the incoming call, we use the routing lemma \ruleref{key-switch-rec-link-r-in}.
Here we give up the $\linkRecintoken{}(\ECallH(\putcId, [\val]))$ token, which determines that the incoming call is a call to $\putcId$ with argument $\val$.
Now we are finally at the \abscall to $\putcPseudo$ (\stepref{putcproof:speccall}), where we can apply its specification \ruleref{spec-putc-pseudo}.
The return from $\putcPseudo$ to $\echoRec$ uses similar rules in the reverse direction.

This proof shows how to perform the switch using the primitive rules provided by \thelogic{}.
To avoid repeating these steps and directly go from \stepref{putcproof:reccall} to \stepref{putcproof:speccall}, one can prove a derived rule:
\begin{mathpar}
  \inferH{rec-to-pseudo}{\fnext{\fnname} \\ \fnname \in \textlog{dom}(\PiSpec) \\ \linkRecouttoken{} \\ \susp{\PiSpec}{\specwait P} \\ \linkRecouttoken{} \wand \susp{\PiRec}{(\recwait \switchin{\PiRec}{\EReturnH}{\valS}\rettok(\valS))} \wand \speccall{\fnname}{\vals}{\val}
    \switchoutS{\PiSpec}{\EReturnH(\val)}
    \specwait P
    }
  {\bicall{\fnname}{\vals}{\val}\rettok(\val)}
\end{mathpar}
This rule shows that for the $\PiRec$ and $\PiSpec$ in this multi-language program, a \thelangRec{} abstract call to a function implemented in \thelangSpec{} resolves to an abstract call in \thelangSpec{} with \thelangRec{} suspended.

\paragraph{What have we seen?}
This section introduced three key concepts of \thelogic{}: \emph{\pinames}, the \emph{switching modality} $\switchout{\Pi}{\Pi'}{\event}P$, and \emph{suspended tokens} $\susp{\Pi'}{P'}$.
When emitting an event, we use \ruleref{key-switch-redirect} to select a suspended \piname as the target of the switch and then resolve the switch using rules like \ruleref{key-switch-rec-link-l-to-r}.
This is how \thelogic{} allows switching between languages inside its logic.
Crucially, all specifications like \ruleref{spec-echo} are parametric over their lane, allowing one to reuse these specifications in different multi-language programs, as we will see in the next sections.

\subsection[Challenge \#3: External Calls (Rec + Pseudo)]{Challenge \#3: \ExtCalls (\thelangRec{} + \thelangSpec{})}\label{sec:key:extcalls}
The previous section showed how \pinames enable reasoning about function calls across different languages.
In this section, we take this principle one step further and show how \pinames can also be used to reason about calls to functions that are \emph{not} implemented as part of the program.
We refer to such calls as \emph{external calls}.

\paragraph{Specifying \ext calls}
To explain external calls, let us come back to our running example and consider the program consisting of all \thelangRec{} functions from \autoref{sec:key:abscalls}:
\[\echolib \eqdef {} \impsynlink{\mainRec}{\impsynlink{\echoRec}{\impsynlink{\getcRec}{\putcRec}}}\]
For this program, $\mainId$, $\echoId$, $\getcId$, and $\putcId$ are internal functions, but $\readId$ and $\writeId$ (called by $\getcRec$ resp. $\putcRec$) are \emph{external} functions since they are not defined as part of $\echolib$.
To reason about these external functions, we follow DimSum (and many other works) and prove a \emph{refinement} between the program $\echolib$ and a specification program.
Concretely, we show that all external calls in our implementation program $\echolib$ can be matched by external calls in the specification program.
The external calls can be seen as the I/O operations of the program, and the specification program describes the allowed I/O behavior of the implementation.
Concretely, for our example we define the following \thelangSpec{} function $\mainPseudo$ as the specification for $\echolib$:
\begin{align*}
  &\mainPseudo() \eqdef {}
  \Ploop{{\Pexists{\Pvalid{\loc}};~ \PcallN{\readId}{\loc, 1};~ \Plet{\Pvar{x}}{\Pload{\loc}}~\Pexists{\Pvalid{\locS}};~ \Passert{\Pvar{x} = \Pload{\locS}};~ \PcallN{\writeId}{\locS, 1};}}
\end{align*}

$\mainPseudo$ abstractly describes the behavior of $\echolib$: In an infinite loop, it calls $\readId$ with  a non-deterministically chosen, valid (\ie allocated in a new underlying heap) location $\loc$. It then calls $\writeId$ with a (potentially different) valid location $\locS$ that stores the read value. This specification captures the ``echo'' behavior at the level of $\readId$ and $\writeId$.
$\mainPseudo$ allows us to state the final goal we want to prove: The (DimSum) refinement $\impsyn{\echolib} \refinesraw \specsyn{\mainPseudo}$. Let us now see how \thelogic{} allows us to prove this refinement by reusing the specifications from \autoref{sec:key:abscalls}.

\paragraph{Binary reasoning via unary specifications}
We prove this refinement in two steps:%
\footnote{\thelogic{} does not require the split into two steps. If one is not interested in a specification, one can also do all reasoning in one step.}
First, we prove a (language-local) specification for $\mainPseudo$.
Second, we perform a cross-language proof (similar to \autoref{sec:key:switching}) to show that the specifications of $\echolib$ (from \autoref{sec:key:abscalls}) and $\mainPseudo$ match.

Importantly, for the first step we reuse the \emph{same \thelogic{} constructs} from prior sections (\eg \abscalls), but this time for programs on the specification side of the refinement.
Concretely, we prove the following specification for $\mainPseudo$:
\begin{mathpar}
  \inferH{spec-main-pseudo}{
    \fninternalSpec{\mainId}{\mainPseudo}\\
    \bilabelSRev{\gotolbl}
    \Exists \loc, \val . \loc \hptsto \val \ast~
    \speccallS{\readId}{\loc, 1}
    \All \valS . \loc \hptsto \valS \wand
    \switchsilent{}
    \Exists \locS . \locS \hptsto \valS \ast
    \speccallS{\writeId}{\locS, 1}
    \locS \hptsto \valS \wand
    \switchsilent{}
    \bigotoS{\gotolbl}
}{\speccall{\mainId}{}{\val}\rettok}
\end{mathpar}
This specification is similar to the specification we have seen in \autoref{sec:key:abscalls}, except that now the caller of $\mainId$ needs to \emph{provide} a location $\loc$ and value $\val$ with a points-to $\loc \hptsto \val$ to prove that the location $\loc$ is $\mathit{valid}$.
The caller gets back the points-to with the read value $\valS$, before having to provide a points-to with the same value to $\writeId$.
(The silent switches $\switchsilent{}P$ are explained below.)

\paragraph{Switching between implementation and specification}
\newcommand{\PiSrc}{\PiSpec}
\newcommand{\PiTgt}{\PiRec}
With the \thelogic{} specifications in hand, it remains to assemble them into the final refinement proof of $\impsyn{\echolib} \refinesraw \specsyn{\mainPseudo}$.
Following \autoref{sec:key:switching}, we first need to determine the \pinames $\PiTgt{}$ for $\echolib$ and $\PiSrc{}$ for $\mainPseudo$.
Recall that \pinames describe where (the module of) a language is situated in the multi-language program.
Here, $\echolib$ is on the implementation side of the refinement, so we define $\PiTgt{}$ as the \emph{implementation \piname} $\tgtPi{\thelangImpSub}$.
Conversely, $\mainPseudo$ is on the specification side of the refinement, so we define $\PiSrc{}$ as the \emph{specification \piname} $\srcPi{\thelangImpSub}$.
\begin{figure}
  \centering
  \begin{mathpar}
    \inferhref{switch-ref-i-in}{key-switch-ref-i-in}
    {\tgtintoken{\thelangImpSub}(\event) \\ \event = \event' \wand \srcintoken{\thelangImpSub}(\event) \wand P}
    {\switchinto{\tgtPi{\thelangImpSub}}{\srcPi{\thelangImpSub}}{\eventS} P}
    \and
    \inferhref{switch-ref-s-in}{key-switch-ref-s-in}
    {\srcintoken{\thelangImpSub}(\event) \\ \tgtouttoken{\thelangImpSub} \wand P}
    {\switchinto{\srcPi{\thelangImpSub}}{\tgtPi{\thelangImpSub}}{\event} P}
    \and
    \inferhref{switch-ref-i-out}{key-switch-ref-i-out}
    {\tgtouttoken{\thelangImpSub} \\ \srcouttoken{\thelangImpSub}(\event) \wand P}
    {\switchout{\tgtPi{\thelangImpSub}}{\srcPi{\thelangImpSub}}{\event} P}
    \and
    \inferhref{switch-ref-s-out}{key-switch-ref-s-out}
    {\srcouttoken{\thelangImpSub}(\event) \\ \All \event. \tgtintoken{\thelangImpSub}(e) \wand P}
    {\switchout{\srcPi{\thelangImpSub}}{\tgtPi{\thelangImpSub}}{\event} P}
    \and
    \inferhref{switch-ref-i-silent}{key-switch-ref-i-silent}
    {\tgtouttoken{\thelangImpSub} \\ \srcouttoken{\thelangImpSub}(\silent) \wand P}
    {\switchoutsilent{\tgtPi{\thelangImpSub}}{\srcPi{\thelangImpSub}} P}
    \and
    \inferhref{switch-ref-s-silent}{key-switch-ref-s-silent}
    {\srcouttoken{\thelangImpSub}(\silent) \\ \tgtouttoken{\thelangImpSub} \wand P}
    {\switchoutsilent{\srcPi{\thelangImpSub}}{\tgtPi{\thelangImpSub}} P}
  \end{mathpar}
  \caption{Refinement switching rules}
  \label{fig:key:refinement-rules}
\end{figure}
To start the refinement proof we use an adequacy theorem:
\begin{theorem}[Adequacy for \thelangRec{} and \thelangSpec{}]\label{lem:key-sim-intro}
  Suppose that for all $\event$, under the separation logic assumptions
  $\tgtintoken{\thelangImpSub}(\event)$ and
  $\susp{\PiSrc}{\specwait\top}$,
  and all $\fninternal{\cdot}{\cdot}$, $\fnext{\cdot}$, $\fninternalSpec{\cdot}{\cdot}$, and $\fnextSpec{\cdot}$ assumptions for $\reclib{1}$ and $\speclib_2$,
  one can prove $\recwait\bot$.
  Then $\impsyn{\reclib{1}} \refinesraw \specsyn{\speclib_2}$.
\end{theorem}

This adequacy theorem lets us prove a DimSum refinement $\impsyn{\reclib{1}} \refinesraw \specsyn{\speclib_2}$ via the logic of \thelogic{}.%
\footnote{Note that \autoref{lem:key-sim-intro} is a specialized adequacy theorem for this example. The general adequacy theorem of \thelogic{} is described in \autoref{sec:model:semantics}.}
After applying \autoref{lem:key-sim-intro}, we start the proof in the \thelangRec{} waiting state $\recwaitraw$ and \thelangSpec{} is suspended in its waiting state $\specwaitraw$.
Additionally, we obtain the $\tgtintoken{\thelangImpSub}(\event)$ token, which states that the first incoming event is $\event$ and we obtain the $\fninternal{\cdot}{\cdot}$ assumptions that tell us how the functions are implemented in the multi-language program.

Once we are inside the logic of \thelogic{}, we can use the routing rules of the $\tgtPi{\thelangImpSub}$ and $\srcPi{\thelangImpSub}$ lanes in \autoref{fig:key:refinement-rules} to handle the refinement reasoning. Let us explain the verification flow by sketching the verification of our running example at a high level:

First, we turn $\recwaitraw$ into a switching modality for an incoming call (using an analogue of \ruleref{key-pseudo-wait-call} for \thelangRec{}).
Then we resolve the switch by switching to \thelangSpec{} (using \ruleref{key-switch-ref-i-in}).
Here, we turn $\specwaitraw$ into a switching modality for an incoming call (using \ruleref{key-pseudo-wait-call}) and resolve it by showing that the two incoming switches match (\ruleref{key-switch-ref-s-in}).
Now we are back in \thelangRec{} and need to prove $\bicall{\mainId}{}{\val}\switchoutS{\PiRec}{\EReturnH(\val)}\recwait\bot$.
Here we can apply \ruleref{spec-main}, \ruleref{spec-echo}, and \ruleref{spec-getc}.%
\footnote{The loop is handled by a coinductive argument that we omit here.}
\ruleref{spec-getc} provides us with a points-to predicate and creates an \abscall to $\readId$. This \abscall is transformed to a switch (using \ruleref{key-switch-rec-call}) since $\readId$ is external.
We then switch to \thelangSpec{} (using \ruleref{key-switch-ref-i-out}), where we can apply \ruleref{spec-main-pseudo}. Here we provide the points-to predicate from \ruleref{spec-getc} and then turn the \abscall of $\readId$ into a switching modality (using an analogue of \ruleref{key-switch-rec-call}). This call to $\readId$ in our \thelangSpec{} specification matches the call emitted by the implementation and thus we can resolve the switch (using \ruleref{key-switch-ref-s-out}) and land back on the implementation side.
Next, we need to give back the points-to predicate to \ruleref{spec-getc}. To obtain this points-to, we perform a \emph{silent switch} to the specification (\ruleref{key-switch-ref-i-silent}), \ie a switch that is not mediated by a visible event.
In the specification, we introduce the magic wand after $\readId$ to obtain the points-to and then use the silent switching modality $\switchsilent{}P$ to switch back to the implementation (\ruleref{key-switch-ref-s-silent}).
These silent switches are always possible during normal \textlog{wp}-based program verification, but need to be included in specifications to denote where switches are possible (similar to basic updates in Iris).
The rest of the verification is handled similarly.
All switches between the $\tgtPi{\thelangImpSub}$ and $\srcPi{\thelangImpSub}$ \pinames are mediated by the tokens $\tgtintoken{\thelangImpSub}(\event)$, $\srcintoken{\thelangImpSub}(\event)$, $\tgtouttoken{\thelangImpSub}$, and $\srcouttoken{\thelangImpSub}(\event)$, which enforce the verification flow: incoming event in $\tgtPi{\thelangImpSub}$ to matching incoming event in $\srcPi{\thelangImpSub}$ to outgoing event in $\tgtPi{\thelangImpSub}$ to matching outgoing event in $\srcPi{\thelangImpSub}$.

\paragraph{What have we seen?}
The main message of the verification sketch is not the exact details of how the different rules are used, but the high-level structure of refinement reasoning in \thelogic{}:
Unlike previous work on refinement reasoning in separation logic~\cite{Liang14TP, ReLoC, Simuliris}, which is based around binary judgments, \thelogic{} is based on unary reasoning for implementation and specification that is linked via switching modalities.
The benefit of \thelogic{}'s approach is that it unifies refinement reasoning and multi-language reasoning:
The refinement specification becomes a new \piname that one can switch to, analogous to cross-language calls shown in \autoref{sec:key:switching}.
This enables \thelogic{} specifications to be modular in whether \abscalls are implemented by cross-language reasoning or refinement reasoning.
In fact, in our running example we reused \ruleref{spec-echo} both when $\putcId$ is implemented using a cross-language call to \thelangSpec{} in \autoref{sec:key:switching} and when $\putcId$ is external and calls are resolved using refinement reasoning in this section.

\subsection[Challenge \#4: View Reconciliation (Rec + Asm + Rec-to-Asm)]{Challenge \#4: View Reconciliation (\thelangRec{} + \thelangAsm{} + \thelangRec{}-to-\thelangAsm{})}
\label{sec:key:wrapper}
\autoref{sec:key:switching} showed how \thelogic{} can verify multi-language programs where all languages (here, \thelangRec{} and \thelangSpec{}) share the same view of the state (here, an abstract heap) and communicate via the same events (here, function calls and returns).
In this section, we consider linking \thelangRec{} with DimSum's \thelangAsm{}, an assembly language that agrees on neither:
instead of an abstract heap, it operates on a flat memory, and
instead of function calls and returns, it has unstructured control flow via jumps. Reasoning about such a program requires translating between the events of the two languages and \emph{reconciling} their different views of the underlying memory.

\paragraph{Implementing $\readId$ in $\thelangAsm$}
In \autoref{sec:key:extcalls}, we treated calls to $\readId$ as external I/O behavior.
In this section, we instead implement $\readId$ in \thelangAsm{} using a system call. A syscall expects an identifier determining the requested operation in register $\reg{8}$ and its arguments in registers $\reg{0}$ to $\reg{7}$.
\[
  \readAsm() \eqdef {} \Amov{\reg{8}}{\READ};~\Asyscall;~\Aret
\]
$\readAsm$ is a thin wrapper around the $\READ$ syscall: it stores the $\READ$ identifier in $\reg{8}$ and leaves the argument registers $\reg{0}$ and $\reg{1}$ untouched, passing the arguments of $\readAsm$'s caller through to the syscall. The $\Asyscall$ instruction triggers the syscall, which reads the specified number of values into the memory address given in $\reg{0}$. Finally, $\readAsm$ returns using the $\Aret$ instruction.

\paragraph{Background: Linking $\echolib$ and $\readAsm$ using DimSum}
Let us now see how we can use $\readAsm$ to resolve the $\readId$ call from $\getcRec$ in $\echolib$.
Like \thelangRec{}, \thelangAsm{} comes with syntactic and semantic linking operators: $\asmsynlinkraw$ links two syntactic \thelangAsm{} programs, and $\asmlinkraw$ links two \thelangAsm{} modules, analogous to the \thelangRec{} operators from \autoref{sec:key:dimsum}.
We can compile $\echolib$ and directly link it to $\readAsm$: $\asmsynlink{\compile{\echolib}}{\readAsm}$.
However, we do not want to reason about the compiled code directly.
Instead, DimSum introduces a \emph{wrapper} $\moditoa{\gapImp}$ that turns a $\thelangRec$ module into a (semantic) $\thelangAsm$ module.
DimSum's compiler correctness statement relates the compiled code to the wrapped original code: ${\refinesS{\asmsyn{\compile{\impprog{\libImp}}}} {\moditoa{\impsyn{\impprog{\libImp}}}}}$.
Thus, we can reason about the wrapped $\echolib$ semantically linked with  $\readAsm$: $\asmlink{\moditoa{\impsyn{\echolib}}}{\asmsyn{\readAsm}}$.
The wrapper mediates between the two languages, translating events according to the calling convention. For this, it maintains an injection $\loc \itoabij \addrraw$ that maps \thelangRec{} locations $\loc$ to corresponding \thelangAsm{} addresses $\addrraw$.
This injection is lifted to values as $\val \rtoabij \aval$.

\paragraph{Reasoning about $\moditoa{\gapImp}$ in \thelogic{}}
To reason about $\moditoa{\gapImp}$, \thelogic{} introduces a new lane $\rtoaPi{\lane}$ and the \emph{exchange} $\rtoaexchange(\exchanged)$
for exchanging heap points-tos $\loc \hptsto \val$ and memory points-tos $\addrraw \mptsto \aval$ across the injection $\loc \rtoabij \addrraw$.
To explain these concepts, let us see how we can reason about $\asmlink{\moditoa{\impsyn{\echolib}}}{\asmsyn{\readAsm}}$ in \thelogic{}. First, we fix the lanes: $\PiRec \eqdef \linkAsmLPiSubst{\Pi}{\rtoaPiStandalone{}}$ and $\PiAsm \eqdef \linkAsmRPi{\Pi}$. The lane $\PiAsm$ is the right component of a linking lane, analogous to \autoref{sec:key:switching}, only now for \thelangAsm{}. The lane $\PiRec$ nests the wrapper lane inside the left linking lane.

The interesting part of the verification starts when we resolve the \abscall to $\getcId$ (from \ruleref{spec-echo}).
For this, we first apply the specification \ruleref{spec-getc}.
This gives us a points-to $\loc \hptsto \val$ and we need to verify the \abscall{} $\bicallSraw{\readId}{\loc, 1}$. Since $\readId$ is not implemented in \thelangRec{}, the \abscall becomes a switch (\ruleref{key-switch-rec-call}), just like in \autoref{sec:key:switching}.
Now the reasoning about the wrapper starts:
We leave $\rtoaPi{\lane}$ using \ruleref{key-r2a-switch-call}.
For this, we need to give up the $\rtoaclosed$ assertion that locks the \thelangAsm{} assertions while executing in \thelangRec{}.
Next, we learn that the \thelangRec{} arguments are related to the values in the corresponding \thelangAsm{} argument registers.
For our example, this means that $\loc \itoabij \rlookup{\regs}{\reg{0}}$ and $\rlookup{\regs}{\reg{1}} = 1$.
Next, we obtain the exchange $\rtoaexchange([])$ and the suspended token for \thelangRec{}.
The wrapper translates the $\ECallH$ to a $\EJumpH$ followed by the wrapper waiting state $\rtoawait{}$.

The $\EJumpH$ is resolved using the routing rules for the $\asmlinkraw$ lanes (analogous to the $\implinkraw$ routing rules in \autoref{sec:key:switching}) and we arrive at $\readAsm$.
Here, we need to turn the $\loc \hptsto \val$ from \ruleref{spec-getc} into an \thelangAsm{} points-to $\rlookup{\regs}{\reg{0}} \mptsto \aval$ to verify the system call to $\READ$. This is where the exchange comes in: \ruleref{key-exchange-rec-to-asm} lets us exchange our heap points-to $\loc \hptsto \val$ for a memory points-to $\rlookup{\regs}{\reg{0}} \mptsto \aval$ since we have $\loc \itoabij \rlookup{\regs}{\reg{0}}$, adding $\loc$ to the list of exchanged locations in $\rtoaexchange([\loc])$.
The $\READ$ syscall gives back an \thelangAsm{} points-to $\rlookup{\regs}{\reg{0}} \mptsto \avalB$ containing the read value $\avalB$.
We exchange this \thelangAsm{} points-to back into a \thelangRec{} points-to using \ruleref{key-exchange-asm-to-rec}, which we use to complete the proof obligation of \ruleref{spec-getc}.
Additionally, exchanging the points-to back clears the list of exchanged locations in $\rtoaexchange([])$, which is necessary before switching back from $\PiAsm$ to $\PiRec$.

\begin{figure}
\begin{mathpar}
  \inferhref{exchange-rec-to-asm}{key-exchange-rec-to-asm}{\loc \notin \exchanged \\ \loc \itoabij \addrraw \\ \rtoaexchange(\exchanged) \\ \loc \hptsto \val }
  {\upd\Exists \aval. \rtoaexchange(\lcons{\loc}\exchanged) \ast \addrraw \mptsto \aval \ast \val \itoabij \aval}
  \and
  \inferhref{exchange-asm-to-rec}{key-exchange-asm-to-rec}
  {\loc \itoabij \addrraw \\ \val \itoabij \aval \\ \rtoaexchange(\lcons{\loc}\exchanged) \\ \addrraw \mptsto \aval }
  {\upd\rtoaexchange(\exchanged) \ast \loc \hptsto \val}
  \and
  \inferhref{r2a-switch-call}{key-r2a-switch-call}
  {\rtoaclosed \\ \All \regs . {\left({\textstyle\Sep\nolimits_{\val\in\vals,\aval\in{\rlookup{\regs}{\reg{0} \ldots}}}} {\val} \itoabij {\aval}\right)} \wand \rtoaexchange([]) \wand \susp{\rtoaPi{\lane}}{P} \wand \switchoutS{\lane}{\EJumpH(\regs)} \rtoawait{\lane}}
  {\switchoutS{\rtoaPi{\lane}}{\ECallH(\fnname, \vals)} P}
\end{mathpar}
  \caption{Rules for reasoning about $\thelangRec{}$ and $\thelangAsm{}$ (simplified)}
  \label{fig:key:wrapper-rules}
\end{figure}

\paragraph{What have we seen?}
This section demonstrated two things:
First, specifications proven for \thelangRec{} functions remain usable when we link with \thelangAsm{} code: the wrapper lane $\rtoaPi{\lane}$ lets us  reason in the \thelangRec{} program logic, reusing the same specifications we have established previously.
Second, \thelogic{} addresses the view reconciliation problem through the \emph{exchange} $\rtoaexchange(\exchanged)$, which lets us exchange points-to assertions of one language for the related points-to assertions of the other.

\subsection[All Together Now (Rec + Pseudo + Asm + Rec-to-Asm)]{All Together Now (\thelangRec{} + \thelangSpec{} + \thelangAsm{} + \thelangRec{}-to-\thelangAsm{})}
To show how the concepts from the previous sections can be combined, let us consider the following ``trilingual'' refinement proving that $\echolib$ combined with $\readAsm$ refines a version of $\mainPseudo$ stated using system calls for $\readId$:
$\asmlink{\moditoa{\impsyn{\echolib}}}{\asmsyn{\readAsm}} \refinesraw \mainPseudoThreeS$.
To prove this refinement, we define $\PiRec \eqdef [\asmlink{\rtoaPiStandalone{}}{\_} \preceq \_]$, $\PiAsm \eqdef [\asmlink{\_}{\bullet} \preceq \_]$, and $\PiSpec \eqdef \srcPi{\thelangAsmSub}$, where $\PiAsm$ is obtained by instantiating $\Pi$ in the $\linkAsmRPi{\Pi}$ lane with the $\tgtPi{\thelangAsmSub}$ lane, and similarly for $\PiRec$.
We can reason about these lanes using the principles from previous sections.
In particular, the rules for the $\linkAsmLPi{\Pi}$ lane ensure that calls from $\echolib$ to $\readId$ are handled by $\readAsm$ in $\PiAsm$, while calls to $\writeId$ are resolved by the refinement to $\mainPseudoThreeS$ in $\PiSpec$.
Crucially, proofs from the previous sections like \ruleref{spec-echo} from \autoref{sec:key:abscalls} or proofs about the $\echolib$ and $\readAsm$ interaction from \autoref{sec:key:wrapper} can be reused since these proofs kept the lanes (partially) abstract. \autoref{sec:case-study} demonstrates this reuse across six variants of the echo example.

\section{Building the Program Logic}\label{sec:model}
We now lay out the formal foundations of \thelogic: the simulation relation and its adequacy with regard to DimSum's refinement (\autoref{sec:model:semantics}), and the switching modality and \pinames (\autoref{sec:model:lanes}). We then present the \pinames from the previous section for refinement (\autoref{sec:model:refinement-lanes}), linking (\autoref{sec:model:linking}), and the wrapper (\autoref{sec:model:wrapper}).

\subsection{Multi-Language Semantics}\label{sec:model:semantics}

\paragraph{Background: DimSum}
Before diving into \thelogic, we recap the core concepts of DimSum~\cite{DimSum}, the underlying multi-language semantics of \thelogic.
DimSum is centered around \emph{modules}: labeled transition systems $\module \eqdef (\mstates,\mstepraw,\mstateinit) \in \modules{\events}$ consisting of a set of states $\mstates$, the initial state $\mstateinit$ and a transition relation ${\mstepraw{}}\!\!\in \powerset{\mstates \times (\labels) \times \powerset{\mstates}}$. Each transition has a label $\lbl\in\labels$, denoting that the transition either emits an outgoing event ($\sendevent{\event}$), synchronizes on an incoming event ($\recvevent{\event}$), or is silent ($\silent$).
The transition relation allows dual non-determinism:
An execution demonically picks a transition $\mstate \mstep{\lbl} \mstateP$ and then angelically continues in all states $\mstate' \in \mstateP$.\footnote{Following DimSum, angelic non-determinism is restricted to silent transitions.}

DimSum's refinement is a termination-insensitive coinductive simulation relation that requires every step of the implementation $\mstate_i \mstepSmall{\lbl}_{\module_i} \mstateP_i$ to be matched by zero or more steps $\mstate_s \msteps{\lbl}_{\module_s} \mstateP_s$ of the specification with matching label $\lbl$.
Then, to account for the angelic non-determinism, each successor $\mstate_s' \in \mstateP_s$ of the specification requires a related successor $\mstate_i' \in \mstateP_i$ of the implementation.
\begin{align*}
  \refines{\module_i}{\module_s} &\eqdef{} \dimsumsim{\module_i}{\mstateinit_{\module_i}}{\module_s}{\mstateinit_{\module_s}}\\
  \dimsumsim{\module_i}{\mstate_i}{\module_s}{\mstate_s} &\eqdefcoind{}
  \All \lbl, \mstateP_i.
    \begin{array}[t]{@{} l}
    \mstate_i \mstepSmall{\lbl}_{\module_i} \mstateP_i \Rightarrow \\
    \Exists \mstateP_s. \mstate_s \msteps{\lbl}_{\module_s} \mstateP_s \wedge
  \All \mstate_s' \in \mstateP_s. \Exists \mstate_i' \in \mstateP_i.
  \dimsumsim{\module_i}{\mstate_i'}{\module_s}{\mstate_s'}
    \end{array}
\end{align*}

\paragraph{Refinement in \thelogic}
To lift DimSum's notion of refinement into the logic of \thelogic, we introduce the \emph{module weakest precondition} $\simgen{\module}{\mstate}{\Phi}$.
It is inspired by Iris' \textlog{wp} connective~\cite{IrisGroundUp}, but applies to DimSum modules.
Intuitively, $\simgen{\module}{\mstate}{\Phi}$ encodes the verification condition of module $\module$ in state $\mstate$ with postcondition $\Phi$.
There are two variants of \simgenname, to distinguish whether $\module$ is on the implementation side of the refinement ($\simgenname^{\tgtsub}$) or on the specification side ($\simgenname^{\srcsub}$):%
\footnote{For simplicity, we omit the basic update modality $\upd$ from this presentation.}
\begin{align*}
  \simtgt{\module}{\mstate}{\Phi} &\eqdefind
    \begin{array}[t]{@{} l}
      \Phi(\tau, \mstate) \lor
      \All \lbl, \Sigma. \mstate \mstepSmall{\lbl}_{\module} \Sigma \wand
        \Exists \mstate' \in \Sigma. \dslater (\Phi(\lbl, \mstate') \lor (\lbl = \tau \land \simtgt{\module}{\mstate'}{\Phi}))
    \end{array}\\
  \simsrc{\module}{\mstate}{\Phi} &\eqdefind
    \begin{array}[t]{@{} l}
      \Phi(\tau, \mstate) \lor
      \Exists \lbl, \Sigma. \mstate \mstepSmall{\lbl}_{\module} \Sigma \ast
        \All \mstate' \in \Sigma. \Phi(\lbl, \mstate') \lor (\lbl = \tau \land \simsrc{\module}{\mstate'}{\Phi})
    \end{array}
\end{align*}
Both versions follow the same structure and distinguish a base case and a step case:
The base case $\Phi(\silent, \mstate)$ directly invokes the postcondition $\Phi$ with the current state $\mstate$ and passes $\silent$ to note that no visible event was emitted.
The step case differs between the versions since the meaning of the non-determinism flips depending on the side of the refinement.
For $\simgenname^{\tgtsub}$, one (demonically) obtains a transition $\mstate \mstepSmall{\lbl}_{\module} \mstateP$ and has to (angelically) pick a resulting state $\mstate' \in \mstateP$. Vice versa for $\simgenname^{\srcsub}$.
Then, one either has to prove $\Phi$ with the resulting state and label, or one can continue in \simgenname if the transition was silent.
$\simgenname^{\tgtsub}$ includes the later modality to allow reasoning about loops. (As discussed in \autoref{sec:introduction}, this later modality can \emph{not} be used for higher-order ghost state due to the presence of angelic non-determinism.) In the following, we omit later modalities to avoid clutter.

We use \simgenname to build \thelogic{}'s simulation relation $\simbin{\module_i}{\mstate_i}{\module_s}{\mstate_s}$:
\begin{align*}
  \simbin{\module_i}{\mstate_i}{\module_s}{\mstate_s} &\eqdefind
  \simtgt{\module_i}{\mstate_i}{\lbl, \mstate_i'.\spac \simsrc{\module_s}{\mstate_s}{\lbl', \mstate_s'.\spac \lbl = \lbl' \land \simbin{\module_i}{\mstate_i'}{\module_s}{\mstate_s'}}}
\end{align*}
This simulation relation applies \simgenname to both the implementation module $\module_i$ and the specification module $\module_s$, and checks that both sides emitted the same label, all in a loop.
Even though this simulation and $\simgenname$ are defined inductively, the simulation is termination insensitive (like DimSum's refinement) due to the later modality in $\simgenname^{\tgtsub}$.
The key property of this simulation is that it is adequate, \ie it implies DimSum's notion of refinement:
\begin{theorem}[Adequacy]\label{lem:adequacy}
  For all $\module_i$, $\module_s$: If $\simbin{\module_i}{\mstateinit_{\module_i}}{\module_s}{\mstateinit_{\module_s}}$ then $\refines{\module_i}{\module_s}$.
\end{theorem}

Notably, the binary simulation ($\simbinraw$) is defined as a step-wise interleaving of the unary \simgenname, whereas in previous approaches the binary simulation is either primitive (\eg in Simuliris~\cite{Simuliris}) or composed from non-interleaved unary connectives (\eg in ReLoC~\cite{ReLoC}).
The interleaving allows \thelogic{} to handle interleaved angelic and demonic non-determinism while the split into unary weakest preconditions enables \thelogic{} to unify cross-language and refinement reasoning via lanes.

\subsection{Lanes and Switching}\label{sec:model:lanes}
Next, we turn to lanes and the switching modalities.

\paragraph{Lanes}
Intuitively, \emph{\pinames} describe where a module is located in the multi-language program.
Formally, we define a \piname $\lane$ as a pair $(\gvar_{\lane}, \lanepost{\lane}) \in \lanes{\ts,\module} \eqdef (\GName, \events_\module \to \mstates_\module \to \iProp)$. The notion of lanes is parameterized by the module $\module$ in the lane and the refinement side $\ts \in \{\tgtsub, \srcsub\}$. Each lane consists of a ghost variable $\gvar_{\lane}$ and a predicate $\lanepost{\lane}$ over the events $\events_\module$ and states $\mstates_\module$ of $\module$. The ghost variable $\gvar_\lane$ stores the state of $\module$ in the assertion $\mstatefrac{\gvar}{\mstate}$ with the usual rules shown in \autoref{fig:gvar}. $\gvar_\lane$ is used to communicate the state of $\module$ to the suspended token, as we will see below.
\begin{figure}
  \centering
\begin{mathpar}
\inferH{gvar-merge}{}{\mstatefrac{\gvar}{\mstate} \ast \mstatefrac{\gvar}{\mstate'} \vdash \mstatefull{\gvar}}
\and
\inferH{gvar-split}{}{\mstatefull{\gvar} \vs \mstatefrac{\gvar}{\mstate} \ast \mstatefrac{\gvar}{\mstate}}
\and
\inferH{gvar-agree}{}{\mstatefrac{\gvar}{\mstate} \ast \mstatefrac{\gvar}{\mstate'} \vdash \mstate = \mstate'}
\end{mathpar}
  \caption{Rules for ghost variables}
  \label{fig:gvar}
\end{figure}
The interesting part of a \piname is the predicate $\lanepost{\lane}$.
It describes the position of the lane in the multi-language program via the postcondition of \simgenname, \ie we verify the module $\module$ in \piname $\lane$ by proving $\simgen{\module}{\mstate}{\lanepost{\lane}}$. (In the following, we just write $\simgen{\module}{\mstate}{\lane}$.)
The postcondition $\lanepost{\lane}$ contains the $\simgenname$ for the larger context that $\module$ runs in.

We require that \pinames allow skipping over silent events, encoded as $\simgen{\module}{\mstate}{\lane} \vdash \lanepost{\lane}(\silent, \mstate)$.
This property is used to prove the introduction rule for silent switches (\ruleref{switch-silent-intro}).

\paragraph{Suspended}
Next, we introduce the \emph{suspended token} $\susp{\lane}{P}$.
Intuitively, $\susp{\lane}{P}$ asserts that the lane $\lane$ is currently suspended (\ie not executing) and we can resume $\lane$ by proving $P$.
Formally, $\susp{\lane}{P}$ is defined as follows:
\[  \susp{\lane}{P} \eqdef \All \mstate. \mstatefrac{\gvar_\lane}{\mstate} \wand P \wand \simgen{\module}{\mstate}{\lane}. \]
This definition captures the above intuition: $\susp{\lane}{P}$ allows us to continue to verify $\module$ (\ie prove $\simgen{\module}{\mstate}{\lane}$) by proving $P$. The state $\mstate$ of $\module$ is tracked using the ghost variable $\gvar_{\lane}$. The other half of $\gvar_{\lane}$ is part of a larger lane containing $\lane$, tying $\gvar_{\lane}$ to the actual state of $\module$.

\paragraph{Switching}
Finally, we introduce the \emph{switching modality} for switching between \pinames:
\[\switchSome{\laneA}{\laneB}{\lbl} P \eqdef \All \mstate. \mstatefrac{\gvar_{\laneA}}{\mstate} \wand \susp{\laneB}{P} \wand \Phi_{\laneA}(\lbl, \mstate).\]
This modality states that we emit the label $\lbl$ in \piname $\laneA$ and afterwards we continue with proving $P$ in \piname $\laneB$.
This modality satisfies a set of useful rules shown in \autoref{fig:switching_rules}:
First, the modality is monotonic (\ruleref{switch-mono}) since $P$ appears in a double negative position. Monotonicity is useful to cancel a switch in an assumption (\eg from a specification) with a switch in the goal without knowing the details of $\laneA$ and $\laneB$, allowing us to treat \pinames as abstract during single-language verification (\eg rule \ruleref{key-switch-rec-call}).
Second, \ruleref{switch-redirect} redirects the target of a switch and, in particular, enables turning a same-lane switch (as \eg used by \ruleref{key-switch-rec-call}) into a switch to a different lane.
As shown in \autoref{sec:key:switching}, this rule is crucial to use single-language specifications in a multi-language program.
The proof of \ruleref{switch-redirect} simply swaps out the suspended token in the switch.
Finally, a silent switch within a single lane can be introduced trivially (\ruleref{switch-silent-intro}).

\begin{figure}
  \centering
    \begin{mathpar}
    \inferH{switch-mono}{\switchSome{\laneA}{\laneB}{\lbl}P \\ P \wand Q}
        {\switchSome{\laneA}{\laneB}{\lbl}Q}
    \and
    \inferH{switch-redirect}{\susp{\laneC}{Q} \\ \susp{\laneB}{P} \wand \switchSome{\laneA}{\laneC}{\lbl}Q}
        {\switchSome{\laneA}{\laneB}{\lbl}P}
    \and
    \inferH{switch-silent-intro}{P}{\switchsilent{\lane}P}
    \and
 \end{mathpar}
  \caption{Properties of the switching modality}
  \label{fig:switching_rules}
\end{figure}

\begin{figure}
\begin{align*}
  \Phi_{\srcside{}}(\lbl', \mstate_s) & \eqdef \All \mstate_i, \lbl. \mstatefrac{\tgtgvar}{\mstate_i} \wand \mstatefrac{\gvar_{\lbl}}{\lbl} \wand{} \lbl' = \lbl \land \simbin{\module_i}{\mstate_i}{\module_s}{\mstate_s}\\
  \Phi_{\tgtside{}}(\lbl, \mstate_i) & \eqdef {} \All \mstate_s. \mstatefrac{\srcgvar}{\mstate_s} \wand \mstatefull{\gvar_\lbl} \wand \simsrc{\module_s}{\mstate_s}{\lbl', \mstate_s'.\spac \lbl' = \lbl \land \simbin{\module_i}{\mstate_i}{\module_s}{\mstate_s'}}
\end{align*}
  \begin{mathpar}
    \inferH{ref-lane-elim}
    {\All \event.
      \tgtintoken{\nolang{X}}(\event) \wand
      \susp{\srcPi{\nolang{X}}}{(\mstatefull{\srcgvar} \wand \simsrc{\module_s}{\mstate_s}{\srcPi{\nolang{X}}})} \wand \mstatefull{\tgtgvar} \wand \simtgt{\module_i}{\mstate_i}{\tgtPi{\nolang{X}}} }
    {\simbin{\module_i}{\mstate_i}{\module_s}{\mstate_s}}
  \end{mathpar}
  \caption{Refinement \pinames (simplified)}
  \label{fig:model:refinement-lanes}
\end{figure}

\subsection{Refinement Lanes}
\label{sec:model:refinement-lanes}

After seeing the general concepts of \thelogic{}, we now introduce the different \pinames that \thelogic{} provides, starting with the refinement lanes $\tgtPi{\nolang{X}} \in \lanes{\tgtsub,\module_i}$ and $\srcPi{\nolang{X}} \in \lanes{\srcsub,\module_s}$.
The definition of these lanes is given in \autoref{fig:model:refinement-lanes}.%
\footnote{The presentation here is simplified in two ways: It omits a mechanism to enforce alternation of incoming and outgoing events and the specification lane omits a disjunct about $\silent$ events to satisfy \ruleref{switch-silent-intro}.}
$\Phi_{\tgtside{}}$ resp. $\Phi_{\srcside{}}$ correspond to the postcondition of $\simgenname^{\tgtsub}$ resp. $\simgenname^{\srcsub}$ in $\simbin{\module_i}{\mstate_i}{\module_s}{\mstate_s}$ (\autoref{sec:model:semantics}).
The main difference is that both \pinames use the \piname ghost variable of the other side (\ie $\srcgvar$ or $\tgtgvar$) to obtain the state of the module in the other lane, and $\Phi_{\srcside{}}$ uses one half of the ghost variable $\gvar_\lbl$ to track the label $\lbl$ emitted by the implementation.
The other half is held by the \tokennames{} $\srcintoken{\nolang{X}}$ and $\srcouttoken{\nolang{X}}$ (\eg $\srcintoken{\nolang{X}}(\event)$ holds $\mstatefrac{\gvar_{\lbl}}{\recvevent{\event}}$), which, together with their counterparts $\tgtintoken{\nolang{X}}$ and $\tgtouttoken{\nolang{X}}$, enforce alternation between incoming and outgoing events.

\paragraph{Lane elimination}
The routing rules of the refinement \pinames were already discussed in \autoref{sec:key:extcalls}.
There is one other important rule: the \emph{\piname elimination} rule \ruleref{ref-lane-elim}.
\PiName elimination shows that proofs about a \piname imply the concept the \piname represents.
For the refinement \pinames, this means that proving $\simgenname^{\tgtsub}$ in the $\tgtPi{\nolang{X}}$ \piname, assuming a suspended $\simgenname^{\srcsub}$ in the $\srcPi{\nolang{X}}$ \piname and a $\tgtintoken{\nolang{X}}(\event)$ token, implies the simulation $\simbin{\module_i}{\mstate_i}{\module_s}{\mstate_s}$.
The elimination rule is straightforward to prove since the definition of the refinement \pinames closely matches the definition of the simulation.
This elimination rule (and the elimination rules for other \pinames) can then be composed with the adequacy theorem (\autoref{lem:adequacy}) to prove a DimSum refinement using \piname-based reasoning.

\subsection{Linking Lanes}\label{sec:model:linking}

\paragraph{Background: Linking in DimSum}
Before we can introduce the linking \pinames $\linkLPi{\lane}{\nolang{X}}$ and $\linkRPi{\lane}{\nolang{X}}$, we need to briefly recap DimSum's semantic linking combinator (see \cite[\S 3.3]{DimSum} for the full definition).
The linking combinator $\modlink{X}{\module_1}{\module_2} \in \modules{\events}$ lets us link two modules $\module_1, \module_2 \in \modules{\events}$ sharing the same type of events $\events$.
The parameter $X = (\linkstates, \linkfn, \linkstateinit)$ consists of a set of linking states $\linkstates$, an initial state $\linkstateinit$ and a deterministic relation $\linkfn{} \subseteq (\modseqdirs \times \linkstates \times \events) \times (\modseqdirs \times \linkstates \times \events)$ that determines how events are routed.
The direction $\modseqdirs \eqdef\{\modseqnone, \modseqleft, \modseqright\}$ represents the left ($\modseqleft$) or right ($\modseqright$) side of the linking, or the environment ($\modseqnone$) containing the linking.
Concretely, $\inlinkfn{\linkfn}{(\modseqdir, \linkstate, \event)}{(\modseqdir', \linkstate', \event')}$ means the event $\event \in \events$ coming from direction $\modseqdir$ is routed to $\modseqdir'$ as the event $\event'$, updating the linking state to $\linkstate'$.
These parameters $X$ are instantiated for different event types $\events$, yielding the \thelangRec{} linking $\implinkraw$ and \thelangAsm{} linking $\asmlinkraw$.
For example, for $\implinkraw$, the linking state tracks a call-stack to determine how returns are resolved.

\begin{figure}
  \centering
  \begin{mathpar}
    \inferH{switch-link-l-to-r}
    {(\modseqleft,s,e) \linkfn{} (\modseqright, s', e') \\ \linkouttoken{\lane}{\nolang{X}}(s) \\
     \linkintoken{\lane}{\nolang{X}}(s',e') \wand P}
    {\switchout{\linkLPi{\lane}{\nolang{X}}}{\linkRPi{\lane}{\nolang{X}}}{\event}P}
    \and
    \inferH{switch-link-r-in}
    {\linkintoken{\Pi}{\nolang{X}}(s,e) \\ \event = \event' \wand \linkouttoken{\Pi}{\nolang{X}}(s) \wand P}
    {\switchinS{\linkRPi{\Pi}{\nolang{X}}}{\eventS}P}
    \and
    \inferH{switch-link-l-to-env}
    {(\modseqleft, s, e) \linkfn{} (\modseqnone, s', e') \\ \linkouttoken{\lane}{\nolang{X}}(s) \\ \switchoutS{\Pi}{e'}{\switchin{\Pi}{\cdot}{e''}\linkrouting{\Pi}{\nolang{X}}(s',e'')}}
    {\switchoutX{\linkLPi{\lane}{\nolang{X}}}{\event}}
    \and
    \inferH{switch-link-route-l}
    {(\modseqnone, s, e) \linkfn{} (\modseqleft, s', e') \\ \susp{\linkLPi{\lane}{\nolang{X}}}{P} \\ \linkintoken{\Pi}{\nolang{X}}(s',e') \wand P}
    {\linkrouting{\Pi}{\nolang{X}}(s,e)}
    \and
    \inferH{link-lane-elim-l}
  {\mstatefrac{\rightgvar}{\mstate_2} \\ \mstatefull{\gvar_\Pi} \\ \linkouttoken{\lane}{\nolang{X}}(s) \wand \simgen{\module_1}{\mstate_1}{\linkLPi{\lane}{\nolang{X}}}}
  {\simgen{\modlink{X}{\module_1}{\module_2}}{(\modseqleft, \linkstate, \mstate_1, \mstate_2)}{\lane}}
  \end{mathpar}
  \caption{Linking rules, implementation side version, $\switchSomeX{\laneA}{\lbl}$ characterized via $(\switchSome{\laneA}{\laneB}{\lbl} P) \dashv\vdash (\susp{\laneB}{P} \wand \switchSomeX{\laneA}{\lbl})$}
  \label{fig:model:linking-rules}
\end{figure}

\paragraph{Linking \pinames}
After establishing the background, we can now introduce the linking lanes $\linkLPi{\lane}{\nolang{X}}$ and $\linkRPi{\lane}{\nolang{X}}$ via their routing rules shown in \autoref{fig:model:linking-rules}.
(For simplicity, we state rules for one side of the linking only and omit symmetric rules for the other side.)
The rules use the ${\linkfn}$ relation to determine how events are routed through the linking. \ruleref{switch-link-l-to-r} routes an event from left to right where it can be received using \ruleref{switch-link-r-in}.
\ruleref{switch-link-l-to-env} routes an event to the environment (\eg to a refinement lane). Then the linking enters the $\linkrouting{\Pi}{\nolang{X}}$ state that is used to dispatch incoming events to the left or right using \ruleref{switch-link-route-l} and a symmetric rule for the right side.
The rules are mediated by the linking tokens introduced in \autoref{sec:key:switching}.
From these rules, one can derive the rules for specific linking combinators like the rules of $\implinkraw$ (\autoref{sec:key:switching}).

The linking lanes come with elimination rules (\ruleref{link-lane-elim-l} and a symmetric rule for the right side), showing that a \simgenname of a linking lane implies a \simgenname of the linking operator in the left resp. right state.

\subsection{Wrapper Lane}\label{sec:model:wrapper}

\paragraph{Background: DimSum wrappers}
Before explaining the wrapper lane, let us give some background on the wrapper combinator that DimSum uses to translate between different event types.
The wrapper takes a module $\module \in \modules{\events_1}$ with events $\events_1$ and translates it to a module $\modprepost{X}{\module} \in \modules{\events_2}$ with events $\events_2$. The translation is described by the parameter $X = (\seplogic, \postrel, \prerel)$.
It contains two relations: $\postrel$ for translating $\events_1$ to $\events_2$ and $\prerel$ for translating $\events_2$ to $\events_1$.
These relations are stated in the separation logic of the wrapper $\seplogic$. (More formally, the wrapper picks a resource algebra $R$, which induces a separation logic $\seplogic \eqdef \UPred(R)$.)
This separation logic is used in the relations to state which memory is shared by the translation and which memory remains private.
For example, the translation from \thelangRec{} to \thelangAsm{} events requires sharing all locations passed as arguments via the injection $\loc \itoabij \addrraw$, while one can retain ownership of all other locations, ensuring that the assembly code cannot access them.
The separation logic relations are encoded into the state transition system of the wrapper using the satisfiability predicate $\satisfiable{P}$~\cite[p. 108]{spies2025phd}, which states that the separation logic proposition $P$ holds for some resource (see \citet{DimSum} for details).

\begin{figure}
\begin{mathpar}
  \inferH{embed-mono}{P \vdash Q}{\wembed{P} \vdash \wembed{Q}}
  \and
  \inferH{embed-sep}{}{\wembed{P \ast Q} \dashv\vdash \wembed{P} \ast \wembed{Q}}
  \and
  \inferH{embed-emp}{}{\wembed{\mathsf{emp}} \dashv\vdash \mathsf{emp}}
  \and
  \inferH{embed-exist}{}{\wembed{\Exists x. P(x)} \dashv\vdash \Exists x. \wembed{P(x)}}
  \and
  \inferH{embed-and}{}{\wembed{P \land Q} \dashv\vdash \wembed{P} \land \wembed{Q}}
  \and
  \inferH{embed-or}{}{\wembed{P \lor Q} \dashv\vdash \wembed{P} \lor \wembed{Q}}
  \and
  \inferH{embed-pure}{}{\wembed{\pprop} \dashv\vdash \pprop}
  \and
  \inferH{embed-forall}{}{\wembed{\All x. P(x)} \vdash \All x. \wembed{P(x)}}
  \and
  \inferH{embed-wand}{}{\wembed{P \wand Q} \vdash \wembed{P} \wand \wembed{Q}}
  \and
  \inferH{embed-pers}{}{\wembed{\always P} \vdash \always \wembed{P}}
  \and
  \inferH{embed-upd}{}{\wembed{\upd P} \ast \wembedtok \vs \wembed{P} \ast \wembedtok}
  \and
  \inferH{embed-close}{}
    {\wembed{P} \ast \wembedtok \vdash \Exists \prepostframe.
      \satisfiable{P \ast \prepostframe} \ast \wembedclosed{\prepostframe}}
  \and
  \inferH{embed-open}{}
    {\satisfiable{Q \ast \prepostframe} \ast \wembedclosed{\prepostframe} \vs \wembed{Q} \ast \wembedtok}
\end{mathpar}
\caption{Laws of the weak embedding}
\label{fig:model:embed-rules}
\end{figure}

\paragraph{Weak embedding}
While the use of separation logic in the wrappers means the wrappers are very expressive, it also creates a challenge: How can we integrate the separation logic of the wrapper $\seplogic$ into the logic of \thelogic{}?
To address this challenge, we introduce the \emph{weak embedding} $\wembed{\cdot}^{\gname} : \sepprop \to \iProp$ that maps propositions from $\sepprop$, the separation logic $\seplogic$, into $\iProp$, the notion of Iris assertions used by \thelogic{}.
(The name $\gname$ distinguishes embeddings of different logics and we usually omit it.)
The embedding has two important properties:
\emph{First}, the separation logic connectives inside $\wembed{\cdot}$ map to the separation connectives of $\iProp$ (see \eg \ruleref{embed-sep}, \ruleref{embed-exist}, \ldots), allowing us to manipulate the assertions of $\seplogic$ like any other assertions.
However, it is not a full equivalence: universal quantifiers, magic wands and the persistence modality can only be lifted out of $\wembed{\cdot}$.
(This is why it is called \emph{weak} embedding.)
\emph{Second}, a weak embedding of $\wembed{P}$ can be turned into $\satisfiable{P \ast F}$ for some $F$ using \ruleref{embed-close}.
This rule is important to integrate with the state transition system of the wrapper, which is defined using $\satisfiable{P \ast F}$.
We call this process \emph{closing} the embedding.
Intuitively, the frame $F$ contains the ownership of all embedded assertions that are framed around this rule.
This is crucial for usability since it means that all embedded assertions that do not need to be given to the wrapper can remain where they are.
For soundness, it is important that we can only perform updates in the embedding when it is not closed.
To encode this obligation, we introduce the token $\wembedtok$ that tracks whether the embedding is open. Performing updates requires presenting this token (\ruleref{embed-upd}).
\ruleref{embed-close} turns $\wembedtok$ into the closed token $\wembedclosed{F}$, which can be turned back into $\wembedtok$ using \ruleref{embed-open}.

\paragraph{Definition of weak embedding}
For Iris experts, the weak embedding is defined as follows:
\[
  \wembed{P}^\gname \eqdef \Exists a. \Sem{P}(a) \ast (a \mincl \epsilon \vee \ownGhost{\gname}{\authfrag a})
  \quad \wembedtok^\gname \eqdef \Exists a. \ownGhost{\gname}{\authfull a}
  \quad \wembedclosed{F}^\gname \eqdef \All P. \satisfiable{P \ast F} \wand\upd (\wembed{P}^\gname \ast \wembedtok^\gname)
\]
The resource $a$ underlying the separation logic $\seplogic$ is tracked in an authoritative construction, where each $\wembed{P}^\gname$ contains a fractional part $\ownGhost{\gname}{\authfrag a}$ such that $P$ holds for the resource $a$.
The combination of all these fractional parts is tracked in the authoritative part $\authfull a$ that is part of $\wembedtok^\gname$.
This $a$ contains a global view of the resource underlying $\seplogic$, which is important to prove \ruleref{embed-close}.
The definition of $\wembedclosed{F}^\gname$ encodes \ruleref{embed-open}.
The $a \mincl \epsilon$ in $\wembed{P}^\gname$ is necessary to prove \ruleref{embed-emp}.
Proving \ruleref{embed-close} additionally requires that the non-core parts of the resource algebra underlying $\seplogic$ can be canceled:
\[ \mval(a \mtimes b) \Rightarrow a \mtimes b \equiv a \mtimes c \Rightarrow \mcore{a} \mtimes b \equiv \mcore{a} \mtimes c \]
All resource algebras that we use satisfy this property.

\begin{figure}
\begin{mathpar}
  \inferH{switch-wrapper-out}
  {\All \event_2.
     \wembed{\event_1 \postrel \event_2} \wand
     \wembedtok \wand
     \switchoutS{\lane}{\event_2} \wrapwait{\lane}}
  {\switchoutX{\wrapPi{\lane}}{\event_1}}
  \and
  \inferH{switch-wrapper-in}
  {\susp{\wrapPi{\lane}}{P} \\
    \switchin{\lane}{\cdot}{\event_2}
     \Exists \event_1.\wembed{\event_1 \prerel \event_2} \ast \wembedtok \ast
     P}
  {\wrapwait{\lane}}
\end{mathpar}
\caption{Wrapper rules}
\label{fig:model:wrapper-rules}
\end{figure}

\paragraph{Wrapper lane}
Now, we can present the wrapper lane $\wrapPi{\lane}$.
Its main rules are shown in \autoref{fig:model:wrapper-rules}.%
\footnote{These rules are simplified to omit some tokens and specialized for the case when the wrapper appears on the implementation side. The specification side rules are the dual of these rules.}
\ruleref{switch-wrapper-out} states that when the module in the wrapper lane emits an event $\event_1$, we obtain an event $\event_2$ translated via $\event_1 \postrel \event_2$ and $\wembedtok$, then emit $\event_2$ in the outer lane $\lane$ and enter the waiting state $\wrapwait{\lane}$.
From this waiting state we can switch back into the wrapper using \ruleref{switch-wrapper-in}.
Here we have to find a matching event $\event_1$ for the incoming event $\event_2$, prove $\event_1 \prerel \event_2$ and give up $\wembedtok$, before continuing in the suspended state $P$ of $\wrapPi{\lane}$.
We omit the elimination rule of $\wrapPi{\lane}$ that is similar to \ruleref{link-lane-elim-l} and relates $\wrapPi{\lane}$ to $\simgenname_{\modprepost{\wrapsub}{\module}}$.

\section{Reasoning about Languages}\label{sec:languages}
After introducing the core \thelogic{} constructs for reasoning about DimSum modules in \autoref{sec:model}, this section describes how we can build program logics using these constructs.
Concretely, \autoref{sec:wp} introduces \thelogic{}'s expression-based weakest precondition connective.
Then, \autoref{sec:rec-asm-pseudo} describes the \thelangRec{}, \thelangAsm{}, and \thelangSpec{} languages.
Finally, \autoref{sec:rec-asm-wrapper} describes the \thelangRec{}-to-\thelangAsm{} wrapper $\moditoa{\gapImp}$---in particular, how its separation logic is connected to the program logics of \thelangRec{} and \thelangAsm{}.

\subsection{Expression-Based Program Logic}\label{sec:wp}
The module weakest precondition \simgenname from \autoref{sec:model:semantics} lets us reason about DimSum modules inside \thelogic{}, but it has an important drawback:
It operates on the whole state of the module.
In this section, we will see how we can obtain a more standard weakest precondition that operates on expressions instead of module states.
For this, we first define a \emph{language} as a five-tuple $\lang \eqdef (\module_\lang, \exprs_\lang, \ctxs_\lang, \exprrelname{\lang}, \SI{\lang})$:
First, we have the module $\module_\lang$ underlying the language.
Second, we have a type of expressions $\expr \in \exprs_\lang$.
Third, we have a type of evaluation contexts $\ctx \in \ctxs_\lang$ that denote subexpressions that can be evaluated (with a fill operation $\ctx[\expr] \in \exprs_\lang$).
Fourth, we have the relation $\exprrelname{\lang} \subseteq (\exprs_\lang \times \mstates_{\module_\lang})$ that relates expressions to states of the underlying module $\module_\lang$.
Finally, we have the state interpretation $\SI{\lang} : \mstates_{\module_\lang} \to \iProp$ that links the state of $\module_\lang$ to the resources of the program logic.
For any language $\lang$, we define the expression weakest precondition $\simgenexprname$:
\begin{align*}
  &\simgenexprsimpl[\lane]{\expr}{\Phi} \eqdefind{}
    \All \mstate, \ctx.\spac \exprrel{\lang}{\ctx[\expr]}{\mstate} \wand \SI{\lang}(\mstate) \wand {}\\
    &\quad\simgenname^{\ts_{\lane}}_{\module_{\lang}}\spac\mstate\spac\big\{\lbl, \mstate'.\spac
      \begin{array}[t]{@{} l}
        \big(\Exists \expr'.\spac \lbl = \silent \ast \exprrel{\lang}{\ctx[\expr']}{\mstate'} \ast \SI{\lang}(\mstate') \ast
\Phi(\expr')\big) \lor {}\\
        \big(\mstatefrac{\gvar_\lane}{\mstate'} \wand \switchSomeS{\lane}{\lbl}
          \Exists \mstate'', \expr'.\mstatefrac{\gvar_\lane}{\mstate''} \ast \exprrel{\lang}{\ctx[\expr']}{\mstate''} \ast \SI{\lang}(\mstate'') \ast
 \simgenexpr{\expr'}{\lane}{\Phi}\big)\big\}
      \end{array}
\end{align*}
Intuitively, $\simgenexprsimpl[\lane]{\expr}{\Phi}$ states that it is safe to execute $\expr$ in lane $\lane$ with postcondition $\Phi$.
This meaning is encoded using the module-based weakest precondition $\simgenname$:
Proving $\simgenexprsimpl[\lane]{\expr}{\Phi}$ means proving $\simgenname~\mstate$ for a state $\mstate$ that contains the expression $\expr$ (encoded via $\exprrel{}{\ctx[\expr]}{\mstate}$) and for which the state interpretation $\SI{}(\mstate)$ holds.
Following $\simgenname$, there are two cases:
Either we reach the postcondition $\Phi$ for an expression $\expr'$ related to the new state $\mstate'$. In this case, we also need to establish $\SI{}(\mstate')$ and show that there was no visible event ($\lbl = \silent$).
Or, we emit $\lbl$ using $\switchSomeS{\lane}{\lbl}$ and then continue with $\simgenexpr{\expr'}{\lane}{\Phi}$ for an expression $\expr'$ related to the final state $\mstate''$.
We use the lane's ghost variable $\gvar_\lane$ to track the module's state around the switch: before switching out we obtain $\mstatefrac{\gvar_\lane}{\mstate'}$ for the new state $\mstate'$, and afterwards we ensure that we continue in the module's state by requiring $\mstatefrac{\gvar_\lane}{\mstate''}$.

We can use this definition to prove standard program logic rules for concrete expressions (see \autoref{sec:rec-asm-pseudo} for examples). In addition, $\simgenexprname$ satisfies some useful generic rules shown in \autoref{fig:sim_gen_expr_rules}:
First, \ruleref{wp-stop} establishes a $\simgenexprname$ by proving the postcondition for the current expression $\expr$. (Note that there is no notion of values and the postcondition is defined on expressions as in Simuliris~\cite{Simuliris}.)
Second, \ruleref{wp-bind} provides a bind rule for evaluation contexts.
Third, \ruleref{wp-switch-silent} lets one perform a silent switch when proving $\simgenexprname$.
This is useful to switch from the implementation to the specification (or back) during the proof.
Finally, the elimination rule \ruleref{wp-elim} allows us to prove a module weakest precondition using an expression weakest precondition.
The postcondition $\lane$ of $\simgenname$ becomes the lane of $\simgenexprname$. The postcondition of $\simgenexprname$ is $\FALSE$ since DimSum programs do not terminate, but wait for new incoming calls after returning from a call.

\begin{figure}
  \centering
  \begin{mathpar}
    \inferH{wp-stop}
    {\Phi(\expr)}
    {\simgenexpr{\expr}{\lane}{\Phi}}
    \and
    \inferH{wp-bind}
    {\simgenexpr{\expr}{\lane}{\expr'.\spac \simgenexpr{\ctx[\expr']}{\lane}{\Phi}}}
    {\simgenexpr{\ctx[\expr]}{\lane}{\Phi}}
    \and
    \inferH{wp-switch-silent}
    {\switchsilent{\lane} \simgenexpr{\expr}{\lane}{\Phi}}
    {\simgenexpr{\expr}{\lane}{\Phi}}
    \and
    \inferH{wp-elim}
    {\exprrel{\lang}{\ctx[\expr]}{\mstate} \\ \mstatefull{\gvar_\lane} \\ \SI{\lang}(\mstate) \\
     \simgenexpr{\expr}{\lane}{\_. \FALSE}}
    {\simgenlane{\module_{\lang}}{\mstate}{\lane}}
  \end{mathpar}
  \caption{Rules for the expression-level weakest precondition}
  \label{fig:sim_gen_expr_rules}
\end{figure}

\subsection{Languages}\label{sec:rec-asm-pseudo}
This section gives a brief overview of the program logics of \thelangRec{}, \thelangAsm{}, and \thelangSpec{}. More details, including the full syntax, semantics, and program logic rules, can be found in \appendixref.

\begin{figure}
  \centering
\begin{mathpar}
   \inferH{rec-add}
   {\post(\mathimp{n_1 + n_2})}
   {\simgenexpr{\mathimp{n_1} + \mathimp{n_2}}{\PiRec}{\post}}
   \and
   \inferH{rec-load}
   {\loc \hptsto \val \\ \loc \hptsto \val \wand \post(\val)}
   {\simgenexpr{\Hload{\loc}}{\PiRec}{\post}}
   \and
   \inferH{rec-store}
   {\loc \hptsto \val \\ \loc \hptsto \valS \wand \post(\valS)}
   {\simgenexpr{\Hstore{\loc}{\valS}}{\PiRec}{\post}}
   \and
   \inferH{switch-rec-call}{\fnext{\fnname} \\ \All \heap. \heapinv(\heap)\wand
      \switchoutS{\PiRec}{\ECallH(\fnname, \vals, \heap)}
      \recwait
      \switchin{\PiRec}{\EReturnH}{\valS, \heapS}\heapinv(\heapS) \ast \rettok(\valS)}
    {\bicall{\fnname}{\vals}{\val}\rettok}
\end{mathpar}
  \caption{\thelangRec{} program logic rules}
  \label{fig:rec-rules}
\end{figure}
\paragraph{\thelangRec{}}
Let us now describe our program logic for \thelangRec{}.
The program logic is based on the following assertions:
First, we have the points-to predicate $\loc \hptsto \val$ and the heap invariant $\heapinv(\heap)$ that ties the points-to assertions to the heap $\heap$. (For Iris experts, the heap invariant contains the authoritative ghost state.)
In fact, since \thelangRec{} uses a block-based memory model, the program logic also provides a block points-to $\prov \hptstoBlock \vals$ stating that the memory block $\prov$ contains the values $\vals$.
The constructions in this paper generalize to blocks, but the presentation sticks to single locations for simplicity.
Second, we have the (duplicable) function name assertions $\fninternal{\fnname}{\mathimp{fn}}$ and $\fnext{\fnname}$, which track the knowledge whether a function name $\fnname$ is implemented as a \thelangRec{} function $\mathimp{fn}$ or not.
These assertions are tied to the global set of \thelangRec{} functions $\mathimp{fns}$ with the function invariant $\fninv(\mathimp{fns})$.
The instantiation of the language interface from \autoref{sec:wp} for \thelangRec{} instantiates the state interpretation with the two invariants $\heapinv(\heap)$ and $\fninv(\mathimp{fns})$, linking the user-facing points-to and function assertions to the global state of the \thelangRec{} program.
From this instantiation it is straightforward to prove the standard program logic rules for \thelangRec{}, like \ruleref{rec-add}, \ruleref{rec-load} and \ruleref{rec-store} in \autoref{fig:rec-rules}.
We define \abscalls using $\simgenexprname$ applied to function calls:
\[
  \bicall{\fnname}{\vals}{\val}{\rettok(\val)} \eqdef \simgenexpr{\fnname(\vals)}{\PiRec}{\rettok}
\]
Using this definition, we prove \ruleref{switch-rec-call} in \autoref{fig:rec-rules}. (Compared to the simplified \ruleref{key-switch-rec-call} from \autoref{sec:key:switching}, this version includes the heap.)

\paragraph{\thelangAsm{}}
The program logic for \thelangAsm{} is based on three kinds of assertions:
First, we have the points-to predicate $\addrraw \mptsto \aval$ and the corresponding invariant $\meminv(\mem)$ about the \thelangAsm{} memory $\mem$.
Second, we have the register points-tos $\reg{} \rptsto \aval$ and the register invariant $\reginv(\regs)$.
Third, we have the instruction assertions $\inssinternal{\insaddr}{\instrs}$ and $\insext{\insaddr}$ and the instruction invariant $\instrinv(\instrmap)$.
Similar to \thelangRec{}, the state interpretation for \thelangAsm{} uses $\meminv(\mem)$, $\reginv(\regs)$, and $\instrinv(\instrmap)$ to link the memory, register, and instruction assertions to the global state of the \thelangAsm{} program.
From this instantiation, it is straightforward to prove the expected program logic rules for \thelangAsm{}.

\paragraph{\thelangSpec{}}
\thelangSpec{} is a shallowly embedded language based on interaction trees (ITrees)~\cite{ITrees}. It provides similar operations to DimSum's \thelangSpecOrig{}. The main difference is that \thelangSpec{} includes a native notion of functions, allowing us to define the waiting assertion $\specwaitraw$ as (the $\simgenexprname$ of) the ITree waiting for arbitrary incoming function calls or returns.
Thus, we obtain rules like \ruleref{key-pseudo-wait-call} (\autoref{sec:key:switching}).

\subsection[Rec-to-Asm Lane]{\thelangRec-to-\thelangAsm{} Lane}\label{sec:rec-asm-wrapper}
After discussing \thelangRec{} and \thelangAsm{} in \autoref{sec:rec-asm-pseudo}, this section describes how \thelogic{} supports cross-language reasoning between \thelangRec{} and \thelangAsm{}, in particular the $\rtoaPi{\lane}$ lane and the exchange $\rtoaexchange(\exchanged)$.

\paragraph{Background: \thelangRec{}-to-\thelangAsm{} wrapper}
Before diving into \thelogic{}, we need to review DimSum's $\moditoa{\gapImp}$ wrapper.
As described in \autoref{sec:model:wrapper}, this wrapper defines a separation logic $\seplogicitoa$, which is used to express the $\postrelitoa$ and $\prerelitoa$ relations to translate between \thelangRec{} and \thelangAsm{} events.
$\seplogicitoa$ contains the following parts:
First, it contains separation logic assertions for the \thelangRec{} heap, in particular the heap invariant $\heapinv(\heap)$ and corresponding points-to predicates $\loc \hptsto \val$.
Second, it contains separation logic assertions for the \thelangAsm{} memory (the memory invariant $\meminv(\mem)$ and points-to predicates $\addrraw \mptsto \aval$).
Finally, it contains a (persistent) injection assertion $\loc \itoabij \addrraw$ stating that the location $\loc$ on the \thelangRec{} heap is mapped to the address $\addrraw$ in the \thelangAsm{} memory.
The injection is mediated by the $\rtoainj(\exchanged)$ invariant that keeps track of the global injection mapping.
The list $\exchanged$ allows us to temporarily remove elements from the injection to obtain the underlying points-to facts (inspired by \citet{Simuliris}) as shown by the rules in \autoref{fig:bij}.
\begin{figure}
  \centering
  \begin{mathpar}
    \inferH{r2a-inj-borrow}
    {}
    {\loc \notin \exchanged \ast \loc \itoabij \addrraw \ast \rtoainj(\exchanged) \vdash \Exists \val, \aval. \rtoainj(\lcons{\loc}\exchanged) \ast \loc \hptsto \val \ast \addrraw \mptsto \aval \ast \val \itoabij \aval}
    \and
    \inferH{r2a-inj-return}
    {}
    {\rtoainj(\lcons{\loc}\exchanged) \ast\loc \itoabij \addrraw \ast \loc \hptsto \val \ast \addrraw \mptsto \aval \ast \val \itoabij \aval \vdash \rtoainj(\exchanged)}
  \end{mathpar}
  \caption{Rules for $\rtoainj(\exchanged)$ (omitting a side condition that $\loc$ is allocated in \ruleref{r2a-inj-borrow})}
  \label{fig:bij}
\end{figure}
To see how this separation logic is used, let us sketch $\postrelitoa$ (the full version is in \citet[Appendix E]{DimSum}):
\[
\ECall{\fnname}{\vals}{\heap} \postrelitoa \EJump{\regs}{\mem} \eqdef \heapinv(\heap) \ast \meminv(\mem) \ast \rtoainj([]) \ast \left(\textstyle\Sep\nolimits_{\val\in\vals,\aval\in{\rlookup{\regs}{\reg{0} \ldots}}} {\val} \itoabij {\aval}\right) \ast \ldots
\]
To translate a $\ECallH$ to a $\EJumpH$, $\postrelitoa$ asserts the invariants for the heap and memory and the full injection $\rtoainj([])$. Additionally, the \thelangRec{} arguments must be related to the arguments stored in the corresponding registers. ($\val \itoabij \aval$ is obtained by lifting $\loc \itoabij \addrraw$ to values.)

\paragraph{Connecting separation logics}
The main challenge when reasoning about $\moditoa{\gapImp}$ in \thelogic{} is to connect all the different separation logic assertions:
After constructing the $\rtoaPi{\lane}$ lane using the generic construction from \autoref{sec:model:wrapper}, we obtain two instances of the \thelangRec{} assertions: $\wembed{\heapinv(\heap)}$ and $\wembed{\loc \hptsto \val}$ as the embedded (\autoref{sec:model:wrapper}) assertions of $\moditoa{\gapImp}$ and $\heapinv(\heap)$ and $\loc \hptsto \val$ as the native \thelangRec{} assertions in \thelogic{} from \autoref{sec:rec-asm-pseudo}.
Similarly, we obtain both $\wembed{\meminv(\mem)}$ / $\wembed{\addrraw \mptsto \aval}$ embedded from $\moditoa{\gapImp}$ and $\meminv(\mem)$ / $\addrraw \mptsto \aval$ natively from \autoref{sec:rec-asm-pseudo}.
In this section, we will see how we connect these assertions and translate between them.
For this, we first introduce the concept of a \emph{trader} before showing how to obtain the \emph{exchange} from \autoref{sec:key:wrapper}.

\paragraph{Trading points-tos}
First, we consider the \thelangAsm{} ghost state.
When a \thelangRec{} function calls an \thelangAsm{} function with a pointer argument, we can use \ruleref{r2a-inj-borrow} to obtain $\wembed{\addrraw \mptsto \aval}$.
However, to use this points-to with the program logic rules for \thelangAsm{}, we need the native $\addrraw \mptsto \aval$. To bridge this gap, we introduce the concept of a \emph{trader}.

A \emph{trader} $\memtrader$ links the embedded and native \thelangAsm{} assertions via the following rule:%
\footnote{Technically, both directions use a basic update instead of an entailment.}
\begin{mathpar}
  \inferH{mem-trade-ptsto}
  {}
  {\memtrader \ast \meminv(\mem) \ast \wembed{\addrraw \mptsto \aval} \dashv\vdash \memtrader \ast \meminv(\mem) \ast \addrraw \mptsto \aval}
\end{mathpar}
This rule allows us to trade an embedded $\wembed{\addrraw \mptsto \aval}$ for a native $\addrraw \mptsto \aval$ (and vice versa) in the presence of the trader $\memtrader$ and the invariant $\meminv(\mem)$.
($\meminv(\mem)$ is part of the state interpretation of the \thelangAsm{} program logic and thus is always available when proving an \thelangAsm{} $\simgenexprname$.)

The trader is defined as follows:
\begin{align*}
  \memtrader \eqdef{} \Exists \mem.
    \wembed{\meminv(\mem)} \ast
    \textstyle\Sep\nolimits_{\addrraw \mapsto \aval \in \mem}
      \big(\wembed{\addrraw \mptsto \aval} \lor \addrraw \mptsto \aval\big)
\end{align*}
It stores one points-to for each memory address, \ie either $\wembed{\addrraw \mptsto \aval}$ or $\addrraw \mptsto \aval$.
Trading a points-to flips which side of the disjunction holds for the traded address (using the exclusivity of $\addrraw \mptsto \aval$ to prune the other case).
Note that the memories of $\wembed{\meminv(\mem)}$ and $\meminv(\memS)$ are not directly kept in sync. Instead, the trader contains the ownership necessary to synchronize them when needed.

While we here only present the trader for \thelangAsm{}, the idea is more general. In fact, \thelogic{} also provides a trader for \thelangRec{} and it does not just allow relating embedded and native assertions, but also assertions of multiple wrappers for more complex proofs. We use this capability to significantly simplify the difficult-to-prove vertical compositionality theorem of \citet{DimSum}.

\paragraph{Exchange}
After seeing how we can use the trader to link the \thelangAsm{} assertions, let us now describe the \emph{exchange} $\rtoaexchange(\exchanged)$ that allows us to exchange $\loc \hptsto \val$ and $\addrraw \mptsto \aval$, encapsulating all reasoning about the embedded separation logic $\seplogicitoa$.
As described in \autoref{sec:key:wrapper}, the parameter $\exchanged$ tracks the locations that have been exchanged since they need to be exchanged back before switching back from \thelangAsm{} to \thelangRec{}.
The exchange $\rtoaexchange(\exchanged)$ is defined as follows:
\begin{align*}
  \rtoaexchange(\exchanged) \eqdef{}
    &
    \wembedtok \ast
    \memtrader \ast
    \wembed{\rtoainj(\exchanged)} \ast
    \Exists \heap.
    \wembed{\heapinv(\heap)} \ast
    \heapinv(\heap) \ast
      \textstyle\Sep\nolimits_{\loc \in \exchanged} \Exists \val. \wembed{\loc \hptsto \val} \ast \loc \hptsto \val
\end{align*}
It consists of four parts:
First, the $\wembedtok$ token that enables updates inside the embedding (\autoref{sec:model:wrapper}).
Second, the trader $\memtrader$.
Third, the global injection $\wembed{\rtoainj(\exchanged)}$ with the elements from $\exchanged$ removed.
Fourth, both invariants for the \thelangRec{} heap with the $\loc \hptsto \val$ assertions for all exchanged locations.

\begin{figure}
  \centering
  \begin{mathpar}
    \inferH{exchange-rec-to-asm}
    {}
    {\loc \notin \exchanged \ast \wembed{\loc \itoabij \addrraw} \ast \rtoaexchange(\exchanged) \ast \loc \hptsto \val \ast \meminv(\mem) \vs \Exists \aval. \rtoaexchange(\lcons{\loc}\exchanged) \ast \addrraw \mptsto \aval \ast \wembed{\val \itoabij \aval} \ast \meminv(\mem)}
    \and
    \inferH{exchange-asm-to-rec}
    {}
    {\wembed{\loc \itoabij \addrraw} \ast \wembed{\val \itoabij \aval} \ast \rtoaexchange(\lcons{\loc}\exchanged) \ast \addrraw \mptsto \aval \ast \meminv(\mem) \vs \rtoaexchange(\exchanged) \ast \loc \hptsto \val \ast \meminv(\mem)}
  \end{mathpar}
  \caption{Exchange rules}
  \label{fig:exchange}
\end{figure}
To explain how the exchange works, let us sketch the proof of its key rules shown in \autoref{fig:exchange}.
These rules are the full version of the rules shown in \autoref{sec:key:wrapper}. In particular, they include $\meminv(\mem)$ and the embedding, which were omitted in \autoref{sec:key:wrapper} for simplicity.
\ruleref{exchange-rec-to-asm} allows us to exchange a heap points-to $\loc \hptsto \val$ for a memory points-to $\addrraw \mptsto \aval$ of the $\addrraw$ mapped to by the injection.
Under the hood, this works by first removing $\loc$ from the injection (\ruleref{r2a-inj-borrow}) to obtain $\wembed{\loc \hptsto \valS}$ and $\wembed{\addrraw \mptsto \aval}$.
Then, we learn that the values of $\loc \hptsto \val$ and $\wembed{\loc \hptsto \valS}$ agree (\ie $\val = \valS$) since the two $\heapinv$ in the exchange hold for the same heap.
Finally, we use the trader to trade $\wembed{\addrraw \mptsto \aval}$ for $\addrraw \mptsto \aval$.
\ruleref{exchange-asm-to-rec} works in reverse: We first trade $\addrraw \mptsto \aval$ for $\wembed{\addrraw \mptsto \aval}$, then we update the $\loc \hptsto \valS$ and $\wembed{\loc \hptsto \valS}$ assertions in the exchange to the desired value $\val$ using $\heapinv$
and, finally, we return the borrow to the injection (\ruleref{r2a-inj-return}).

Overall, this shows that while the reasoning about the $\moditoa{\gapImp}$ might seem complex with its four different kinds of separation logic assertions, we can hide all this complexity behind the abstractions of the exchange and the trader. A user of \thelogic{} only needs to interact with the separation logic assertions of \thelangRec{} and \thelangAsm{} and the high-level concept of the exchange, without needing to know how the $\moditoa{\gapImp}$ wrapper works under the hood.

\section{Examples} \label{sec:case-study}
This section illustrates \thelogic{} based on two examples: First, an extended version of the $\echoId$ example from \autoref{sec:keyideas}, showing how \thelogic{} supports reusing specifications across many different verifications. Second, an example that demonstrates \thelogic{}'s ability for higher-order cross-language reasoning.
Finally, we describe how the second example can be integrated with DimSum's verified compiler.

\begin{figure}
  \centering
  \begin{tikzpicture}[
      remember picture,
      node distance=2mm,   %
      ref/.style={inner sep=1pt},
      lbl/.style={inner sep=1pt, font=\small\sffamily},
      topedge/.style={draw, out=-90, in=90, looseness=0.4, shorten <=1pt, shorten >=1pt},
      botedge/.style={draw, out=90, in=-90, looseness=0.4, shorten <=1pt, shorten >=1pt},
      ]
    \def\pspace{4mm}
    \def\botspace{12mm}
    \def\vertspace{14mm}
    \def\botvertspace{8mm}
    \node[ref] (g1) {(1) $\tikzmarknode{g1l}{\recbullet} \refinesraw \tikzmarknode{g1r}{\pseudobullet}$};
    \node[ref, base right=of g1] (g2) {(2) $\tikzmarknode{g2l}{\recbullet} \refinesraw \tikzmarknode{g2r}{\specbullet}$};
    \node[ref, base right=of g2] (g3) {(3) $\tikzmarknode{g3l}{\recbullet} \recoplus \tikzmarknode{g3m}{\pseudobullet} \refinesraw \tikzmarknode{g3r}{\specbullet}$};
    \node[ref, base right=of g3] (g4) {(4) $\tikzmarknode{g4l}{\recbullet} \recoplus \tikzmarknode{g4m}{\pseudobullet} \refinesraw \tikzmarknode{g4r}{\specbullet}$};
    \node[ref, base right=of g4] (g5) {(5) $\lceil\tikzmarknode{g5l}{\recbullet}\rceil_{\rtoasub} \asmoplus \tikzmarknode{g5m}{\asmbullet} \refinesraw \tikzmarknode{g5r}{\specbullet}$};
    \node[ref, base right=of g5] (g6) {(6) $\lceil\tikzmarknode{g6l}{\recbullet}\rceil_{\rtoasub} \asmoplus \tikzmarknode{g6m}{\asmbullet} \refinesraw \tikzmarknode{g6r}{\specbullet}$};

    \coordinate (rowc) at ($(g1.west)!0.5!(g6.east)$);

    \node[lbl, above=\vertspace of rowc] {%
      \tikzmarknode{putcrec}{$\putcRec$}\hspace{\pspace}%
      \tikzmarknode{getcrec}{$\getcRec$}\hspace{\pspace}%
      \tikzmarknode{mainechorec}{$\impsynlink{\mainRec}{\echoRec}$}\hspace{\pspace}%
      \tikzmarknode{putcpseudotwo}{$\putcPseudoTwo$}\hspace{\pspace}%
      \tikzmarknode{putcpseudo}{$\putcPseudo$}\hspace{\pspace}%
      \tikzmarknode{readasm}{$\readAsm$}\hspace{\pspace}%
      \tikzmarknode{putcasm}{$\putcAsm$}\hspace{\pspace}%
      \tikzmarknode{getcasm}{$\getcAsm$}%
    };

    \node[lbl, below=\botvertspace of rowc] {%
      \tikzmarknode{mainpseudo}{$\mainPseudo$}\hspace{\botspace}%
      \tikzmarknode{mainbufpseudo}{$\mainPseudoTwo$}\hspace{\botspace}%
      \tikzmarknode{mainsyspseudo}{$\mainPseudoThree$}%
    };

    \draw[topedge, \reclinecolor] (putcrec)  edge (g1l) edge (g2l);
    \draw[topedge, \reclinecolor] (getcrec.south)   edge (g1l);
    \draw[topedge, \reclinecolor] (getcrec.south)   edge (g2l);
    \draw[topedge, \reclinecolor, out=-95] (getcrec.south)   edge (g3l);
    \draw[topedge, \reclinecolor, out=-85, in=100] (getcrec.south)   edge (g4l.north);
    \draw[topedge, \reclinecolor, out=-85, looseness=0.35] (getcrec.south)   edge (g5l.north);
    \draw[topedge, \reclinecolor] (mainechorec.south)   edge (g1l);
    \draw[topedge, \reclinecolor] (mainechorec.south)   edge (g2l);
    \draw[topedge, \reclinecolor] (mainechorec.south)   edge (g3l);
    \draw[topedge, \reclinecolor, out=-95] (mainechorec.south)   edge (g4l);
    \draw[topedge, \reclinecolor, out=-85, in=90] (mainechorec.south)   edge (g5l);
    \draw[topedge, \reclinecolor, out=-85, in=90, looseness=0.3] (mainechorec.south)   edge (g6l);
    \draw[topedge, \pseudolinecolor] (putcpseudotwo) edge (g3m);
    \draw[topedge, \pseudolinecolor] (putcpseudo)  edge (g4m);
    \draw[topedge, \asmlinecolor] (readasm) edge (g5m);
    \draw[topedge, \asmlinecolor] (putcasm) edge (g5m) edge (g6m);
    \draw[topedge, \asmlinecolor] (getcasm) edge (g6m);
    \draw[botedge, \pseudolinecolor] (mainpseudo)  edge (g1r) edge (g3r);
    \draw[botedge, \pseudolinecolor] (mainbufpseudo)  edge (g2r) edge (g4r);
    \draw[botedge, \pseudolinecolor, out=90] (mainsyspseudo.north)  edge (g5r);
    \draw[botedge, \pseudolinecolor, out=80] (mainsyspseudo.north)  edge (g6r);
  \end{tikzpicture}

  \vspace{0.25cm}
  \begin{tabular}{lll}
    \midrule
    $\recbullet$: \thelangRec{} module & $\pseudobullet$: \thelangSpec{}/\thelangSpecOrig{} module & $\asmbullet$: \thelangAsm{} module\\
    \midrule
  \end{tabular}

  \caption{Echo programs}
  \label{fig:casestudy-echo}
\end{figure}

\paragraph{Echo}
To demonstrate how \thelogic{} specifications can be reused across implementations, we verify six multi-language programs involving $\echoId$, composed in different ways.
The refinements are shown in \autoref{fig:casestudy-echo}. Each of the six refinement proofs involves the functions connected by lines to its components, illustrating how functions are reused across verifications. Crucially, we prove a \emph{single} \thelogic{} specification for each function and reuse that specification across \emph{all} programs that involve the function.
We prove the following refinements: (1) \thelangRec{} implementation refines $\mainPseudo$, (2) \thelangRec{} implementation refines a buffered $\mainPseudoTwo$, (3) \thelangRec{} implementation of $\echoId$ and $\getcId$ linked with unbuffered \thelangSpec{} implementation of $\putcId$ refines $\mainPseudo$,
(4) \thelangRec{} implementation of $\echoId$ and $\getcId$ linked with buffered \thelangSpec{} implementation of $\putcId$ refines $\mainPseudoTwo$,
(5) \thelangRec{} implementation of $\echoId$ and $\getcId$  linked with \thelangAsm{} implementation of $\readId$ and $\putcId$ refines $\mainPseudoThree$,
and (6) \thelangRec{} implementation of $\echoId$ linked with \thelangAsm{} implementation of $\getcId$ and $\putcId$ refines $\mainPseudoThree$.
This example shows that the same (\thelangRec{}) specification can be reused when the implementation (a) is in the same language (\thelangRec{}), (b) is in a different language with the same events (\thelangSpec{}), (c) is in a language with different events behind a wrapper (\thelangAsm{}), and (d) is treated as external to the verification and handled via refinement reasoning. (Some implementations use DimSum's \thelangSpecOrig{} language. For the purposes of this example, \thelangSpecOrig{} is equivalent to \thelangSpec{}.)

\newcommand{\readBufferId}{{process\_message}} %
\newcommand{\readBufferAsm}{\asmprog{\readBufferId}_{\thelangAsmSubLong}}
\newcommand{\mainBufferRec}{\imp{\Hid{main\_buffer}_{\thelangImpSubLong}}}
\newcommand{\hashRec}{\imp{\Hid{hash}_{\thelangImpSubLong}}}
\newcommand{\mainBufferPseudo}{\specorigprog{\Hid{main\_buffer}}}

\paragraph{Higher-order cross-language reasoning}
To demonstrate \thelogic{}'s capabilities for reasoning about higher-order functions across languages, we verified a handwritten assembly routine $\readBufferAsm$ that reads a dynamically sized message from a socket and processes it using a callback.
(Such callbacks are common in the context of web servers.)
Concretely, $\readBufferAsm$ uses a syscall to receive bytes until it reads a zero byte and stores them in a dynamically allocated buffer on the stack.
It then invokes the callback to process the received message and deallocates the buffer afterwards.
Thus, $\readBufferAsm$ abstracts away the interaction with the system and all dynamic memory management.
We call $\readBufferAsm$ from a \thelangRec{} main function that passes a \thelangRec{} callback computing a hash of the content of the buffer and prove that the combined program refines a \thelangSpec{} specification: \[\asmlink{\moditoa{\impsyn{\impsynlink{\mainBufferRec}{\hashRec} }}}{\asmsyn{\readBufferAsm}} \refinesraw \mainBufferPseudo\]

We independently verify $\readBufferAsm$, $\mainBufferRec$, and $\hashRec$ in the \thelangAsm{} resp. \thelangRec{} sublogic, using abstract calls to abstract over the function pointer invocation in $\readBufferAsm$.
Then, we combine the proofs by resolving the abstract call via the wrapper lane, and we use the exchange to convert the dynamically allocated memory in \thelangAsm{} to \thelangRec{} points-tos by extending the injection.
This example shows that \thelogic{} can reason about multi-language programs that share dynamically allocated memory and function pointers across a heterogeneous language boundary.
The heterogeneous language boundary is necessary since $\readBufferAsm$ interleaves memory operations with system calls and therefore cannot be represented using \thelangRec{} events.

\paragraph{Integration with DimSum's verified compiler}
To demonstrate that programs verified with \thelogic{} are compatible with verified compilation, we combine the refinement from proven in the previous example with DimSum's verified \thelangRec{}-to-\thelangAsm{} compiler. This integration is seamless since the \thelogic{} proof uses the same wrapper as the compiler correctness statement.
\section{Related Work}
\label{sec:related-work}
We first compare with multi-language program logics for specific language pairs, then with program logics with reasoning principles for external calls, before discussing related work more broadly.

\paragraph{Multi-language program logics for specific language pairs}
The most closely related work is Melocoton~\cite{Melocoton}, which provides a multi-language program logic for OCaml and C.
The OCaml/C FFI considered by Melocoton involves a garbage collector and thus is more complex than the \thelangRec{}/\thelangAsm{} interaction considered in this paper.
However, Melocoton only considers the specific OCaml/C setting and does not aim to be a general framework for multi-language reasoning like \thelogic{}.
In Melocoton, the verification of the OCaml and C code is separated, with function specifications mediating at the boundary. In contrast, \thelogic{} does not enforce such a boundary, but allows one to freely switch between languages in a single proof using the switching modality.
Also, Melocoton does not consider refinement reasoning and is not connected to a verified compiler.

Iris-Wasm~\cite{IrisWasm} provides a program logic for the interaction of WebAssembly and its host-language. Since these languages share basically the same memory model, Iris-Wasm can avoid the view reconciliation problem and use one set of separation logic assertions for the whole verification. In contrast, \thelogic{} introduces exchanges (\autoref{sec:rec-asm-wrapper}) to address the view reconciliation problem.

\paragraph{Program logics with FFIs}
There are various program logics that allow reasoning about code interacting with external functions via an FFI~\cite{VSTFFI, AdamsLightbulb, Cito, CitoNewer, Disel, Sandboxing, Islaris, VeriFFI}.
However, these works only provide a program logic for one side of the FFI, not one spanning across both sides of the FFI.

\citet{VeriFastIO2, VeriFastIO} show how to encode specifications for external functions using nested Hoare triples. These nested Hoare triples serve a similar purpose to the \abscalls of \thelogic. However, nested Hoare triples require the use of separation logic tokens for encoding sequencing and prophecy variables to predict the return value of an abstract function. In contrast, \abscalls implicitly express sequencing and provide the result via a binder to the following specification.

\paragraph{Multi-language semantics}
There is a large body of work on multi-language semantics not based on program logics~\cite{Multilanguage, FunTAL, MultilanguageCompiler, CompCertO, Pilsner, 700cc}. Most of this work focuses on compiler verification or type soundness of interoperation. We only compare with the most closely related approaches.

\thelogic{} is based on the multi-language semantics provided by DimSum~\cite{DimSum}. DimSum focuses on constructing multi-language programs and only provides a basic simulation-based proof technique to prove refinements. In contrast, \thelogic{} provides a program logic for reasoning about DimSum-based multi-language programs that enables small-footprint style specifications and reasoning.

\citet{SemanticSoundness} prove sound interoperability of languages with different memory models by translating them to a common target language. This is in contrast to \thelogic{}, which reasons about interoperation of languages independent of compilation.

\citet{GITrees} introduce GITrees as a step-indexed denotational semantics for higher-order programs. They prove type safety of cross-language interoperability by denoting the languages into GITrees.
However, they do not provide a source-level program logic for the languages nor do they reason about languages with different views on the same shared state.

\citet{ThreeDimensionalRefinement} provide a multi-language semantics rich enough to embed the semantics of CompCertO~\cite{CompCertO, DirectRefinement}. Like \thelogic{}, the work addresses the limitations of single-language program logics~\cite[Section 1.1]{ThreeDimensionalRefinement}, but it takes a different approach: They equip the refinement relation with an algebraic structure to enable compositional reasoning in the vertical, horizontal, and spatial dimensions, whereas \thelogic{} provides principles for building multi-language program logics.

\paragraph{Staged specifications}
\thelogic{}'s \abscalls are closely related to the staged specification mechanism of \citet{HSSL}.
However, they use this mechanism to reason about higher-order imperative programs (and later \citet{HSSLEffects} to reason about algebraic effects), while \thelogic{} uses \abscalls to express calls to unknown functions in a multi-language program.

\paragraph{Refinement and separation logic}
CRIS~\cite{CRIS} and its predecessor CCR~\cite{CCR} combine refinement reasoning and separation logic. %
Concretely, CRIS introduces the concept of \emph{imaginary specifications}: executable specifications with ownership assertions realized in the operational semantics via dual non-determinism.
\thelogic{} takes the opposite approach: instead of operationalizing ownership, it lifts operational concepts---in particular, function calls---to the logic as \abscalls.
Thus, \thelogic{} stays closer to standard Iris-style logics for individual languages.
Also, as far as we know, CRIS has not been explored as a target for denoting multi-language programs.

Simuliris~\cite{Simuliris} provides an Iris-based program logic for proving compiler optimizations.
Like \thelogic{}, Simuliris integrates refinement reasoning into Iris, but the two differ in style and scope: Simuliris proves refinements through a binary simulation judgment relating two programs, whereas \thelogic{} uses unary reasoning linked via switches.

\paragraph{Post-crash modality}
The trader (\autoref{sec:rec-asm-wrapper}) has similarities to Perennial's post-crash modality~\cite{Perennial, NextGenModality}. The post-crash modality is used to exchange old assertions for new assertions after a crash, while the trader allows bidirectional exchange between embedded and native assertions.

\begin{acks}
  We would like to thank the anonymous reviewers for their constructive feedback and suggestions, which significantly improved the presentation of this paper.

\end{acks}

\section*{Disclosure of Use of AI in Research}
In a few instances, we employed AI agents to assist with Rocq proofs. All generated proofs were manually reviewed.

\section*{Data Availability Statement}
\ifarxiv
The Rocq development for this paper can be found in the supplementary material~\cite{Artifact}.
\else
The Rocq development and appendix for this paper can be found in the supplementary material~\cite{Artifact, Appendix}.
\fi
{
\interlinepenalty=10000
\bibliography{bib}
}

\appendix
\ifappendixincluded
\clearpage
\section{\thelangRec{} Semantics and Program Logic}\label{sec:appendix:rec}
We reproduce the syntax and semantics of \thelangRec{} from~\citet{DimSumAppendix} and give the full definition of its program logic in \thelogic{}.
\subsection{Syntax and Semantics}
\autoref{fig:appendix:rec-grammar} shows the syntax of \thelangRec{}, adapted from~\citet{DimSum} by adding function pointers as values. Besides the static expressions $\ExprImp$, there are runtime expressions $\RunExpr$, which are inserted during execution: $\Halloc{\Repeat{(x, n)}}{\rexpr}$ and $\Hfree{\Repeat{(\loc, n)}}{\rexpr}$ allocate and free stack frames, $\RetExt{\boolval}{\rexpr}$ returns to the environment once $\rexpr$ has been evaluated, and $\WaitingImp{\boolval}$ is the state in which the module waits for the environment.

\begin{figure}[h]
  \begin{align*}
    \Loc \ni \loc \eqdef{}& \{\compBlockId: \Prov,\, \compOffset : \mathbb{Z}\} \qquad
    \FnName \ni \fnname \eqdef{} \mathsf{String} \qquad \ImpHeap \ni \heap \eqdef{} \Loc \fpfn \ImpVal \\
    \ValImp \ni \val \eqdef{}& \intval : \mathbb{Z} \bnfor \boolval : \Bool \bnfor \loc : \Loc \bnfor \fnname : \FnName\\
    \Binop \ni \binop \eqdef{} & + \bnfor < \bnfor == \bnfor \leq \\
    \ExprImp \ni \exprimp \eqdef{}&
        \val \bnfor x \bnfor\exprimpA \binop \exprimpB \bnfor  \Hlet{x}{\exprimpA}~\exprimpB \bnfor \Hifelse{\exprimpA}{\exprimpB}{\exprimpC} \bnfor \Hcall{\exprimpA}{\Repeat{\exprimpB}} \bnfor \Hload{\exprimp} \bnfor \Hstore{\exprimpA}{\exprimpB} \\
    \RunExpr \ni \rexpr \eqdef{}& \cdots \bnfor \Halloc{\Repeat{(x, n)}}{\rexpr} \bnfor \Hfree{\Repeat{(\loc, n)}}{\rexpr} \bnfor \RetExt{\boolval}{\rexpr} \bnfor \WaitingImp\boolval \\
    \LibraryRec \ni \libImp \eqdef{}& (\Hfndef{\fnname}{\Repeat{x}}\Repeat{\Hlocal{y}{n};}\spac\exprimp), \libImp \bnfor \emptyset\\
    \ImpEvents \ni \impevent{} \eqdef{}&
      \ECallSend{\fnname}{\vals}{\heap} \bnfor \ECallRecv{\fnname}{\vals}{\heap} \bnfor \EReturnSend{\val}{\heap} \bnfor \EReturnRecv{\val}{\heap}
  \end{align*}
  \caption[]{Grammar and events of~\thelangRec{}, adapted from~\cite{DimSum, DimSumAppendix}.}\label{fig:appendix:rec-grammar}
\end{figure}

\autoref{fig:appendix:rec-sem} shows the small-step semantics of \thelangRec{} modules.

\begin{figure}
  \begin{mathpar}
    \inferH{rec-binop}{}{(\valA \binop \valB, \heap, \libImp) \mstep{\silent}_\thelangImpSub{} \setComp{(\val, \heap, \libImp)}{\mathsf{eval}_{\binop}(\valA, \valB, \val)}}

    \hspace{-5pt}\inferhref{rec-load}{appendix-rec-load}{}{(\Hload{\valA}, \heap, \libImp) \mstep\silent_\thelangImpSub{} \setComp{(\valB, \heap, \libImp)}{\Exists \loc. \valA = \loc \land \heap(\loc) = \valB}}

    \inferhref{rec-store}{appendix-rec-store}{}{(\Hstore{\valA}{\valB}, \heap, \libImp) \mstep\silent_\thelangImpSub{} \setComp{(\valB, \heap[\loc \mapsto \valB], \libImp)}{\Exists \loc. \valA = \loc \land \mathsf{heap\_alive}(\heap, \loc)}}

    \inferH{rec-if}{}{(\Hifelse{\val}{\exprimpA}{\exprimpB}, \heap, \libImp) \mstep\silent_\thelangImpSub{} \setComp{(\exprimp, \heap, \libImp)}{\Exists \boolval. \val = \boolval \land \ite{\boolval}{\exprimp = \exprimpA}{\exprimp = \exprimpB}}}

    \inferH{rec-let}{}{(\Hlet{\var}{\val}~{\exprimp}, \heap, \libImp) \mstep\silent_\thelangImpSub{} \set{(\exprimp[\val/\var], \heap, \libImp)}}

    \inferH{rec-var}{}{(\Hvar, \heap, \libImp) \mstep\silent_\thelangImpSub{} \emptyset}

    \inferH{rec-alloc}{\mathsf{heap\_alloc\_list}(\Repeat{n}, \Repeat{\loc}, \heapA, \heapB)}{(\Halloc{\Repeat{(y, n)}}\exprimp, \heapA, \libImp) \mstep\silent_\thelangImpSub{} \setComp{(\Hfree{\Repeat{(\loc, n)}}{(\exprimp[\Repeat{\loc}/\Repeat{y}])}, \heapB, \libImp)}{\All m \in \Repeat{n}. m > 0}}

    \inferH{rec-free}{}{(\Hfree{\Repeat{(\loc, n)}}{\val}, \heapA, \libImp) \mstep\silent_\thelangImpSub{} \setComp{(\val, \heapB, \libImp)}{\mathsf{heap\_free\_list}(\Repeat{(\loc, n)}, \heapA, \heapB)}}

    \inferH{rec-start}{\fnname \in \libImp}{(\WaitingImp{\boolval}, \heap, \libImp) \mstep{\ECallRecv{\fnname}{\vals}{\heap'}}_\thelangImpSub{} \set{(\RetExt{\boolval}{\Hcall{\fnname}{\vals}}, \imp{\heap'}, \libImp)}}

    \inferH{rec-call-internal}{(\Hfndef{\fnname}{\Repeat{x}}\Repeat{\Hlocal{y}{n}};\spac\exprimp) \in \libImp }{(\Hcall{\fnname}{\vals}, \heap, \libImp) \mstep{\silent}_{\thelangImpSub} \setComp{(\Halloc{\Repeat{(y, n)}}{(\exprimp[\vals/\Repeat{x}])}, \heap, \libImp)}{\lengthOf{\Repeat{x}} = \lengthOf{\vals}}}

    \inferH{rec-call-ub}{\val \notin \FnName}{(\Hcall{\val}{\vals}, \heap, \libImp) \mstep{\silent}_{\thelangImpSub} \emptyset}

    \inferH{rec-call-external}{\fnname \notin \libImp}{(\Hcall{\fnname}{\vals}, \heap, \libImp) \mstep{\ECallSend{\fnname}{\vals}{\heap}}_{\thelangImpSub} \set{(\WaitingImp{\trueImp}, \heap, \libImp)}}

    \inferH{rec-ret-incoming}{}{(\WaitingImp{\trueImp}, \heap, \libImp) \mstep{\EReturnRecv{\val}{\heap'}}_\thelangImpSub{} \set{(\val, \imp{\heap'}, \libImp)}}

    \inferH{rec-ret-return}{}{(\RetExt{\boolval}{\val}, \heap, \libImp) \mstep{\EReturnSend{\val}{\heap}}_\thelangImpSub{}\!\! \set{(\WaitingImp{\boolval}, \heap, \libImp)}}

    \hspace{-6.5pt}\inferH{rec-eval-ctx}
    {(\rexpr, \heap, \libImp) \mstep{\lbl}_\thelangImpSub{} \mstateP}
    {(\impctx[\rexpr], \heap, \libImp)
     \mstep{\lbl}_\thelangImpSub{}\!\!
     \setComp{(\impctx[\imp{\rexpr'}], \imp{\heap'}, \imp{\libImp'})}
     {(\imp{\rexpr'}, \imp{\heap'}, \imp{\libImp'}) \in \mstateP}
    }
  \end{mathpar}

  \caption{Operational semantics of \thelangRec{}, adapted from~\cite{DimSumAppendix}.}\label{fig:appendix:rec-sem}
\end{figure}

\subsection{Program Logic}

\autoref{fig:appendix:rec-logic} shows the program logic of \thelangRec{}, where $\maybelater[\ts_\lane] P \eqdef \mathit{if}~\ts_\lane = \tgtsub~\mathit{then}~\later P~\mathit{else}~P$, since only steps on the target side produce a later modality.
As mentioned in \ifappendixincluded\autoref{sec:rec-asm-pseudo}\else the paper\fi, the rules \ruleref{rec-wp-binop}, \ruleref{rec-wp-let}, \ruleref{rec-wp-if}, \ruleref{rec-wp-load}, and \ruleref{rec-wp-store} are familiar from any standard program logic. Since the corresponding steps involve no non-determinism, we can state them independently of whether we reason about the implementation or the specification.
For steps that do involve non-determinism, we provide separate rules for the implementation (suffix \rulenamestyle{-i}) and the specification (suffix \rulenamestyle{-s}).
As mentioned in \ifappendixincluded\autoref{sec:rec-asm-pseudo}\else the paper\fi, since \thelangRec{} uses a block-based memory, we use block points-tos $\prov \hptstoBlock \mathit{b}$, with blocks mapping offsets to values. Locations are pairs $(\prov, \mathit{o})$ of a block and an offset $\mathit{o}$. We write $0^{n}$ for the block mapping offsets $0, \ldots, n-1$ to $0$, and call $\prov \hptstoBlock \emptyset$ a deallocation token.
In \ruleref{rec-wp-alloc-i} and \ruleref{rec-wp-alloc-s}, when the body returns, it has to give back ownership of the blocks with their original domain (which the implementation has to prove and the specification gets to assume), and in turn the postcondition receives deallocation tokens $\prov_i \hptstoBlock \emptyset$.

The remaining rules \ruleref{rec-wp-call-external}, \ruleref{rec-wait-ret}, and \ruleref{rec-wait-call} are for reasoning about the interaction with the environment. In \ruleref{rec-wait-call}, we use the notation
$\switchinChoice{\lane}{\ECallH}{\fnname, \vals, \heap}{\Exists \mathimp{fn}. \fninternal{\fnname}{\mathimp{fn}}}$ to reflect the demonic nature of DimSum's interaction semantics. On the specification side, we have to provide $\fninternal{\fnname}{\mathimp{fn}}$, proving that $\fnname$ is an internal function; on the implementation side, we get to assume it. For the full details, see the Rocq development~\cite{Artifact}.

The waiting modality $\recwait P$ is defined as follows:
\begin{align*}
  \recwait[\lane] P \eqdef{}& \All \mathimp{fns}, \impctx, \heap.\spac \fninv(\mathimp{fns}) \wand \mstatefull{\gvar_{\lane}} \wand {}\\
  &\quad \big(P \wand \switchin{\lane}{\EReturnH}{\val, \heapS} \mstatefull{\gvar_{\lane}} \wand
    \simgenaux[\ts_{\lane}]{}{(\impctx[\val], \heapS, \mathimp{fns})}{\lane}\big) \wand {}\\
  &\quad \simgenaux[\ts_{\lane}]{}{(\impctx[\WaitingImp{\trueImp}], \heap, \mathimp{fns})}{\lane}
\end{align*}
Intuitively, $\recwait P$ states that the module is waiting for the environment: it can handle incoming calls (\ruleref{rec-wait-call}) or continue with $P$ (\ruleref{rec-wait-ret}), which has to handle the incoming return.

\begin{figure}
  \begin{mathpar}
    \inferH{rec-wp-binop}
    {\mathsf{eval}_{\binop}(\valA, \valB, \valC) \\ \maybelater[\ts_\lane]\post(\valC)}
    {\simgenexpr{\valA \binop \valB}{\lane}{\post}}

    \inferH{rec-wp-let}
    {\maybelater[\ts_\lane]\simgenexpr{\exprimp[\val/\var]}{\lane}{\post}}
    {\simgenexpr{\Hlet{\var}{\val}~\exprimp}{\lane}{\post}}

    \inferH{rec-wp-if}
    {\maybelater[\ts_\lane]\simgenexpr{\ite{\boolval}{\exprimpA}{\exprimpB}}{\lane}{\post}}
    {\simgenexpr{\Hifelse{\boolval}{\exprimpA}{\exprimpB}}{\lane}{\post}}

    \inferH{rec-wp-load}
    {\loc \hptsto \val \\ \loc \hptsto \val \wand \maybelater[\ts_\lane]\post(\val)}
    {\simgenexpr{\Hload{\loc}}{\lane}{\post}}

    \inferH{rec-wp-store}
    {\loc \hptsto \val \\ \loc \hptsto \valS \wand \maybelater[\ts_\lane]\post(\valS)}
    {\simgenexpr{\Hstore{\loc}{\valS}}{\lane}{\post}}

    \inferHS{rec-wp-alloc-i}
    {$\ts_\lane = \tgtsub$}
    {\All i.\spac n_i > 0 \\
     \All \Repeat{\prov}.\spac
       (\mathop{\textstyle\Sep}\nolimits_i \prov_i \hptstoBlock 0^{n_i}) \wand
       \later \simgenexpr{\exprimp[\Repeat{(\prov, 0)}/\Repeat{y}]}{\lane}{\begin{array}[c]{@{}l@{}}\val.\spac
         \Exists \Repeat{\mathit{b}}.\spac (\mathop{\textstyle\Sep}\nolimits_i \prov_i \hptstoBlock \mathit{b}_i \ast
         \dom(\mathit{b}_i) = \dom(0^{n_i})) \ast {} \\ \quad
         ((\mathop{\textstyle\Sep}\nolimits_i \prov_i \hptstoBlock \emptyset) \wand \post(\val))\end{array}}}
    {\simgenexpr{\Halloc{\Repeat{(y, n)}}{\exprimp}}{\lane}{\post}}

    \inferHS{rec-wp-alloc-s}
    {$\ts_\lane = \srcsub$}
    {\All \Repeat{\prov}.\spac
       (\mathop{\textstyle\Sep}\nolimits_i \prov_i \hptstoBlock 0^{n_i}) \wand
       \simgenexpr{\exprimp[\Repeat{(\prov, 0)}/\Repeat{y}]}{\lane}{\begin{array}[c]{@{}l@{}}\val.\spac
         \Exists \Repeat{\mathit{b}}.\spac (\mathop{\textstyle\Sep}\nolimits_i \prov_i \hptstoBlock \mathit{b}_i) \ast {} \\ \quad
         ((\mathop{\textstyle\Sep}\nolimits_i \dom(\mathit{b}_i) = \dom(0^{n_i}) \ast \prov_i \hptstoBlock \emptyset) \wand \post(\val))\end{array}}}
    {\simgenexpr{\Halloc{\Repeat{(y, n)}}{\exprimp}}{\lane}{\post}}

    \inferHS{rec-wp-call-internal-i}
    {$\ts_\lane = \tgtsub$}
    {\fninternal{\fnname}{(\Hfndef{\fnname}{\Repeat{x}}\Repeat{\Hlocal{y}{n}};\spac\exprimp)} \\
     \lengthOf{\vals} = \lengthOf{\Repeat{x}} \\
     \later \simgenexpr{\Halloc{\Repeat{(y, n)}}{(\exprimp[\vals/\Repeat{x}])}}{\lane}{\post}}
    {\simgenexpr{\Hcall{\fnname}{\vals}}{\lane}{\post}}

    \inferHS{rec-wp-call-internal-s}
    {$\ts_\lane = \srcsub$}
    {\fninternal{\fnname}{(\Hfndef{\fnname}{\Repeat{x}}\Repeat{\Hlocal{y}{n}};\spac\exprimp)} \\
     \lengthOf{\vals} = \lengthOf{\Repeat{x}} \wand \simgenexpr{\Halloc{\Repeat{(y, n)}}{(\exprimp[\vals/\Repeat{x}])}}{\lane}{\post}}
    {\simgenexpr{\Hcall{\fnname}{\vals}}{\lane}{\post}}

    {\mprset{sep=1em}
    \inferH{rec-wp-call-external}
    {\fnext{\fnname} \\
     \All \heap. \heapinv(\heap) \wand \maybelater[\ts_\lane]
      \switchoutS{\lane}{\ECallH(\fnname, \vals, \heap)}
      \recwait[\lane]
      \switchin{\lane}{\EReturnH}{\valS, \heapS}
      \heapinv(\heapS) \ast \post(\valS)}
    {\simgenexpr{\Hcall{\fnname}{\vals}}{\lane}{\post}}}

    \hspace{-10pt}\inferH{rec-wait-ret}
    {P}
    {\recwait[\lane] P}

    \inferH{rec-wait-call}
    {\switchinChoice{\lane}{\ECallH}{\fnname, \vals, \heap}{\Exists \mathimp{fn}. \fninternal{\fnname}{\mathimp{fn}}}
       \heapinv(\heap) \ast
       \simgenexpr{\Hcall{\fnname}{\vals}}{\lane}{\val.\spac
         \All \heapS. \heapinv(\heapS) \wand
         \switchoutS{\lane}{\EReturnH(\val, \heapS)}
         \recwait[\lane] P}
    }
    {\recwait[\lane] P}
  \end{mathpar}
  \caption{Program logic rules of \thelangRec{}.}\label{fig:appendix:rec-logic}
\end{figure}
\section{\thelangAsm{} Semantics and Program Logic}\label{sec:appendix:asm}

We reproduce the syntax and semantics of \thelangAsm{} from~\cite{DimSumAppendix} and give the full definition of its program logic in \thelogic{}.

\subsection{Syntax and Semantics}

The instructions $\instr$ of \thelangAsm{} are sequences of \emph{micro-instructions} (\autoref{fig:appendix:asm-grammar}):
$\storeRegAsm{\regMeta}{\regs}{\aval}\instr$ updates the register $\regMeta$ with the result of a function $\regs \mapsto \aval$ applied to the current register map.
$\readMemAsm{\regMetaA}{\regMetaB}{\aval}{\avalS}\instr$ applies $\aval \mapsto \avalS$ to the value of $\regMetaB$ to obtain an address and loads the value stored there into $\regMetaA$, analogously for $\storeMemAsm{\regMetaA}{\regMetaB}{\aval}{\avalS}\instr$.
$\syscallAsm\instr$ performs a syscall, and $\jumpAsm$ jumps to the address stored in $\pc$.
The assembly instructions used in the \ifappendixincluded main text\else paper\fi{} are derived from these micro-instructions:
\begin{mathpar}
  \Amov{\reg{}}{\aval} \eqdef \storeRegAsm{\reg{}}{\regs}{\aval}\nextAsm

  \Asyscall \eqdef \syscallAsm\nextAsm

  \Aret \eqdef \storeRegAsm{\pc}{\regs}{\regs(\reg{30})}\jumpAsm
\end{mathpar}
where $\nextAsm \eqdef \storeRegAsm{\pc}{\regs}{\rlookup{\regs}{\pc} + 1}\jumpAsm$.

\begin{figure}
  \begin{align*}
    \AsmHeap \ni \mem \eqdef{}& \mathbb{Z}\! \fpfn\! \AsmVal \cup \set{\unmapped} \quad
    \AsmRegs \ni \regs \eqdef{} \AsmRegName\! \to\! \AsmVal{} \quad
    \AsmRegName \ni \regMeta \eqdef{} \set{\reg{0}, \ldots, \reg{30}, \SP, \pc} \\
    \AsmVal \ni \aval \eqdef{}& \mathbb{Z} \\
    \Instr \ni \instr \eqdef{}& \syscallAsm\instr \bnfor \storeRegAsm{\reg{}}{\regs}{\aval}\instr \bnfor \readMemAsm{\regMetaA}{\regMetaB}{\aval}{\avalS}\instr \bnfor \storeMemAsm{\regMetaA}{\regMetaB}{\aval}{\avalS}\instr \bnfor \jumpAsm\\
    \ExecState \ni \rinstr \eqdef{}& \WaitingAsm \bnfor \RunningInstrAsm{\instr}{\regs}{\mem} \bnfor \WaitingSyscallInstrAsm{\instr}{\regs} \bnfor \HaltedAsm\\
    \LibraryAsm \ni \libAsm \eqdef{}& \mathbb{Z} \fpfn \Instr\\
    \AsmEvents \ni \asmevent{} \eqdef{}&
      \EJumpSend{\regs}{\mem} \bnfor \EJumpRecv{\regs}{\mem} \bnfor \ESyscallSend{\avals}{\mem} \bnfor \ESyscallRetRecv{\aval}{\mem}
  \end{align*}
  \caption[]{Grammar and events of~\thelangAsm{}, adapted from~\cite{DimSum, DimSumAppendix}.}\label{fig:appendix:asm-grammar}
\end{figure}

\autoref{fig:appendix:asm-sem} shows the small-step semantics of \thelangAsm{} modules.

\begin{figure}[h]
  \begin{mathpar}
    \inferH{asm-upd}{}
    {(\RunningInstrAsm{\storeRegAsm{\regMeta}{\regs}{f(\regs)}\instr}{\regs}{\mem}, \libAsm)
     \mstep{\silent}_\thelangAsmSub{}
     \set{(\RunningInstrAsm{\instr}{\rset{\regs}{\regMeta}{f(\regs)}}{\mem}, \libAsm)}}

    \inferH{asm-ldr}{\addrraw = f(\rlookup{\regs}{\regMetaB})}
    {(\RunningInstrAsm{\readMemAsm{\regMetaA}{\regMetaB}{\aval}{f(\aval)}\instr}{\regs}{\mem}, \libAsm)
     \mstep{\silent}_\thelangAsmSub{}\!\!
     {\begin{cases}
       \set{(\HaltedAsm, \libAsm)} & \mem(\addrraw) = \unmapped, \cr
       \setComp{(\RunningInstrAsm{\instr}{\rset{\regs}{\regMetaA}{\avalS}}{\mem}, \libAsm)}{\mem(\addrraw) = \avalS} & \text{otherwise}
     \end{cases}}}

    \inferH{asm-str}{\addrraw = f(\rlookup{\regs}{\regMetaB}) \\ \avalB = \rlookup{\regs}{\regMetaA}}
    {(\RunningInstrAsm{\storeMemAsm{\regMetaA}{\regMetaB}{\aval}{f(\aval)}\instr}{\regs}{\mem}, \libAsm)
     \mstep{\silent}_\thelangAsmSub{}\!\!
     {\begin{cases}
       \set{(\HaltedAsm, \libAsm)} & \mem(\addrraw) = \unmapped, \cr
       \setComp{(\RunningInstrAsm{\instr}{\regs}{\rset{\mem}{\addrraw}{\avalB}}, \libAsm)}{\addrraw \in \dom(\mem)} & \text{otherwise}
     \end{cases}}}

    \inferH{asm-syscall}{}
    {(\RunningInstrAsm{\syscallAsm\instr}{\regs}{\mem}, \libAsm)
     \mstep{\ESyscallSend{[\rlookup{\regs}{\reg{0}},\ldots,\rlookup{\regs}{\reg{8}}]}{\mem}}_\thelangAsmSub{}
     \set{(\WaitingSyscallInstrAsm{\instr}{\regs}, \libAsm)}}

    \inferH{asm-syscall-ret}{}
    {(\WaitingSyscallInstrAsm{\instr}{\regs}, \libAsm)
     \mstep{\ESyscallRetRecv{\aval}{\mem'}}_\thelangAsmSub{}
     \set{(\RunningInstrAsm{\instr}{\rset{\regs}{\reg{0}}{\aval}}{\mem'}, \libAsm)}}

    \inferH{asm-jump-internal}{\rlookup{\regs}{\pc} = \insaddr \\ \insaddr \in \asmdom\libAsm}
    {(\RunningInstrAsm{\jumpAsm}{\regs}{\mem}, \libAsm)
     \mstep{\silent}_\thelangAsmSub{}\!
     \set{(\RunningInstrAsm{\libAsm(\insaddr)}{\regs}{\mem}, \libAsm)}}

    \inferH{asm-jump-external}{\rlookup{\regs}{\pc} = \insaddr \\ \insaddr \notin \asmdom\libAsm}
    {(\RunningInstrAsm{\jumpAsm}{\regs}{\mem}, \libAsm)
     \mstep{\EJumpSend{\regs}{\mem}}_\thelangAsmSub{}\!
     \set{(\WaitingAsm, \libAsm)}}

    \inferH{asm-incoming}{\rlookup{\regs}{\pc} = \insaddr \\ \libAsm(\insaddr) = \instr}
    {(\WaitingAsm, \libAsm)
     \mstep{\EJumpRecv{\regs}{\mem}}_\thelangAsmSub{}\!
     \set{(\RunningInstrAsm{\instr}{\regs}{\mem}, \libAsm)}}
  \end{mathpar}
  \caption{Operational semantics of \thelangAsm{}.}\label{fig:appendix:asm-sem}
\end{figure}

\subsection{Program Logic}

The program logic rules of \thelangAsm{} in \autoref{fig:appendix:asm-wp-rules} are unsurprising.
The rules use the instruction assertions $\insinternal{\insaddr}{\instr}$, stating that address $\insaddr$ contains the instruction $\instr$, and $\insext{\insaddr}$, stating that $\insaddr$ is not part of the \thelangAsm{} program; the assertion $\inssinternal{\insaddr}{\instrs}$ used in \ifappendixincluded\autoref{sec:rec-asm-pseudo}\else the paper\fi{} is shorthand for $\mathop{\textstyle\Sep}\nolimits_k \insinternal{\insaddr + k}{\instr_k}$.
\ruleref{asm-wp-ldr} and \ruleref{asm-wp-str} require the points-tos of the involved registers and of the accessed memory location, and return the updated assertions.
\ruleref{asm-wp-upd} looks more involved, since $f$ may read arbitrary registers: we have to provide the points-tos of the registers $\Repeat{\regMetaB}$ that $f$ reads. The remaining rules handle jumps and syscalls. Note that \ruleref{asm-wp-syscall} does not enter the waiting state $\asmwait$, since a module waiting for a syscall only accepts the syscall return. The waiting assertion $\asmwait$ is defined similarly to the waiting modality $\recwaitraw$ of \thelangRec{}, without the continuation $P$, since \thelangAsm{} does not have well-bracketed calls.
The \absjump $\bijump$ is defined as follows: $\bijump \eqdef \simgenexprasm{\jumpAsm}{\PiAsm}$.

\begin{figure}[h]
  \begin{mathpar}
    \inferH{asm-wp-upd}
    {(\mathop{\textstyle\Sep}\nolimits_{(\regMeta, \asm{w}) \in (\Repeat{\regMetaB}, \Repeat{\avalB})} \regMeta \rptsto \asm{w}) \land
     (\regMetaA \rptsto \aval \ast \maybelater[\ts_\lane] (\All \regs. \regs(\Repeat{\regMetaB}) = \Repeat{\avalB} \wand \regMetaA \rptsto f(\regs) \wand \simgenexprasm{\instr}{\lane}))}
    {\simgenexprasm{\storeRegAsm{\regMetaA}{\regs}{f(\regs)}\instr}{\lane}}

    {\mprset{sep=1.35em}
    \inferH{asm-wp-ldr}
    {\regMetaA \rptsto \avalA \\ \regMetaB \rptsto \avalB \\ f(\avalB) \mptsto \avalS \\
     \maybelater[\ts_\lane] (\regMetaA \rptsto \avalS \wand \regMetaB \rptsto \avalB \wand f(\avalB) \mptsto \avalS \wand \simgenexprasm{\instr}{\lane})}
    {\simgenexprasm{\readMemAsm{\regMetaA}{\regMetaB}{\aval}{f(\aval)}\instr}{\lane}}
    }

    {\mprset{sep=1.35em}
    \inferH{asm-wp-str}
    {\regMetaA \rptsto \avalA \\ \regMetaB \rptsto \avalB \\ f(\avalB) \mptsto \avalS \\
     \maybelater[\ts_\lane] (\regMetaA \rptsto \avalA \wand \regMetaB \rptsto \avalB \wand f(\avalB) \mptsto \avalA \wand \simgenexprasm{\instr}{\lane})}
    {\simgenexprasm{\storeMemAsm{\regMetaA}{\regMetaB}{\aval}{f(\aval)}\instr}{\lane}}
    }

    \inferH{asm-wp-jump-internal}
    {\pc \rptsto \insaddr \\ \insinternal{\insaddr}{\instr} \\
     \maybelater[\ts_\lane] (\pc \rptsto \insaddr \wand \simgenexprasm{\instr}{\lane})}
    {\simgenexprasm{\jumpAsm}{\lane}}

    {\mprset{sep=1em}
    \inferH{asm-wp-syscall}
    {{\begin{array}[b]{@{}l@{}}
       (\mathop{\textstyle\Sep}\nolimits_{(\regMeta, \asm{w}) \in (\Repeat{\regMetaB}, \Repeat{\avalB})} \regMeta \rptsto \asm{w}) \land
       \maybelater[\ts_\lane] (\All \regs, \mem. \regs(\Repeat{\regMetaB}) = \Repeat{\avalB} \wand \meminv(\mem) \wand
       \switchoutS{\lane}{\ESyscallH([\rlookup{\regs}{\reg{0}},\ldots,\rlookup{\regs}{\reg{8}}], \mem)} \cr
       \qquad\ (\Exists \avalA. \reg{0} \rptsto \avalA) \ast
       \All \aval, \mem'. \reg{0} \rptsto \aval \wand
       \switchinS{\lane}{\ESyscallRetH(\aval, \mem')}
       \meminv(\mem') \ast \simgenexprasm{\instr}{\lane})
     \end{array}}}
    {\simgenexprasm{\syscallAsm\instr}{\lane}}}

    \inferH{asm-wp-jump-external}
    {\pc \rptsto \insaddr \\ \insext{\insaddr} \\
     \All \mem, \regs. \meminv(\mem) \wand \reginv(\regs) \wand \pc \rptsto \insaddr \wand
       \maybelater[\ts_\lane] \switchoutS{\lane}{\EJumpH(\regs, \mem)} \asmwait[\lane]}
    {\simgenexprasm{\jumpAsm}{\lane}}

    \inferH{asm-wait-jump}
    {\switchinChoice{\lane}{\EJumpH}{\regs, \mem}{\Exists \instr. \insinternal{\rlookup{\regs}{\pc}}{\instr}}
     \meminv(\mem) \ast \reginv(\regs) \ast \simgenexprasm{\instr}{\lane}}
    {\asmwait[\lane]}
  \end{mathpar}
  \caption{Program logic rules of \thelangAsm{}.}\label{fig:appendix:asm-wp-rules}
\end{figure}

\section{\thelangSpec{} Semantics and Program Logic}\label{sec:appendix:pseudo}

\subsection{Syntax and Semantics}

\newcommand{\CmdsPseudo}{\textlog{Cmd}}
\newcommand{\cmdPseudo}{\textlog{c}}
\newcommand{\cmdPseudoA}{\textlog{c}_1}
\newcommand{\cmdPseudoB}{\textlog{c}_2}
\newcommand{\selPseudo}[1]{\textlog{sel}_{#1}}
\newcommand{\implPseudo}[1]{\textlog{cmds}_{#1}}
\newcommand{\stateinv}{\textlog{inv}_{S}}
\newcommand{\heapinvS}{\textlog{inv}^{S}_{\textlog{h}}}
\thelangSpec{} is a shallowly embedded language based on interaction trees (ITrees)~\cite{ITrees}.
A \thelangSpec{} program is an ITree $t$ whose effects are visible events, angelic and demonic choice, and access to a module state $s$.
\autoref{fig:appendix:pseudo-syntax} shows the commands $\CmdsPseudo_{E,S}(R)$, parameterized by the events $E$, the module state $S$, and the result type $R$.

\begin{figure}
  \begin{align*}
    \CmdsPseudo_{E,S}(R) \ni \textlog{c} \eqdef{}& \Pret{(\valSpec : R)} \bnfor x \leftarrow \cmdPseudoA; \cmdPseudoB \bnfor \Pcmd{vis}(\event : \taggedevents) \bnfor \Pcmd{angelic}(X) \bnfor \Pcmd{demonic}(X) \\ &\!\!\!\bnfor \Pcmd{get} \bnfor \Pcmd{set}(s : S) \\
    \LibrarySpec_{E,S} \ni \speclib \eqdef{}& (\selPseudo{\speclib} : E \pfn A,\ \implPseudo{\speclib} : A \to \CmdsPseudo_{E,S}(E))
  \end{align*}
  \caption{Syntax of \thelangSpec{}.}\label{fig:appendix:pseudo-syntax}
\end{figure}

From the basic commands, we derive $\Pcmd{assume}~\pprop \eqdef{} \_ \leftarrow \Pcmd{angelic}(\pprop); \Pret{()}$, which encodes angelically \emph{assuming} a proposition $\pprop$, and $\Pcmd{assert}~\pprop \eqdef{} \_ \leftarrow \Pcmd{demonic}(\pprop); \Pret{()}$, the dual notion of demonically \emph{asserting} $\pprop$.
Further, we derive the commands $\Pcmd{wait}_{\speclib} : \CmdsPseudo_{E,S}(E)$ and $\Pcmd{call}_{\speclib} : E \to \CmdsPseudo_{E,S}(E)$, both returning the next incoming event, which give \thelangSpec{} a notion of functions and waiting states, like the other DimSum modules. The $\Ploop{\cmdPseudo}$ command is defined by (co-)recursion.

\begin{figure}
  \begin{mathpar}
    \inferH{pseudo-angelic}{}
    {(x \leftarrow \Pcmd{angelic}(X); t, s) \mstep{\silent}_{\thelangSpecSub} \setComp{(t[x'/x], s)}{x' \in X}}

    \inferH{pseudo-demonic}{x' \in X}
    {(x \leftarrow \Pcmd{demonic}(X); t, s) \mstep{\silent}_{\thelangSpecSub} \set{(t[x'/x], s)}}

    \inferH{pseudo-vis}{}
    {(\Pcmd{vis}(\event); t, s) \mstep{\event}_{\thelangSpecSub} \set{(t, s)}}

    \inferH{pseudo-get}{}
    {(x \leftarrow \Pcmd{get}; t, s) \mstep{\silent}_{\thelangSpecSub} \set{(t[s/x], s)}}

    \inferH{pseudo-set}{}
    {(\Pcmd{set}(s'); t, s) \mstep{\silent}_{\thelangSpecSub} \set{(t, s')}}

  \end{mathpar}

  \begin{mathpar}
    \inferH{pseudo-wait-incoming}{}
    {(x \leftarrow \Pcmd{wait}_{\speclib}; t, s) \mstep{?\lbl}_{\thelangSpecSub}\!
        \set{(t[\lbl/x], s)} \cup
        \setComp{(\lbl' \leftarrow \implPseudo{\speclib}(a);~ \Pcmd{vis}(!\lbl');~ x \leftarrow \Pcmd{wait}_{\speclib}; t, s)}{\selPseudo{\speclib}(\lbl) = a}}

    \inferH{pseudo-call-internal}{\selPseudo{\speclib}(\lbl) = a}
    {(x \leftarrow \Pcmd{call}_{\speclib}(\lbl); t, s) \mstep{\silent}_{\thelangSpecSub} \set{(x \leftarrow \implPseudo{\speclib}(a); t, s)}}

    \inferH{pseudo-call-external}{\lbl \notin \dom(\selPseudo{\speclib})}
    {(x \leftarrow \Pcmd{call}_{\speclib}(\lbl); t, s) \mstep{!\lbl}_{\thelangSpecSub} \set{(x \leftarrow \Pcmd{wait}_{\speclib}; t, s)}}
  \end{mathpar}
  \caption[]{Operational semantics of \thelangSpec{}.}\label{fig:appendix:pseudo-sem}
\end{figure}

\autoref{fig:appendix:pseudo-sem} shows the operational semantics of \thelangSpec{}, \ie the interpretation of ITrees into modules, in particular the effects for non-determinism (\ruleref{pseudo-angelic} and \ruleref{pseudo-demonic}) and for accessing state (\ruleref{pseudo-get} and \ruleref{pseudo-set}). For interacting with the environment, \ruleref{pseudo-wait-incoming} receives an incoming event $\lbl$ and angelically either passes it to the continuation or, handles it with $\implPseudo{\speclib}$, emits the resulting event, and returns back to waiting.
\ruleref{pseudo-call-internal} and \ruleref{pseudo-call-external} are derived rules for the $\Pcmd{call}$ construct.

\subsection{Program Logic}

\autoref{fig:appendix:pseudo-wp-rules} shows the program logic of \thelangSpec{}.
\ruleref{pseudo-wp-get} and \ruleref{pseudo-wp-set} access the module state via the state invariant $\stateinv(s)$.
The rules for $\Pcmd{angelic}$ and $\Pcmd{demonic}$ show the duality between implementation and specification: on the implementation side, we get to pick an angelic value and have to handle all demonic ones; on the specification side, it is the other way around. The rules for interacting with the environment are \ruleref{pseudo-wp-call-external}, \ruleref{pseudo-wait-i}, and \ruleref{pseudo-wait-s}.

The waiting modality is the weakest precondition of $\Pcmd{wait}_{\speclib}$, with a postcondition $\post$ on the received event: $\specwait[\lane] \post \eqdef \simgenexpr{\Pcmd{wait}_{\speclib}}{\lane}{\post}$.

\begin{figure}[h]
  \begin{mathpar}
    \inferHS{pseudo-wp-angelic-i}{$\ts_\lane = \tgtsub$}
    {\Exists x \in X.\spac \later \post(x)}
    {\simgenexpr{\Pcmd{angelic}(X)}{\lane}{\post}}

    \inferHS{pseudo-wp-angelic-s}{$\ts_\lane = \srcsub$}
    {\All x \in X.\spac \post(x)}
    {\simgenexpr{\Pcmd{angelic}(X)}{\lane}{\post}}

    \inferHS{pseudo-wp-demonic-i}{$\ts_\lane = \tgtsub$}
    {\All x \in X.\spac \later \post(x)}
    {\simgenexpr{\Pcmd{demonic}(X)}{\lane}{\post}}

    \inferHS{pseudo-wp-demonic-s}{$\ts_\lane = \srcsub$}
    {\Exists x \in X.\spac \post(x)}
    {\simgenexpr{\Pcmd{demonic}(X)}{\lane}{\post}}

    \inferH{pseudo-wp-get}
    {\stateinv(s) \\ \stateinv(s) \wand \post(s)}
    {\simgenexpr{\Pcmd{get}}{\lane}{\post}}

    \inferH{pseudo-wp-set}
    {\stateinv(s) \\ \stateinv(s') \wand \post(())}
    {\simgenexpr{\Pcmd{set}(s')}{\lane}{\post}}

    \inferH{pseudo-wp-call-internal}
    {\selPseudo{\speclib}(\lbl) = a \\ \simgenexpr{\implPseudo{\speclib}(a)}{\lane}{\post}}
    {\simgenexpr{\Pcmd{call}_{\speclib}(\lbl)}{\lane}{\post}}

    \inferH{pseudo-wp-call-external}
    {\lbl \notin \dom(\selPseudo{\speclib}) \\
     \switchoutS{\lane}{\lbl}
     \specwait[\lane] \post}
    {\simgenexpr{\Pcmd{call}_{\speclib}(\lbl)}{\lane}{\post}}

    \inferHS{pseudo-wait-i}{$\ts_\lane = \tgtsub$}
    {\later \switchin{\lane}{}{\lbl}
      \big(\post(\lbl) \lor \Exists a.\spac \selPseudo{\speclib}(\lbl) = a \ast
        \simgenexpr{\implPseudo{\speclib}(a)}{\lane}{\lbl'.\spac \switchoutS{\lane}{\lbl'} \specwait[\lane] \post}\big)}
    {\specwait[\lane] \post}

    \inferHS{pseudo-wait-s}{$\ts_\lane = \srcsub$}
    {\switchin{\lane}{}{\lbl}
      \big(\post(\lbl) \land \All a.\spac \selPseudo{\speclib}(\lbl) = a \wand
        \simgenexpr{\implPseudo{\speclib}(a)}{\lane}{\lbl'.\spac \switchoutS{\lane}{\lbl'} \specwait[\lane] \post}\big)}
    {\specwait[\lane] \post}
  \end{mathpar}
  \caption{Program logic rules of \thelangSpec{}.}\label{fig:appendix:pseudo-wp-rules}
\end{figure}

\subsection{\thelangSpec{} for \thelangRec}

\autoref{fig:appendix:pseudo-rec-lib} show how we instantiate the \thelangSpec{} interface for specifying \thelangRec{} programs that we used throughout the \ifappendixincluded main text\else paper\fi.
We derive \thelangRec{}-like commands: $\PcallN{\fnname}{\vals}$ passes the current heap to the call and installs the returned heap; $\Pcmd{alloc}$, $\Pload{\loc}$, and $\Pstore{\loc}{\val}$ operate on the heap in the module state; $\Pcmd{scramble}$ demonically picks a new heap; and $\Pvalid{\loc}$ asserts that $\loc$ is allocated. The last two commands are useful for writing specifications and from them we derive $\Pexists{\Pvalid{\loc}}$, which first scrambles the heap and demonically picks a valid location $\loc$.
We define \abscalls: $\speccall{\fnname}{\vals}{\val}{\post(\val)} \eqdef \simgenexpr{\PcallN{\fnname}{\vals}}{\PiSpec}{\post}$.

\begin{figure}
  \begin{align*}
    E \eqdef{}& \ImpEvents \qquad S \eqdef{} \ImpHeap \\
    \selPseudo{\speclib}(\ECallH(\fnname, \vals, \heap)) \eqdef{}& (\fnname, \vals, \heap) \qquad \text{if } \fnname \in \dom(c) \\
    \implPseudo{\speclib}(\fnname, \vals, \heap) \eqdef{}&
      \Pcmd{set}(\heap);~ \val \leftarrow c(\fnname)(\vals);~ \heap' \leftarrow \Pcmd{get};~
      \Pret{\EReturnH(\val, \heap')} \\[1ex]
    \PcallN{\fnname}{\vals} \eqdef{}& \heap \leftarrow \Pcmd{get};~
      \EReturnH(\val, \heap') \leftarrow \Pcmd{call}_{\speclib}(\ECallH(\fnname, \vals, \heap));~
      \Pcmd{set}(\heap');~ \Pret{\val} \\
    \Pcmd{alloc}(n) \eqdef{}& \heap \leftarrow \Pcmd{get};~ \Plet{\loc}{\mathsf{heap\_fresh}(\heap)}~
      \Pcmd{set}(\mathsf{heap\_alloc}(\heap, \loc, n));~ \Pret{\loc} \\
    \Pload{\loc} \eqdef{}& \heap \leftarrow \Pcmd{get};~ \Pcmd{assume}~(\loc \in \dom(\heap));~ \Pret{\heap(\loc)} \\
    \Pstore{\loc}{\val} \eqdef{}& \heap \leftarrow \Pcmd{get};~ \Pcmd{assume}~(\loc \in \dom(\heap));~
      \Pcmd{set}(\heap[\loc \mapsto \val]);~ \Pret{\val} \\
    \Pcmd{scramble} \eqdef{}& \heap \leftarrow \Pcmd{demonic}(\ImpHeap);~ \Pcmd{set}(\heap) \\
    \Pvalid{\loc} \eqdef{}& \heap \leftarrow \Pcmd{get};~ \Pcmd{assert}~(\loc \in \dom(\heap)) \\
    \Pexists{\Pvalid{\loc}} \eqdef{}& \Pcmd{scramble};~ \loc \leftarrow \Pcmd{demonic}(\Loc);~ \Pvalid{\loc}
  \end{align*}
  \caption{\thelangSpec{} library for \thelangImp{} built from $c : \FnName \fpfn (\ValImp^{*} \to \CmdsPseudo_{E,S}(\ValImp))$, and derived \thelangImp{} commands.}\label{fig:appendix:pseudo-rec-lib}
\end{figure}

\autoref{fig:appendix:pseudo-rec-wp-rules} shows the resulting program logic rules.
For simplicity, we take the module state to be the heap; the development allows the heap to be a part of a larger module state, accessed via a lens-like interface.
We define a heap invariant $\heapinvS \eqdef \Exists \heap. \stateinv(\heap) \ast \heapinv(\heap)$ tying the ghost state of the \thelangSpec{} module to the heap's ghost state.
With it, \ruleref{pseudo-rec-wp-load}, \ruleref{pseudo-rec-wp-store}, and \ruleref{pseudo-rec-wp-alloc} mirror the corresponding \thelangRec{} rules.
\ruleref{pseudo-rec-wp-scramble-s} allows us to realign the heaps, and \ruleref{pseudo-rec-wp-valid-s} can be verified by providing the relevant points-to. The call rules (\ruleref{pseudo-rec-wp-call-internal}, \ruleref{pseudo-rec-wp-call-external}, \ruleref{pseudo-rec-wait-i}, and \ruleref{pseudo-rec-wait-s}) are derived from the generic rules of \thelangSpec{}, using a waiting modality specialized to \thelangRec{}: its postcondition $\post$ is on the value and heap of the incoming return, and it owns the state invariant $\stateinv(s)$.

\begin{figure}
  \begin{mathpar}
    \inferH{pseudo-rec-wp-load}
    {\heapinvS \\ \loc \hptsto \val \\ \heapinvS \wand \loc \hptsto \val \wand \post(\val)}
    {\simgenexpr{\Pload{\loc}}{\lane}{\post}}

    \inferH{pseudo-rec-wp-store}
    {\heapinvS \\ \loc \hptsto \val \\ \heapinvS \wand \loc \hptsto \valS \wand \post(\valS)}
    {\simgenexpr{\Pstore{\loc}{\valS}}{\lane}{\post}}

    \inferH{pseudo-rec-wp-alloc}
    {\heapinvS \\ \All \prov. \heapinvS \wand \prov \hptstoBlock 0^n \wand \post((\prov, 0))}
    {\simgenexpr{\Pcmd{alloc}(n)}{\lane}{\post}}

    \inferHS{pseudo-rec-wp-scramble-s}{$\ts_\lane = \srcsub$}
    {\stateinv(\heap_0) \\ \heapinv(\heap) \\ \heapinvS \wand \post(())}
    {\simgenexpr{\Pcmd{scramble}}{\lane}{\post}}

    \inferHS{pseudo-rec-wp-valid-s}{$\ts_\lane = \srcsub$}
    {\heapinvS \\ \loc \hptsto \val \\ \heapinvS \wand \loc \hptsto \val \wand \post(())}
    {\simgenexpr{\Pvalid{\loc}}{\lane}{\post}}

    \inferH{pseudo-rec-wp-call-internal}
    {\fninternalSpec{\fnname}{\mathsf{fn}} \\ \stateinv(\heap) \\
     \stateinv(\heap) \wand \simgenexpr{\mathsf{fn}(\vals)}{\lane}{\val.\spac \Exists \heapS. \stateinv(\heapS) \ast (\stateinv(\heapS) \wand \post(\val))}}
    {\simgenexpr{\PcallN{\fnname}{\vals}}{\lane}{\post}}

    \inferH{pseudo-rec-wp-call-external}
    {\fnextSpec{\fnname} \\ \stateinv(\heap) \\
     \switchoutS{\lane}{\ECallH(\fnname, \vals, \heap)}
     \specwait[\lane] (\valS, \heapS.\spac \stateinv(\heapS) \wand \post(\valS))}
    {\simgenexpr{\PcallN{\fnname}{\vals}}{\lane}{\post}}

    \inferHS{pseudo-rec-wait-i}{$\ts_\lane = \tgtsub$}
    {\later \switchin{\lane}{}{\lbl}
     \left(\begin{array}{@{}l@{}}
       (\Exists \val, \heap.\spac \lbl = \EReturnH(\val, \heap) \ast \post(\val, \heap)) \lor {}\cr
       (\Exists \fnname, \vals, \heap, \mathsf{fn}.\spac \lbl = \ECallH(\fnname, \vals, \heap) \ast \fninternalSpec{\fnname}{\mathsf{fn}} \ast {}\cr
       \quad \stateinv(\heap) \wand
       \simgenexpr{\mathsf{fn}(\vals)}{\lane}{\val.\spac
         \Exists \heapS. \stateinv(\heapS) \ast
         \switchoutS{\lane}{\EReturnH(\val, \heapS)}
         \specwait[\lane] \post})
     \end{array}\right)}
    {\specwait[\lane] \post}

    \inferHS{pseudo-rec-wait-s}{$\ts_\lane = \srcsub$}
    {\switchin{\lane}{}{\lbl}
     \left(\begin{array}{@{}l@{}}
       (\All \val, \heap.\spac \lbl = \EReturnH(\val, \heap) \wand \post(\val, \heap)) \land {}\cr
       (\All \fnname, \vals, \heap, \mathsf{fn}.\spac \lbl = \ECallH(\fnname, \vals, \heap) \wand \fninternalSpec{\fnname}{\mathsf{fn}} \wand {}\cr
       \quad \stateinv(\heap) \wand
       \simgenexpr{\mathsf{fn}(\vals)}{\lane}{\val.\spac
         \Exists \heapS. \stateinv(\heapS) \ast
         \switchoutS{\lane}{\EReturnH(\val, \heapS)}
         \specwait[\lane] \post})
     \end{array}\right)}
    {\specwait[\lane] \post}
  \end{mathpar}
  \caption[]{Program logic rules of \thelangSpec{} for \thelangImp{}.} \label{fig:appendix:pseudo-rec-wp-rules}
\end{figure}
 \fi

\end{document}